\documentclass[12pt,titlepage]{report}
\usepackage[margin=1in]{geometry}
\usepackage{setspace} 
\usepackage{bm}
\usepackage{mathptmx} 

\usepackage{amsmath}
\usepackage{amssymb}

\usepackage{graphicx}
\graphicspath{{FIGS//}}

\usepackage[square,numbers,sort&compress]{natbib} 

\begin{document}
\begin{center}
\pagenumbering{roman}
\thispagestyle{empty}
{\Huge \bf ASPECTS OF SYMMETRIES  
\vskip 0.3cm
IN}
\vskip 0.3cm
{\Huge \bf FIELD AND STRING THEORIES}
\vskip 6.0 true cm
{\bf THESIS SUBMITTED FOR THE DEGREE OF \\
DOCTOR OF PHILOSOPHY (SCIENCE) \\
OF THE \\
UNIVERSITY OF CALCUTTA}
\vskip 5.0 true cm
 {\large \bf ANIRBAN SAHA}\\
DEPARTMENT OF PHYSICS, \\
UNIVERSITY OF CALCUTTA, \\
INDIA \\
2008 
\end{center}

\newpage
\thispagestyle{empty}
\vskip 10.0cm
\begin{center}
{\Huge \bf  Dedicated}
\vskip 0.5 cm
{\Huge \bf to }
\vskip 0.5cm
{ \Huge \bf   My Parents}
\vskip 0.5 cm
\end{center}
\newpage
\vskip 2.0cm
\begin{center}
{\large \bf ACKNOWLEDGEMENTS}
\end{center}

With great pleasure, I express my deep sense of gratitude to Dr.~Pradip~Mukherjee for his invaluable guidance and constant encouragement throughout the course of my Ph.D work. It is also my duty and joy to thank the family of Dr. Mukherjee for their hospitality during my visits for professional discussions. 

During a significant portion of time of my Ph. D research I was a CSIR scholar (JRF 2 years, SRF 1 year) in the Department of Physics, Presidency College, Kolkata. The work presented in this dissertation was supported in part by grants from the Council of Scientific and Industrial Research (CSIR), Government of India. I am grateful to Prof.~A.~Chatterjee and Prof.~M.~Ray, ex-Principals of Presidency College for allowing me to carry out my research there. In completing parts of this project I have used freely the library and computer facility of Satyendra Nath Bose National Centre for Basic Sciences (SNBNCBS). I would like to thank Prof.~S.~Dattagupta, the the-then Director of the center. I am also thankful to the Director, Inter University Centre for Astronomy and Astrophysics (IUCAA), Pune for awarding me a visiting associateship for the period 2007-2010. The centralised facilities and excellent academic atmosphere of IUCAA have been a great opportunity during my research work there. 

I acknowledge with sincere gratitude, the invaluable advice I received from Prof.~Rabin~Banerjee and am grateful to him for always being there to help me in my work. I also thank Dr.~Biswajit~Chakraborty for his help during the initial days of this project. It is my pleasure to thank my friends Dr.~K.~Kumar, Dr.~S.~Gangopadhyay, Mr.~A.~Ghosh~Hazra, Mr.~S.~Samanta and Mr.~S.~Kulkarni from SNBNCBS, for their help and support made my research an experience I cherish much.

Finally and most importantly, I express my whole hearted gratitude to my family members whose support and patience enabled me to pursue the studies which finally culminated in this thesis. I dedicate this thesis to them.
\vskip 1.0cm 
\noindent Date: \hskip 9.0cm Anirban Saha \\
Place: 
\newpage
\pagestyle{myheadings}
\markright{List of publications on which this thesis is based}
\vskip 2.0cm
{\large \bf List of publications on which this thesis is based:}
\begin{enumerate}
\item
 {\it Interpolating action for strings and membranes: A Study of symmetries in the constrained Hamiltonian approach}, R.~Banerjee, P.~Mukherjee, A.~Saha, Phys. Rev. {\bf D 70} 026006, 2004, [hep-th/0403065].
\item
{\it Bosonic $p$-brane and A-D-M decomposition}, R.~Banerjee, P.~Mukherjee, A.~Saha, Phys. Rev. {\bf D 72} 066015, 2005, [hep-th/0501030].
\item
{\it Noncommutativity in interpolating string: A study of gauge symmetries in noncommutative framework}, S.~Gangopadhyay, A.~Ghosh Hazra, A.~Saha, Phys. Rev. {\bf D 74} 125023, 2006, [hep-th/0701012].
\item
{\it Gauge invariances vis-a-vis diffeomorphisms in second order metric gravity: A new Hamiltonian approach}, P.~Mukherjee, A.~Saha, [hep-th/0705.4358], Communicated to journal.
\item
 {\it A New approach to the analysis of a noncommutative Chern-Simons theory},  P.~Mukherjee, A.~Saha,  Mod.Phys.Lett.{\bf{A 21}} (2006) 821, [hep-th/0409248].  
\item
{\it On the question of regular solitons in a noncommutative Maxwell-Chern-Simons-Higgs model}, P.~Mukherjee, A.~Saha,  Mod.Phys.Lett.{\bf A 22} 1113 (2007) , [hep-th/0605123].
\item
{\it Schwinger model in noncommutating space-time},  A.~Saha, A.~Rahaman, P.~Mukherjee; Phys. Lett. {\bf B 638} 292, 2006, [hep-th/0603050], Erratum-{\it ibid.} {\bf B 643} (2006) 383.
\item
{\it A Note on the noncommutative correction to gravity}, P.~Mukherjee, A.~Saha, Phys. Rev. {\bf D 74} 027702, 2006, [hep-th/0605287]. 
\item
{\it Seiberg-Witten map and Galilean symmetry violation in a non-commutative planar system}, B.~Chakraborty, S.~Gangopadhyay, A.~Saha, Phys. Rev. {\bf D 70} 107707 (2004), [hep-th/0312292].
\item
{\it Time-space noncommutativity in gravitational quantum well scenario}, A.~Saha, Eur. Phys. J. {\bf C 51} 199-205, 2007, [hep-th/0609195].
\end{enumerate}

\newpage
\pagestyle{myheadings}
\markright{Content}

\tableofcontents

\clearpage
\label{sec:intro}
\chapter{Introduction and Overview}
\label{intro}
\pagestyle{myheadings}
\markright{Introduction and Overview}
\pagenumbering{arabic}

Symmetries have always played a pivotal role in classifying and understanding the basic structures of physical theories. Our present understanding of the fundamental interactions of nature except gravity is based on the theory of quantized fields, i.e. Quantum field theories (QFT). It is also more or less apparent that for a successful inclusion of gravity in a unified theory the notion of the point particle must be replaced by some extended structure like string \cite{pol}. The string theory is, however in the formative stage. In both field and string theories symmetry considerations provide the basic method of modeling the interactions.
Apart from the space-time symmetries these theories are endowed with internal symmetries. Internal symmetry is realised by certain continuous transformations of the fields which leave the Lagrangean invariant. These  transformations are called global gauge transformations (GT) which do not involve space-time. Since the gauge transformation parameter (GTP) is constant, the GT must be the same at all points of space-time. So when the GT is performed at one point of space-time, it has to be performed at all other space points at the same time. But this contradicts the spirit of relativity. To get around this conceptual difficulty, one abandons the requirement that the GTP is constant and considers it as an arbitrary function of space-time. This makes the GT vary from point to point and the idea of `local' gauge transformation is introduced. 

Localisation of the gauge symmetries dictates the presence of gauge fields, which, in turn, mediate the interactions. Gauge symmetries have become the central theme in the investigations of fundamental interactions of nature. Accordingly, manifestations and impacts of gauge symmetries in various contexts of current interest will be the focal point of the present thesis. Notwithstanding this emphasis, we also explore non-relativistic space-time invariances in a $2+1$-dimensional field theoretic model. 

  We begin with the topic of diffeomorphism invariance and their connection with gauge invariances for some generally covariant theories, {\it viz.} some fundamental stringy models and second order metric gravity \cite{bms1, bms2, sgaghas, diff}. As is well known there are different formulations for the action of strings and higher branes. The direct generalisation from the relativistic particle is the Nambu--Goto (N--G) string which poses difficulty in quantisation due to the presence of derivatives under the square root. An alternative redundant description which is analytically easier to handle is the Polyakov string where extra degrees of freedoms are introduced in terms of the world-sheet metric. It is easy to show that the Polyakov model can be reduced to the N--G model by solving out these extra degrees of freedom by using their equations of motion. However, the opposite connection i.e. the transition from N--G to Polyakov model is not adequately explored. The same comments apply in case of the membrane or in general higher branes. Using the diffeomorphism invariance of N--G action we have constructed an interpolating Lagrangean \cite{rbbcsg} which provides a proper platform to investigate this transition from N--G action to Polyakov action for membrane and generic $p$-brane. We have found that the diffeomorphism invariance of N--G action is instrumental to the construction of the induced metric of the Polyakov action. Interestingly, it turns out that the metric automatically anticipates the famous Arnowitt--Deser--Misner (A--D--M) decomposition of General relativity \cite{adm}. Naturally a detailed analysis of the reparametrisation symmetry and their interconnections with the gauge symmetry have also been presented using a purely Hamiltonian algorithm \cite{brr}.  
  
  Another interesting issue in case of the open string in presence of a background Neveu--Schwarz two-form field is the emergence of noncommutative structure in the space-time coordinates of D-branes where the string end-points are attached \cite{SW}. Different approaches have been taken to obtain such results \cite{chu, ar, br} for interacting strings in the conformal gauge. However, a gauge-independent analysis in \cite{rbbcsg} led to a new type of noncommutativity which is manifest at all points of the string in contrast to the conformal gauge results where it appears only at the boundaries. In this scheme noncommutativity appears even for the free string case, though it vanishes in the conformal gauge, maintaining compatibility with the earlier results \cite{chu, ar, br}. We follow the gauge-independent analysis of \cite{rbbcsg} to construct the interpolating theory of string propagating in the Neveu--Schwarz two-form field background \cite{sgaghas}. Our analysis of interpolating string shows that compatibility with the stringy boundary conditions demands crucial modifications in the Poission bracket structures which in turn leads to noncommutativity among the string coordinates. These modifications altered the constraint algebra of the theory considerably from what they were in our earlier analysis of interpolating strings \cite{bms1}. So we re-investigate the interconnection of gauge symmetry and diffeomorphism incorporating the modified Poission brackets and constraint algebra. 
  
  Again the issue of the interconnection between gauge symmetry and diffeomorphism is a basic content of theories of gravity. We therefore apply the same Hamiltonian method \cite{brr} also to the second order metric gravity \cite{diff} to establish a one to one mapping between the gauge and diffeomorphism parameters. This makes the underlying unity of the gauge and diffeomorphism invariances of the theory explicit. The unique feature of our analysis is to establish a mapping between the two sets of transformation parameters using a purely Hamiltonian Dirac methodology. 

Our investigation of symmetries in open string brought us in connection with noncommutativity among the space-time  coordinates.  The idea of a underlying  noncommutative structure of the space-time was suggested very early on, in the pioneering days of quantum field theory (QFT), by Heisenberg and was formalised by Snyder in \cite{sny}. The motivation was to provide an effective ultra-violet cutoff to control the divergences which has plagued theories such as quantum electrodynamics (QED). However, this development was quickly overshadowed by the renormalisation programme of QFT which successfully predicted numerical values for physical observables in QED.
It was revived in 1980s when Connes formulated an operator algebric description of the Yang--Mills theory on a large class of NC spaces \cite{connes0}. This found application in geometrical interpretation of the standard model \cite{stm}, eventually gravity was also included \cite{ncg}.  However this approach suffered many weaknesses (e.g. it could not account for quantum radiative corrections) and eventually died out. The most concrete evidence and motivation of NC geometry came from string theoretic results, the first appearance can be attributed to Witten's paper on string field theory back in 1986 \cite{witten1}. Several works followed, where various duality symmetries of closed string theory were studied using the techniques of NC geometry\cite{sdnc}.
Gradually, the possibility of an underlying NC space-time became apparent in closed string theory \cite{snc}. 
In the early nineties it was found that open string end-points live on fixed hypersurfaces in space-time, known as D-branes \cite{D-branes1}. Their low-energy effective field theory was found to have configuration space which is described in terms of noncommutating matrix-valued space-time coordinate fields \cite{D-branes2}. The fact that quantum field theory on NC space-time arises naturally in string theory and matrix theory led to an tremendous upsurge in formulating and analysing different field theories, including gauge theories over NC space-time  \cite{connes1, szabo}. 

Weyl's prescription to associate a quantum operator to a classical phase space function provides a systematic way to describe NC space and study the field theories defined thereon. We shall refer to this approach as operator approach in this thesis. We shall, however, use an alternative approach to analyse NC field theories in the thesis, one which relies heavily on the aspect of gauge symmetry of the actions. A very significant aspect of gauge symmetry in the NC theories has been discovered by Seiberg and Witten in \cite{SW}, namely the correspondence between the NC gauge transformation and commutative gauge transformation. This can be established by the Seiberg--Witten (SW) maps. Using these maps one can derive a commutative equivalent theory describing the original NC gauge theory perturbatively, in terms of commutative variables. The basic idea is to define the fields over phase space with ordinary multiplication replaced by the Gronewald--Moyal product \cite{szabo}. Then  the original theory can be mapped to an equivalent commutative theory in the framework of perturbative expansion in the NC parameter, using the Seiberg--Witten-type maps \cite{SW, bichl} for the fields. Later, closed form S--W maps were found for certain actions \cite{rbhsk, hsk} which provided a commutative equivalent description of the models exact to all order in the NC parameter. In the thesis we have discussed NC extensions of various gauge field theories in the commutative equivalent approach. The results of our study manifest aspects of gauge symmetries in the context of NC space-time. 

We begin with the application of the exact SW maps. The models we considered were a NC Chern--Simons (NCCS) coupled scalar field theories with and without Maxwell term. In the (non-perturbative) operator approach both the Maxwell and NCCS gauge field coupled to fundamental scalar fields produce B--P--S type solitons \cite{hkl, poly, prnjp}. A solution generating technique, which was developed in \cite{hkl}, has been used in \cite{prnjp} . The static soliton solution of the theory diverges as $\theta \to 0$. These soliton solutions thus essentially belong to the singular sector. A relevant  question, therefore, is whether there is some non-trivial non-perturbative solutions depending on the NC parameter and vanishing continuously along with it. Clearly, this can not be answered by the operator approach. The closed form S--W maps, on the other hand, essentially possess smooth commutative limit. Therefore, investigation of such NC scalar field theories in the exact commutative equivalent approach to check whether they exhibit solitons with smooth commutative limit becomes relevant. In \cite{pmas1} we have specifically addressed this issue by analyzing a $U(1)_{\star}$ Chern--Simons (C--S) coupled scalar field theory in $\left(2+1\right)$ dimension where the scalar field is in the adjoint representation of the gauge group. We also addressed the coupling of adjoint matter with the Maxwell field in addition to the C--S term in \cite{pmas2}. Our selection of these models is further motivated by the fact that in the commutative limit the scalar field decouples from the gauge interaction. In other words any non-trivial result of  our analysis comes from the NC features only. Using the exact SW maps \cite{rbhsk, hsk} we have derived exact commutative equivalent versions of these models. Detailed expressions of different forms of energy momentum tensor have been derived and the existence of solitonic modes has been investigated. It was observed that a satisfactory EM tensor can not be obtained from the canonical prescriptions. A symmetric {\it and} gauge invariant EM tensor is constructed by varying the action with respect to a background metric and finally keeping the metric flat. This has been identified as the physical EM tensor for our model. It turns out in our analysis that minimum configuration of the static energy functional corresponds to trivial solutions. In the static limit there is no coupling between the matter and the gauge field. Hence there is {\it no} B--P--S soliton in the model with a smooth commutative limit $\left(\theta \to 0\right)$.

Noncommutativity, specifically among the spatial coordinates, is well analysed and understood. However, NC structure involving time coordinates remains a contentious issue.
As example of NC gauge theories where time-space noncommutativity is an essential element we have advanced the NC extension of $\left(1+1\right)$-dimensional bosonised Schwinger model \cite{asarpm1}. We considered bosonised form \cite{SCH} of the vector Schwinger model obtained via a gauge invariant regularisation (GIR) scheme \cite{old8} and constructed the corresponding NC model. Using the standard $\star$-product formalism the original commutative regularised model is lifted to the NC platform where the $U\left(1\right)$ gauge symmetry is replaced by $U\left(1\right)_{\star}$ gauge symmetry. Invoking the appropriate Seiberg--Witten (SW) transformations the NC model is then mapped to an equivalent commutative model. The model exhibits emergence of a massive boson interacting with a background generated by the NC geometry. Our analysis thus reveals the presence of a background interaction term which is manifest only in the tiny length scale $\sim \sqrt \theta$. We argued that the theory, in the reduced phase space, can be formulated as a perturbative quantum field theory which is formally similar to the KG theory with a classical source. This ensures the unitarity and causality requirement which is a welcome result in the context of theories with time-space noncommutativity.

By using the correspondence between noncommutative and commutative gauge symmetries we have found non-trivial effects in connection with the Schwinger model as mentioned above. This effect depends on the time-space NC parameter. A natural question  about its size thus arises. While there are several estimates of spatial NC parameters available in the literature \cite{carol, cst, mpr, bert0, RB, stern2} corresponding figures about the time-space noncommutativity is lacking. We have considered the gravitational well problem \cite{nes1} in a NC field theoretic setting in \cite{ani} where such an estimate is obtained using the GRANIT experimental data \cite{nes2, nes3}. Since this exercise, although relevant, is somewhat different from the main body of our thesis we present the analysis in Appendix {\bf A}.

Another topic of considerable current interest is the construction of a theory of gravity in NC space time. Various authors have approached the problem from different angles. In \cite{Chamseddine:2000si} for example  a  deformation of Einstein's gravity was studied using a  construction  based on gauging the noncommutative SO(4,1) de Sitter group and the SW map \cite{SW} with subsequent contraction to ISO(3,1). Another approach of a noncommutative gravitational theory was proposed in \cite{Aschieri:2005yw}. Very recently noncommutative gravity has been connected with stringy perspective\cite{ag}. In all these works the leading order noncommutative effects appear in the second order in the NC parameter $\theta$. Since gravity can be viewed effectively as a gauge theory the commutative equivalent approach also seems to be a promising one. Indeed, a minimal theory of NC gravity \cite{cal} has been constructed recently based on this approach where the NC correction appears as a series expansion in the NC parameter. The leading order correction is reported to be linear in $\theta$ in this work. It, therefore, seems that the result of \cite{cal} is in contradiction with others \cite{Chamseddine:2000si, Aschieri:2005yw, ag}. In this thesis we have extended the commutative equivalent formulation of \cite{cal} to show that the first order correction term reported therein actually vanishes, and thus settled the controversy \cite{pmas3}.

After exploiting the correspondence of NC gauge symmetry and commutative gauge symmetry in the analysis of different NC theories we come back to the problem of analysing symmetry. In our studies of symmetries we have so far not addressed invariances under space-time transformations. This issue is particularly important in connection with the nonrelativistic NC models. A constant noncommutativity explicitly violates Lorentz symmetry \cite{carol}. However the issue of Galilean invariance is non-trivial \cite{hms, hms1, bcsgas}. So we investigate the Galilean symmetry of a Schrodinger field theory coupled to background gauge field in a NC setting where only the spatial sector of the NC algebra is involved. Our analysis shows a symmetry violation in the boost sector resulting in a non-closure of the Galilean algebra \cite{bcsgas}.

The organisation of the thesis is as follows. In Chapter \ref{string} and \ref{gravity} we discuss the internal symmetries of some basic (both free and interacting) stringy models and gravity respectively using a Hamiltonian framework. We shall analyse some noncommutative (NC) gauge field theories coupled to scalar fields in a commutative equivalent approach in Chapter \ref{nc}. This includes Chern--Simons (CS) as well as Maxwell type of gauge fields interacting with adjoint scalar matter in section \ref{chernmax}. The issue of time-space noncommutativity is taken up next in section \ref{ncsch} in the context of $\left(1+1\right)$ dimensional bosonised Schwinger model. A brief investigation on the issue of a first order NC correction to pure gravity theory follows in section \ref{ncgr} where we consider gravity as an NC gauge theory and work in our commutative equivalent approach.  In Chapter \ref{ncq} we discuss the space-time symmetries of a noncommutative quantum mechanical model where the Schrodinger matter is considered to interact with a  background gauge field. Finally we conclude in  Chapter \ref{conclusion} where some possible extensions and further work are indicated. We have also included an appendix where an upper-bound on the time-space NC parameter is set.
\clearpage
\label{sec:string}
\chapter{Gauge symmetry and Reparametrisation in basic stringy models}
\label{string}
\pagestyle{myheadings}
\markright{Gauge symmetry and Reparametrisation in basic stringy models}
\section{Introduction}
\label{stringintro}
\pagestyle{myheadings}
\markright{Introduction}
As discussed in the introductory chapter, we begin with the studies of gauge symmetry and their connection with diffeomorphism invariances in some basic stringy models. String theory was introduced as a candidate for the fundamental theory uniting all the basic interactions at the Planck scale \cite{pol}. At least five different string theories emerged equally viable, namely the Type I, Type IIA, Type IIB, Heterotic SO(32) and Heterotic $E_8\times E_8$ string theories. This, however, raised doubt about the claimed unique status of string theory. Significant progress in the understanding of different ramifications of string theories has been achieved in the last decade with the discovery of the dualities \cite{du} mapping one theory into another, thereby indicating their essential unity. It is now definitely believed that the different perturbative sectors of string theories occupy different corners of some yet unknown M-theory \cite{mr}. Higher dimensional extended objects like membranes are expected to be instrumental in understanding this new theory. A characteristic feature of these structures is that they are loaded with various symmetries. It is useful to gain an understanding of these symmetries from different points of view.

Apart from the symmetries of the space-time in which the string or the higher brane is embedded, they have diffeomorphism ( diff.) invariance which arises from reparametrization invariance of their world `volume'. The latter can be considered as gauge symmetries implemented by the first class constraints of the theory. The constrained Hamiltonian analysis due to Dirac \cite{dir} is a natural methodology for such problems. This method in a gauge independent setting \cite{hrt, rb} has been proved to be very useful in the analysis of space-time symmetries of different field theoretic models, a fact demonstrated by numerous applications in the literature \cite{bcasm, rbpm, hms, hms1}. The investigation of the diff. invariance and the gauge symmetries, including their correspondence in the stringy context, from a gauge-independent constrained Hamiltonian approach is therefore strongly suggested. In the present chapter we will give a detailed and comprehensive analysis of the same. Initially, the treatment will be specific to the particular choice of $p=1 (strings), p=2 (membranes)$ only, which do not admit an obvious generalisation to the arbitrary $p$-dimensional case. We shall therefore take up the task of generalising our method to $p$-dimensional case, once the string and membrane action and its internal symmetries are thoroughly analysed. Since the essential features of the diff. invariance are contained in the  bosonic version of the strings (or membranes) we will only consider such models.

The action for a string can be chosen in analogy with the relativistic particle as the proper area of the world sheet swept out by the dynamical string. This gives the Nambu--Goto ( N--G ) formalism which, however, poses problems in quantization. A redundant description, where the world sheet metric coefficients are considered as independent fields, has been shown by Polyakov to be particularly suitable for quantisation. The ensuing action is known as the Polyakov action. The equivalence between the two approaches is established on shell by solving the independent metric in the Polyakov action. The classical correspondence is assumed to lead to equivalent results at the quantum level \cite{pol}. Understanding this correspondence from different viewpoints will, naturally, be useful. In this context the opposite question of the transition from the N--G to the Polyakov action appears. Recently, this issue has been addressed by a degrees of freedom matching in \cite{rbbcsg}. Again, the Nambu--Goto and the Polyakov actions have their counterparts for the dynamical membranes \cite{t} and in general for the higher branes. Our analysis will reveal the transition from the N--G action to the Polyakov action for the membrane and for the generic $p-$brane. Note that there is a distinctive feature of the Polyakov actions for the higher branes as opposed to the strings. This is the absence of the Weyl invariance observed in the string action. We provide the reasons for the presence of the Weyl invariance in Polyakov strings as well as for its absence in the higher branes in our analysis. 

In \cite{rbbcsg} the correspondence between the two forms for the string action was demonstrated by constructing a Lagrangean description which interpolates between the N--G and the Polyakov form. The interpolating theory thus offers a unified picture for understanding different features of the basic structures including their various symmetry properties. In this sense, therefore. it is more general than either the N--G or Polyakov formulations. An added advantage is that it illuminates the passage from the N--G form to the Polyakov form, which is otherwise lacking. Also the Polyakov string action has the additional Weyl invariance which the N--G string action does not have. The interpolating action, which does not presuppose Weyl invariance, explains the emergence of the Weyl invariance in a natural way. The facilities of the interpolating Lagrangean formalism revealed in the string problem highlight the utility of the generalization of this formalism to the higher branes. Our first step towards this generalization is to analyse the membrane as a simple example. Unlike the string case, the reduction of the interpolating Lagrangean to the Polyakov form for the membranes is plagued with the problem of the extra degrees of freedom of the induced metric. Our analysis in this chapter will show that the emergence of the cosmological term in the membrane action is precisely from the accounting of these extra degrees of freedom. Later in this chapter, we shall show that this mismatch of degrees of freedom and the consequent emergence of the cosmological term is a generic result for the arbitrary $p$-dimensional branes.

Since the interpolating action formalism offers a composite scenario for discussing different features of such basic structures as strings, membranes etc, a thorough understanding of the gauge symmetries occurring in these actions is desirable, if not essential. We investigate them systematically using the Hamiltonian approach of \cite{brr}. The complete equivalence between the gauge and the diff. symmetries is established by providing an explicit map between the corresponding parameters. We will first discuss the string action in section \ref{stringreview} since its interpolating action formalism, though methodologically similar with the other branes, contains the above-mentioned unique features that demands separate treatment. Then in section \ref{stringsym} we will analyse the internal symmetries of the string. Our next goal will be to construct the interpolating membrane formalism, to demonstrate its connection to the N--G and Polyakov membrane actions and analyse its various internal symmetries respectively in sections \ref{membranereview}, \ref{membranereduction} and \ref{membranesym}. Finally, the results of the symmetry analysis of string and membrane actions will be cast in a compact generalised form when we carry over to the $p$-brane case in section \ref{pbrane} and \ref{pbrane4}. An interesting aspect of our interpolating Lagrangean theory is the natural emergence of the Arnowitt--Deser--Misner (ADM) representation used in general relativity \cite{adm}. The ADM representation uses the lapse and shift variables which owe their origin to the presence of first class constraints. The interpolating Lagrangean formalism, built as it is on the constraints which are implemented by Lagrange multipliers ( that are the analogues of the lapse and shift variables ), is seen to elucidate the connection with the ADM representation. We shall explore this connection in detail in section \ref{pbrane3}. 

Imposing string boundary conditions for open string propagating in a Neveu--Schwarz (NS) background suggests modifications in the basic Poission brackets (P.B) of the model \cite{rbbcsg}. We shall address this issue by constructing the interpolating string formalism in a NS background in section \ref{ncstring}. There we will show that compatibility with the string boundary conditions demands a modified P.B structure which in turns lead to a noncommutative structure of space-time. This naturally demands a separate analysis of gauge symmetries of open string action with the modified P.B's taken into account. This will be covered in section \ref{ncstring5}.

Before proceeding with the above undertakings let us first summaries the Hamiltonian algorithm of \cite{brr} of abstracting the most general gauge transformation generator for a theory where all the constraints are first-class. We shall apply this methodology throughout the present and the next chapter whenever we are to discuss the internal symmetries.
\section{The Hamiltonian Algorithm}
\label{Algo}
\pagestyle{myheadings}
\markright{The Hamiltonian Algorithm}
Let us consider a theory with first class constraints only. The set of constraints $\Omega_{a}$ is assumed to be classified as
$\left(\Omega_{a}\right) = \left(\Omega_{a_1};\Omega_{a_2}\right)$
where $a_1$ and $a_2$ indicate the set of primary and secondary constraints respectively. The total Hamiltonian is
\begin{equation}
H_{T} = H_{c} + \Sigma\lambda^{a_1}\Omega_{a_1}
\label{216}
\end{equation}
where $H_c$ is the canonical Hamiltonian and $\lambda^{a_1}$ are Lagrange multipliers enforcing the primary constraints. The most general expression for the generator of gauge transformations is obtained according to the Dirac conjecture \cite{dir} as
\begin{equation}
G = \Sigma \epsilon^{a}\Omega_{a}
\label{217}
\end{equation}
where $\epsilon^{a}$ are the gauge parameters. Note that all the first-class constraints appear in $G$. However, only $a_1$ of the parameters $\epsilon^{a}$ are independent, the number being equal to the number of primary first-class constraints \cite{HTZ}. By demanding the commutation of an arbitrary gauge  variation with the total time derivative,(i.e. $\frac{d}{dt}\left(\delta q \right) = \delta \left(\frac{d}{dt} q \right) $) we arrive at the following equations \cite{brr}
\begin{eqnarray}
\delta\lambda^{a_1} = \frac{d\epsilon^{a_1}}{dt} -\epsilon^{a}\left(V_{a}^{a_1} +\lambda^{b_1}C_{b_1a}^{a_1}\right) \label{218} \\
  0 = \frac{d\epsilon^{a_2}}{dt} -\epsilon^{a}\left(V_{a}^{a_2} + \lambda^{b_1}C_{b_1a}^{a_2}\right) \label{219}
\end{eqnarray}
Here the coefficients $V_{a}^{a_{1}}$ and $C_{b_1a}^{a_1}$ are the structure functions of the involutive algebra, defined as
\begin{eqnarray}
\{H_c,\Omega_{a}\} = V_{a}^b\Omega_{b}
\label{2110a}\\
\{\Omega_{a},\Omega_{b}\} = C_{ab}^{c}\Omega_{c}
\label{2110}
\end{eqnarray}
Solving (\ref{219}) it is possible to choose $a_1$ independent gauge parameters from the set $\epsilon^{a}$ and express $G$ of (\ref{217}) entirely in terms of them. The other set (\ref{218}) gives the gauge variations of the Lagrange multipliers. It can be shown that these equations are not independent conditions but appear as internal consistency conditions. In fact the conditions (\ref{218}) follow from (\ref{219}). These set of equations (\ref{218}), (\ref{219}) and (\ref{2110}) will be used in their continuous form in our analysis of internal symmetries of various stringy models in the present chapter and in context of second order metric gravity in chapter \ref{gravity}. 

\section{Review of Interpolating Lagrangean formalism of the bosonic string}
\label{stringreview}
\pagestyle{myheadings}
\markright{Review of Interpolating Lagrangean formalism of the bosonic string}
We start with a brief review of the N--G and Polyakov action formalisms of the the free bosonic string and the mechanism to formulate the interpolating Lagrangean. The string is a one-dimensional object which will be assumed to be embedded in the D-dimensional Minkowski space-time with mostly positive metric $\eta_{\mu\nu}$. The string sweeps out a world-sheet which may be parametrized by two parameters $\tau$ and $\sigma$. The N--G action for the free bosonic string is obtained from the integrated proper area of the world-sheet
\begin{eqnarray}
S_{\mathrm{NG}} = \int d\tau d\sigma {\cal {L}}_{NG}=- \int d\tau d\sigma \left[\left(\dot{X}.X^{\prime}\right)^{2}
- \dot{X}^{2} X^{\prime 2}\right]^{\frac{1}{2}}
\label{111}
\end{eqnarray}
where $\dot{X}^{\mu} = \frac{\partial X^{\mu}}{ \partial\tau}$ and ${X}^{\prime \mu} = \frac{\partial X^{\mu}}{\partial\sigma}$. The string tension is kept implicit for convenience. The string action is invariant under the world-sheet reparametrization
\begin{eqnarray}
\tau \mapsto \tau^{\prime} = \tau^{\prime}\left(\tau,\sigma\right), \qquad
\sigma \mapsto \sigma^{\prime} = \sigma^{\prime}\left(\tau,\sigma\right)
\label{112}
\end{eqnarray}
while the fields $X^{\mu}$ behave as world-sheet scalars 
\begin{equation}
X^{\prime\mu}\left(\tau^{\prime},\sigma^{\prime}\right) =
X^{\mu}\left(\tau,\sigma\right)
\label{113}
\end{equation}
The canonical momenta corresponding to the basic fields $X^{\mu}$  are
\begin{eqnarray}
\Pi_{\mu} = \frac{\partial {\cal {L}}_{NG}}{\partial {\dot X}^{\mu}} =\frac{X^{\prime 2} \dot X_{\mu} - X^{\prime}_{\mu}
\left(\dot{X}.X^{\prime}\right)}{\left[\left(\dot{X}.X^{\prime}\right)^{2} -
  \dot{X}^{2} X^{\prime 2}\right]^{\frac{1}{2}}}
\label{114}
\end{eqnarray}
     From the definition (\ref{114}) we get the primary constraints
for the N--G string
\begin{eqnarray}
\Omega_{1} = \Pi_{\mu}X^{\prime \mu} \approx 0, \qquad
\Omega_{2} = \Pi^{2} + X^{\prime 2} \approx 0
\label{115}
\end{eqnarray}
Using the nontrivial Poisson's brackets of the theory 
\begin{eqnarray}
 \{X^{\mu}\left(\tau,\sigma\right),
 \Pi_{\nu}\left(\tau,\sigma^{\prime}\right)\} = \eta_{\nu}^{\mu}
 \delta\left(\sigma - \sigma^{\prime}\right)
\label{116}
\end{eqnarray}
it is easy to work out the algebra of the constraints
\begin{eqnarray}
\left\{ \Omega_{1}\left(\sigma\right),\Omega_{1}\left(\sigma^{\prime}\right)
\right\} & = & \left(\Omega_{1}\left(\sigma\right) +
\Omega_{1}\left(\sigma^{\prime}\right)\right)\partial_{\sigma}
\delta\left(\sigma - \sigma^{\prime}\right) \label{117} \\
\left\{ \Omega_{1}\left(\sigma\right),\Omega_{2}\left(\sigma^{\prime}\right)
\right\} & = & \left(\Omega_{2}\left(\sigma\right) +
\Omega_{2}\left(\sigma^{\prime}\right)\right)\partial_{\sigma}
\delta\left(\sigma - \sigma^{\prime}\right) \label{118} \\
\left\{ \Omega_{2}\left(\sigma\right),\Omega_{2}\left(\sigma^{\prime}\right)
\right\} & = & 4 \left(\Omega_{1}\left(\sigma\right) +
\Omega_{1}\left(\sigma^{\prime}\right)\right)
\partial_{\sigma}\delta\left(\sigma - \sigma^{\prime}\right) \label{119}
\end{eqnarray}
Clearly, the constraint algebra (\ref{117}, \ref{118}, \ref{119}) is weakly involutive, so that the set (\ref{115}) is first class. The canonical Hamiltonian 
\begin{eqnarray}
{\cal {H}}_{c} = \Pi_{\mu}\dot X^{\mu} - {\cal{L}}_{NG}
\label{1110}
\end{eqnarray}
vanishes when we substitute the appropriate expressions, as expected for a reparametrization invariant theory. The total Hamiltonian is thus expressed as a linear combination of the first-class constraints (\ref{115}),
\begin{equation}
{\cal{H}}_{T} = - \rho \Omega_{1} - \frac{\lambda}{2}\Omega_{2}
\label{1111}
\end{equation}
where $\rho$ and $\lambda$ are Lagrange multipliers. Conserving the primary constraints no new secondary constraints emerge. The total set of constraints of the N--G theory thus consists of the first class system (\ref{115}).

  To construct the string interpolating action \cite{rbbcsg} we write the N--G Lagrangean in the first order form \cite{ka}
\begin{equation}
{\cal{L}}_{I} = \Pi_{\mu} \dot X^{\mu} - {\cal{H}}_{T}
\label{1113}
\end{equation}
Substituting ${\cal{H}}_{T}$ from (\ref{1111}), ${\cal{L}}_{I}$ becomes
\begin{equation}
{\cal{L}}_{I} = \Pi_{\mu}\dot X^{\mu} + \rho \Pi_{\mu}X^{\prime \mu}
 + \frac{\lambda}{2}\left(\Pi^{2} + X^{\prime 2} \right)
\label{1114}
\end{equation}
where $\lambda$ and $\rho$, though originally introduced as Lagrange multipliers, will be treated as independent
fields. Since $\Pi_{\mu}$ is really an auxiliary field we will eliminate it from (\ref{1114})using its Euler-Lagrange equation 
\begin{equation}
\dot X^{\mu}+ \rho X^{\prime \mu} + \lambda \Pi^{\mu}  = 0
\label{1115}
\end{equation}
This gives the Lagrangian ${\cal{L}}_{I}$
\begin{equation}
{\cal{L}}_{I} = -\frac{1}{2 \lambda}\left[\dot X^{2} +
2 \rho \dot X_{\mu}X^{\prime \mu} + \left( \rho^{2} - \lambda^{2}\right)
X^{\prime 2}\right]
\label{1116}
\end{equation}
which we call the Interpolating Lagrangean of the bosonic string. The justification of the name was established in \cite{rbbcsg} by showing that passing to the appropriate limits one can derive the N--G action and the Polyakov action from (\ref{1116}). We shall briefly recount the same before proceeding with the symmetry issues.

    The reproduction of the N--G action from the interpolating action we need only to eliminate the extra fields $\rho$ and $\lambda$ from (\ref{1116}) by using the corresponding solutions of the E-L equations for $\rho$ and $\lambda$ following from (\ref{1116}) itself. These are 
\begin{eqnarray}
\rho & = & - \frac{\dot X^{\mu} X^{\prime}_{\mu}}{X^{\prime 2}} \label{1117}\\
{\rm and} \qquad 
\lambda^{2} & = & \frac{\left(\dot{X}.X^{\prime}\right)^{2} - \dot{X}^{2} X^{\prime 2}}{X^{\prime 2} X^{\prime 2}}
\label{1118b}
\end{eqnarray}
From (\ref{1118b}) $\lambda$ is determined modulo a sign which can be fixed by demanding the consistency of (\ref{114}) with (\ref{1115}). Accordingly,
\begin{equation}
\lambda = -\frac{\left[\left(\dot{X}.X^{\prime}\right)^{2}
- \dot{X}^{2} X^{\prime 2}\right]^{\frac{1}{2}}}{X^{\prime 2}}
\label{1118}
\end{equation}
Now, substituting $\rho$ and $\lambda$ from (\ref{1117}) and (\ref{1118}) in ${\cal{L}}_{I}$ we get back the N--G form.

 The reproduction of the Polyakov action from (\ref{1116}) is not so straightforward. The Polyakov action for the free bosonic string is 
\begin{equation}
S_{\mathrm{P}} = -\frac{1}{2}\int d^2\xi {\sqrt - g} g^{ij} \partial_{i} X^{\mu}
\partial_{j} X_{\mu}
\label{1112}
\end{equation}
Here, apart from $X_{\mu}$, $g_{ij}$ are also considered as independent fields while $\xi^i$ collectively denote the parameters, $\xi^0 = \tau$ and $\xi^1 = \sigma$. Also note that $g = {\rm{det}}g_{ij}$ and $g^{ij}$ is the inverse of $g_{ij}$ \footnote{The same notation is maintained throughout this chapter, i.e. for membrane, p-brane and interacting string case}. It is equivalent to the N--G action in the sense that solving $ g^{ij} $ in (\ref{1112}) from its equation of motion one can reproduce the N--G action. Evidently, the Polyakov action is a more redundant description of the string than the N--G action. The reparametrization invariance under (\ref{112}) is ensured by the transformations (\ref{113}) along with the transformations
\begin{equation}
g^{\prime}_{ij}(\xi^{\prime}) = \frac{\partial\xi^{k}}{\partial\xi^{\prime i}}
                   \frac{\partial\xi^{l}}{\partial\xi^{\prime j}} g_{kl}(\xi)
\label{113A}
\end{equation}
 Looking at the transformation relations under (\ref{113}) and (\ref{113A}) it is apparent that the reparametrization invariance is synonymous with general covariance on the world-sheet with $X^{\mu}$ transforming as scalar fields. Apart from the reparametrization invariance the Polyakov string has the Weyl invariance
\begin{equation}
g^{\prime}_{ij}(\xi) = \exp\left(\Lambda\left(\xi\right)\right)g_{ij}(\xi)
\label{113B}
\end{equation}
where $\Lambda (\xi)$ is any arbitrary well behaved function of $\xi$. Though there are three different metric coefficients $g_{ij}$, due to the existence of this scale (Weyl) invariance only two of them are really independent. The Weyl invariance is special to the Polyakov string, the higher branes do not share it. Clearly, in the Polyakov action of the string there are only two independent fields apart from $X_{\mu}$, namely the two independent components of the metric. These two components can also be fixed by the two reparametrization symmetries. Usually the light-cone metric ${\rm {diag}}(1, -1)$ is employed in the gauge fixed calculation. However here we work in the gauge independent approach, otherwise the interplay of gauge and diff. symmetries is lost. That the metric is completely determinable is manifested in our approach by the exact matching of the number of independent metric components with the number of extra fields in the interpolating Lagrangean (\ref{1116}). It will thus be possible to map the interpolating Lagrangean to the Polyakov form in a unique manner. We take the following Ansatz \cite{rbbcsg}\footnote {Such a representation was also discussed by Giddings \cite{gid}}
\begin{equation}
g^{ij} = \left(-g\right)^{-\frac{1}{2}}
\left(\begin{array}{cc}
\frac{1}{\lambda}&\frac{\rho}{\lambda}\\
\frac{\rho}{\lambda}&\frac{\rho^2 - \lambda^2}{\lambda}
\end{array}\right)
\label{1119}
\end{equation}
With this choice the Interpolating Lagrangean (\ref{1116}) reduces to 
$\quad {\cal{L}}_{I} = -\frac{1}{2}{\sqrt - g} g^{ij} \partial_{i} X^{\mu} \partial_{j} X_{\mu}\quad $
which, clearly, is the Polyakov form of the string action. To verify the consistency of the construction (\ref{1119}) we note that
\begin{equation}
{\rm det}g^{ij} = \left( - g\right)^{-1}\left[ \frac{\rho^{2} -
\lambda^{2}}{\lambda^{2}} - \frac{\rho^{2}}{\lambda^{2}}\right] = \frac{1}{g}
\label{1121}
\end{equation}
as it should be because $g^{ij}$ is the inverse matrix of $g_{ij}$. Further, from the identification (\ref{1119}) we find
\begin{eqnarray}
g^{00} = \left( - g \right)^{-\frac{1}{2}}\frac{1}{\lambda} \qquad
{\rm and }\qquad 
g^{01}  =  g^{10} = \left( - g \right)^{-\frac{1}{2}}\frac{\rho}{\lambda}
\label{1123}
\end{eqnarray}
From the above equations we can solve $\rho$ and $\lambda$ in terms of
$g^{00}$ and $g^{01}$ as
\begin{eqnarray}
\rho =\frac{g^{01}}{g^{00}}, \qquad 
\lambda = \frac{1}{\left( \sqrt{- g }\right)\left( g^{00}\right)}
\label{1124}
\end{eqnarray}
Finally the mapping (\ref{1119}) also yields
\begin{equation}
g^{11} = \frac{\rho^{2} -\lambda^{2}}{\lambda}
\left( - g \right)^{-\frac{1}{2}}
\label{1125}
\end{equation}
After substituting the solutions for $\rho$ and $\lambda$, the resulting expression becomes
\begin{equation}
\left(g^{01} \right)^{2} -  \left(\sqrt{ - g } \right)^{-2} = g^{11} g^{00}
\label{1126}
\end{equation}
which, after a simple rearrangement, is shown to be the same as the condition (\ref{1121}). This completes the consistency check of the construction (\ref{1119}). This concludes our review of the interpolating Lagrangean formalism and how it consistently leads to Polyakov string action starting from the N--G version. We now move on to analyse the constraint structure and internal symmetries of the different string actions mentioned in this section.

\section{Constraint structure and Gauge symmetry of string}
\label{stringsym}
\pagestyle{myheadings}
\markright{Constraint structure and Gauge symmetry}
In this section we will discuss the gauge symmetries of the different versions of the string actions and find their exact correspondence with the reparametrization invariances. We begin the analysis with the N--G action (\ref{111}). Here the only fields are $X^{\mu}$. The generator of the gauge transformations of (\ref{111}) is obtained from the constraints $\Omega_{i}, \quad \left( i = 1, 2 \right)$ given by (\ref{115}) as
\begin{equation}
G =\int d\sigma \alpha_{i}\Omega_{i}
\label{3112}
\end{equation}
where $\alpha_{i}$ are the independent gauge parameters. The transformations of $X^{\mu}$  under (\ref{3112}) results in the following
\begin{equation}
\delta X_{\mu} = \left\{X_{\mu}, G \right\}
 = \left( \alpha_{1} X_{\mu}^{\prime} + 2 \alpha_{2} \Pi_{\mu} \right)
\label{3114}
\end{equation}
Substituting $\Pi_{\mu}$ from (\ref{114}) in the above we get the appropriate gauge transformation of $X_{\mu}$ that leave (\ref{111}) invariant,
\begin{equation}
\delta X_{\mu}
 = \left(\alpha_1 - \frac{2\alpha_2\left(\dot{X}.X^{\prime}\right)}
{\left[\left(\dot{X}.X^{\prime}\right)^{2} -
  \dot{X}^{2} X^{\prime 2}\right]^{\frac{1}{2}}}
\right)X^{\prime}_{\mu} + \frac{ 2 \alpha_{2}X^{\prime 2}}
{\left[\left(\dot{X}.X^{\prime}\right)^{2} -
  \dot{X}^{2} X^{\prime 2}\right]^{\frac{1}{2}}}
\dot{X}_{\mu}
\label{31141}
\end{equation}
Identifying the coefficients of $\dot{X}$ and $X^{\prime}$ respectively with $\Lambda_{0}$ and $\Lambda_{1}$ we get
\begin{equation}
\delta X_{\mu}
 = \Lambda_{0} \dot{X}_{\mu} + \Lambda_{1} X^{\prime}{}_{\mu}
\label{3117}
\end{equation}
Note that $\Lambda_{0}$ and $\Lambda_{1}$ are arbitrary functions of the parameters $\xi_i$. Using (\ref{113}) we observe that these gauge variations (\ref{3117}) coincide with the variations due to the reparametrization
\begin{eqnarray}
\tau^{\prime} =  \tau - \Lambda_{0}, \qquad \sigma^{\prime}& = & \sigma - \Lambda_{1}
\label{31143}
\end{eqnarray}
The complete mapping of the gauge transformations with the reparametrizations is thus established for the N--G string.

We then take up the interpolating Lagrangean (\ref{1116}). It contains additional fields $\rho$ and $\lambda$ apart from $X_{\mu}$.  Let the canonical momenta corresponding to the independent fields $X_{\mu}$, $\rho$ and $\lambda$ be denoted by $\Pi_{\mu}$, $\Pi_{\rho}$ and $\Pi_{\lambda}$ respectively. By definition
\begin{eqnarray}
\Pi_{\mu} =  -\frac{1}{\lambda}\left(\dot{X}_{\mu} + \rho X^{\prime}_{\mu}\right), \qquad 
\Pi_{\rho} = 0, \qquad 
\Pi_{\lambda} = 0
\label{211}
\end{eqnarray}
In addition to the Poisson brackets similar to (\ref{116}) we now also have
\begin{eqnarray}
 \{\rho\left(\tau,\sigma\right),  \Pi_{\rho}\left(\tau,\sigma^{\prime}\right)\} =
  \delta\left(\sigma - \sigma^{\prime}\right), \qquad
 \{\lambda\left(\tau,\sigma\right), \Pi_{\lambda}\left(\tau,\sigma^{\prime}\right)\} =
  \delta\left(\sigma - \sigma^{\prime}\right)
\label{212}
\end{eqnarray}
The canonical Hamiltonian following from (\ref{1116}) is
\begin{equation}
{\cal{H}}_c = -\rho\Pi_{\mu}X^{\prime\mu} - \frac{\lambda}{2}\left(\Pi^2 +X^{\prime 2}\right) 
\label{213}
\end{equation}
which reproduces the total Hamiltonian (\ref{1111}) of the N--G action. From the definition of the canonical momenta we identify the primary constraints
\begin{eqnarray}
\Pi_{\rho} \approx  0, \qquad \Pi_{\lambda} \approx 0
\label{214}
\end{eqnarray}
Conserving these we find that two new secondary constraints emerge which are nothing but the constraints $\Omega_{1}$ and $\Omega_{2}$ of equation (\ref{115}). The primary constraints of the  N--G action appear as secondary constraints in this formalism. No further constraints appear. The system of constraints for the Interpolating Lagrangean thus comprises of the set (\ref{115}) and (\ref{214}). These constraints are all first class and therefore generate gauge transformations on ${\cal{L}}_I$. But the number of independent gauge parameters is equal to the number of independent primary first class constraints i.e. two. So we can apply the  systematic procedure of \cite{brr} to abstract the most general local symmetry transformations of the Lagrangean; i.e. we could proceed to construct the generator of gauge transformations from (\ref{3112}) by including the whole set of first class constraints (\ref{115}, \ref{214}). Using (\ref{219}) the dependent gauge parameters can be eliminated. After finding the gauge generator in terms of the independent gauge parameters, the variations of the fields $X^{\mu}$, $\rho$ and $\lambda$ can be worked out. However, looking at the intermediate first order form (\ref{1114}) we understand that the variations of the fields $\rho$ and $\lambda$ can be calculated alternatively, (using (\ref{218})) from the N--G theory where they appear as Lagrange multipliers. We adopt this alternative procedure. The generator of gauge transformations has already been given in (\ref{3112}). So the gauge variations of $X^{\mu}$ is again given by (\ref{3114}). Next, we relabel $\rho$ and $\lambda$ by $\lambda_{1}$ and $\lambda_{2}$, where
\begin{equation}
\lambda_{1} = \rho \qquad \rm{and} \qquad \lambda_{2} = \frac{\lambda}{2}
\label{313}
\end{equation}
The  variations of $\lambda_{i}$ are obtained from 
\begin{equation}
\delta\lambda_{i}\left( \sigma\right) = -  \dot \alpha_{i}
- \int d\sigma^{\prime} d\sigma^{\prime \prime} C_{kj}{}^{i}
\left(\sigma^{\prime}, \sigma^{\prime \prime}, \sigma\right)
\lambda_{k}\left(\sigma^{\prime}\right) \alpha_{j}
\left(\sigma^{\prime \prime}\right)
\label{314}
\end{equation}
where
$ C_{kj}{}^{i} \left(\sigma^{\prime}, \sigma^{\prime \prime}, \sigma\right)$ are given by
\begin{equation}
\left\{ \Omega_{\alpha}\left(\sigma\right),\Omega_{\beta}
\left(\sigma^{\prime}\right)\right\} =  \int d\sigma^{\prime \prime}
C_{\alpha \beta}{}^{\gamma}\left(\sigma, \sigma^{\prime}, \sigma^{\prime
\prime}\right) \Omega_{\gamma}\left(\sigma^{\prime \prime}\right)
\label{315}
\end{equation}
Note that equations (\ref{314}) and (\ref{315}) are the continuous forms of equations (\ref{218}) and (\ref{2110}).
Also observe that the structure function $ V_{a}{}^{b}$ does not appear in (\ref{314}) since $H_{c} = 0$ for the N--G theory. The nontrivial structure functions $C_{\alpha \beta}{}^{\gamma}\left(\sigma, \sigma^{\prime}, \sigma^{\prime \prime}\right)$ are obtained from the constraint algebra (\ref{117} - \ref{119}) as
\begin{eqnarray}
C_{1 1}{}^{1}\left(\sigma ,\sigma^{\prime},\sigma^{\prime \prime}\right) & = &
\partial_{\sigma}\delta\left(\sigma - \sigma^{\prime}\right)
\left(\delta\left(\sigma - \sigma^{\prime \prime}\right) +
\delta\left(\sigma^{\prime} - \sigma^{\prime \prime}\right) \right) \label{316} \\
C_{1 2}{}^{2}\left(\sigma, \sigma^{\prime}, \sigma^{\prime \prime}\right) & = &
\partial_{\sigma}\delta\left(\sigma - \sigma^{\prime}\right)
\left(\delta\left(\sigma - \sigma^{\prime \prime}\right) +
\delta\left(\sigma^{\prime} - \sigma^{\prime \prime}\right) \right) \label{317} \\
C_{2 1}{}^{2}\left(\sigma, \sigma^{\prime}, \sigma^{\prime \prime}\right) & = &
\partial_{\sigma}\delta\left(\sigma - \sigma^{\prime}\right)
\left(\delta\left(\sigma - \sigma^{\prime \prime}\right) +
\delta\left(\sigma^{\prime} - \sigma^{\prime \prime}\right) \right) \label{318} \\
C_{2 2}{}^{1}\left(\sigma, \sigma^{\prime}, \sigma^{\prime \prime}\right) & = &
4 \partial_{\sigma}\delta\left(\sigma - \sigma^{\prime}\right)
\left(\delta\left(\sigma - \sigma^{\prime \prime}\right) +
\delta\left(\sigma^{\prime} - \sigma^{\prime \prime}\right) \right) \label{319}
\end{eqnarray}
all other $ C_{\alpha b}{}^{\gamma}$'s are zero. Using the expressions of the structure functions (\ref{316} - \ref{319}) in equation(\ref{314}) we can easily derive the required gauge variations of $\lambda_{1}$ and $\lambda_{2}$ 
\begin{eqnarray}
\delta \lambda_{1} &=& - \dot \alpha_{1}
+ \left(\alpha_{1}\partial_{1}\lambda_{1}
 - \lambda_{1}\partial_{1}\alpha_{1} \right) 
+ 4 \left(\alpha_{2}\partial_{1}\lambda_{2} - \lambda_{2}\partial_{1}
\alpha_{2}\right)\nonumber\\
\delta \lambda_{2} &=& -\dot \alpha_{2}
+\left(\alpha_{2}\partial_{1}\lambda_{1} - \lambda_{1} \partial_{1}\alpha_{2}
\right)
+\left(\alpha_{1}\partial_{1}\lambda_{2} - \lambda_{2} \partial_{1}\alpha_{1}
\right)
\label{GM41}
\end{eqnarray}
and therefore, from the correspondence (\ref{313}) we get the gauge variations of $\rho$
and $\lambda$ as
\begin{eqnarray}
\delta \rho &=& - \dot \alpha_{1}
+ \left(\alpha_{1}\partial_{1}\rho
 - \rho\partial_{1}\alpha_{1} \right) + 2 \left(\alpha_{2}\partial_{1}\lambda - \lambda \partial_{1}\alpha_{2}\right)
\nonumber\\
\delta \lambda &=& -2 \dot \alpha_{2}
+2 \left(\alpha_{2}\partial_{1}\rho - \rho \partial_{1}\alpha_{2}
\right)
+\left(\alpha_{1}\partial_{1}\lambda - \lambda \partial_{1}\alpha_{1}\right)
\label{GM51}
\end{eqnarray}
Note that the above expressions for the gauge variations of $\rho$ and $\lambda$ can also be obtained from their definitions (\ref{1117}), (\ref{1118}) and the expression of gauge variation of $X^{\mu}$ (\ref{31141}). We get from equation (\ref{1117})
\begin{eqnarray}
\delta \rho & = & - \delta \left( \frac{\partial_{0} X^{\mu}\partial_{1} X_{\mu}}{\partial_{1} X^{\mu}\partial_{1} X_{\mu}}\right)
            = - \left\{ \frac{\partial_{1} X_{\nu}}{\partial_{1} X^{\mu}\partial_{1} X_{\mu}}\right\}\partial_{0}\left(\delta X^{\nu}\right) 
+ \left\{2\frac{\partial_{0} X^{\mu}\partial_{1} X_{\mu} \partial_{1} X_{\nu}}
{\left(\partial_{1} X^{\mu}\partial_{1} X_{\mu}\right)^{2}} - \frac{\partial_{0} X_{\nu}}{\partial_{1} X^{\mu}\partial_{1} X_{\mu}} \right\}\partial_{1}\left(\delta X^{\nu}\right)\nonumber\\
\label{GM61}
\end{eqnarray}
which relates the gauge variation of $\rho$ with that of $X^{\mu}$. Using the definitions (\ref{1117}) and (\ref{1118}) the gauge variation of $X^{\mu}$ given by (\ref{31141}) can be reduced to the following convenient form 
\begin{eqnarray}
\delta X^{\mu} = \left( \alpha_{1} - 2 \frac{\alpha_{2}\rho}{\lambda}\right)\partial_{1}X^{\mu} - \frac{2 \alpha_{2}}{\lambda}\partial_{0} X^{\mu}
\label{GM71}
\end{eqnarray}
Substituting $\delta X^{\mu}$ from (\ref{GM71}) in (\ref{GM61}) we recover, after some simplification, the same expression for $\delta \rho$ as in (\ref{GM51}). Similarly, $\delta \lambda $ can also be computed directly from the definition of $\lambda$. We first note that $\rho$ and $\lambda$ can be related as
\begin{eqnarray}
\rho^{2} - \lambda^{2} = \left( \frac{\partial_{0} X^{\mu}\partial_{0} X_{\mu}}{\partial_{1} X^{\mu}\partial_{1} X_{\mu}}\right)
\label{GM81}
\end{eqnarray}
which follows from equation (\ref{1117}) and (\ref{1118}). From the relation (\ref{GM81}) we can easily derive that 
\begin{eqnarray}
2 \rho \delta \rho - 2 \lambda \delta \lambda  =  \left\{ \frac{2 \partial_{0} X_{\nu}}{\partial_{1} X^{\mu}\partial_{1} X_{\mu}}\right\}\partial_{0}\left(\delta X^{\nu}\right) - \left\{ \frac{2 \partial_{0} X^{\mu}\partial_{0} X_{\mu}\partial_{1} X^{\nu}}{\left(\partial_{1} X^{\mu}\partial_{1} X_{\mu}\right)^{2}}\right\}\partial_{1}\left(\delta X^{\nu}\right)
\label{GM91}
\end{eqnarray}
Equation (\ref{GM91}) enables us to find $\delta \lambda$ from the known expressions of $\delta X^{\mu}$ and $\delta \rho$. The resulting expression of $\delta \lambda$ is identical with that given in (\ref{GM51}). This observation again confirms our remark about (\ref{218}) that those are really internal consistency conditions \cite{brr}.

In the above we have found out the full set of symmetry transformations of the fields in the interpolating Lagrangean (\ref{1116}). Clearly, the same set of transformations apply to the first order form (\ref{1115}).
In the latter, $\Pi_{\mu}$ is introduced as an additional field. Its appropriate gauge transformation is not difficult to find
\begin{equation}
\delta \Pi_{\mu} = \left\{\Pi_{\mu}, G \right\}
 = \left( \Pi_{\mu} \alpha_{1} + 2 X_{\mu}^{\prime}\alpha_{2} \right)^{\prime}
\label{3113}
\end{equation}
The symmetry transformations (\ref{GM51}) were earlier given in \cite{ka}. But the results were found there by inspection\footnote { For easy comparison identify $ \alpha_{1} = \eta $ and $ 2 \alpha_{2} = \epsilon $ }.
In our approach the appropriate transformations are obtained systematically by a general method applicable to a whole class of actions.

     At this stage we concentrate our attention on the Polyakov action (\ref{1112}) where the set of basic fields contain $g_{ij}$ apart from the fields $X^{\mu}$. We can take it as an independent example for the application of our analysis based on (\ref{219}). Working out the full set of constraints we can construct the gauge generator $G$ according to (\ref{3112}). Since the set of constraints $\Omega_{i}$ include all the first class constraints, both primary and secondary, we then have to invoke (\ref{219}) to solve the dependent gauge parameters to get the desired form of $G$ in terms of the independent number of gauge parameters. The gauge variation of $g_{ij}$ can then be computed by the usual procedure 
\begin{equation}
\delta g_{ij} = \left\{ g_{ij}, G \right\}
\label{hati}
\end{equation}
However, a particular usefulness of the interpolating Lagrangean formalism can be appreciated now. It is not required to find the gauge variations of $g_{ij}$ from scratch. The identification (\ref{1119}) allows us to find the required gauge variations from the corresponding transformations of $\rho$ and $\lambda$. This possibility is actually a consequence of the essential unity of the nature of gauge symmetries in different versions of the string actions. We have already indicated this for the N--G model. The Polyakov action offers a more important platform to test this proposition. Indeed, the complete equivalence between gauge symmetry and reparametrisation symmetry can be demonstrated from the Polyakov action by comparing the variations of $\rho$ and $\lambda$ from the alternative set of transformations using the identification (\ref{1119}). For this we shall need a map connecting the gauge parameters $\alpha_{i}$ (introduced earlier in equation (\ref{3112})) with reparametrisation parameters $\left( \Lambda_{0}, \Lambda_{1}\right)$ (of equation (\ref{31143})). We can relate $\left( \Lambda_{0}, \Lambda_{1}\right)$ with $\alpha_{i}$'s by demanding that the symmetry transformations on $X^{\mu}$ agree from both the approaches.

     To this end we proceed as follows. From the Lagrangean corresponding to (\ref{1112}) we find
\begin{equation}
\Pi^{\mu} = - \sqrt{- g } g^{00} \dot X^{\mu} - \sqrt{- g } g^{01}
 X^{\prime \mu}
\label{3118}
\end{equation}
Substituting $\dot X^{\mu}$ from (\ref{3118}) in (\ref{3117}) we get after some calculation
\begin{equation}
\delta X^{\mu} = \Lambda^{0} \frac{\sqrt{- g } }{g g^{00} }
 \Pi^{\mu}  + X^{\mu \prime}\left(\Lambda^{1} - \frac{g^{01} }{g^{00} }
 \Lambda^{0}\right)
\label{3119}
\end{equation}
Comparing the above expression of $\delta X^{\mu}$ with that of (\ref{3114}) we find the mapping
\begin{eqnarray}
\Lambda^{0} = -2 \sqrt{- g } g^{00} \alpha_{2}    =
- \frac{ 2 \alpha_{2}}{\lambda}, \qquad 
\Lambda^{1} = \alpha_{1} - 2 \sqrt{- g } g^{01} \alpha_{2}    = \alpha_{1} -
\frac{ 2 \rho \alpha_{2}}{\lambda}
\label{3120a}
\end{eqnarray}
With this mapping the gauge transformation on $X_{\mu}$ in both the formalism agree.

We have already noted how $g_{ij}$ behave under reparametrization (see equations (\ref{113A})). Considering infinitesimal transformation (\ref{31143}) we can write the variation $g_{ij}$ as
\begin{eqnarray}
\delta g_{ij} & = & D_{i}\Lambda_{j} + D_{j}\Lambda_{i} \label{3122}\\
{\rm where,} \qquad 
D_{i}\Lambda_{j} & = & \partial_{i} \Lambda_{j} - \Gamma_{ij}{}^{k} \Lambda_{k} \label{3123}
\end{eqnarray}
$\Gamma_{ij}{}^{k}$ being the usual Christoffel symbols \cite{jvh}. 
Since the metric $g_{ij}$ is associated with $\rho$ and $\lambda$ by the correspondence (\ref{1119}), equation (\ref{3122}) will enable us to derive the reparametrisation variations of $\rho$ and $\lambda$. With the help of the mapping (\ref{3120a}) it will then be possible to express these variations in terms of the parameters $\alpha_{1}$ and $\alpha_{2}$. We have already expressed $\rho$ and $\lambda$ in terms of $g^{ij}$ (see equation(\ref{1124})). To use (\ref{3122}) directly, similar expressions involving the inverse matrix $g_{ij}$ are required. It is easy to find that $\rho$ can be expressed as
\begin{equation}
\rho = -\frac{g_{01}}{g_{11}}
\label{3124}
\end{equation}
The transformations (\ref{3122}) then lead to 
\begin{equation}
\delta \rho = - \partial_{0} \Lambda_{1} + \rho \left( \partial_{0}
\Lambda_{0} - \partial_{1} \Lambda_{1} \right)  - \left( \rho^{2}
 - \lambda^{2} \right) \partial_{1} \Lambda_{0} + 2 \rho^{2} \partial_{1}
 \Lambda_{0} - \Lambda_{k} \partial_{k} \rho
\label{3124a}
\end{equation}
$\Lambda_{i}$ in the last equation can be substituted by $\alpha_{i}$ using the mapping (\ref{3120a}). We find that the resulting expression is identical with the corresponding variation, given in (\ref{GM51}) of $\rho$ under gauge transformation.

      A similar comparison can be done for $\delta \lambda$ also. The ratio
\begin{equation}
\frac{g^{11}}{g^{00}} =\left(\rho^2 - \lambda^2\right)
\label{3125}
\end{equation}
obtained from (\ref{1119}) may be taken as the starting point. We can reduce (\ref{3125}) to $ \frac{g_{00}}{g_{11}} =\left(\rho^2 - \lambda^2\right) $. Now using (\ref{3122}) and the mapping (\ref{3120a}) we get the expression of $\delta \lambda$. Again we find exact matching with (\ref{GM51}). The mapping (\ref{3120a}) thus establishes complete equivalence of the gauge transformations generated by the first class constraints with the diffeomorphisms of the string. 
\section{The Interpolating membrane}
\label{membranereview}
\pagestyle{myheadings}
\markright{The Interpolating membrane}
In the above we have elaborated the interpolating Lagrangean formalism for free strings and studied the gauge symmetry from alternative approaches to establish the correspondence of the gauge transformations generated by the first class constraints and reparametrization symmetry on the world sheet. The analysis, based on the constraints, is applicable in  general. This will be illustrated by taking the bosonic membrane as a first concrete example. Later, the results of this section will be generalized for an arbitrary $p$-brane which we shall take up next. 

The membrane is a two dimensional object which sweeps out a three dimensional world volume in the D dimensional space-time in which it is embedded. We will denote the parameters parameterizing this world volume by $\tau$, $\sigma_{1}$ and $\sigma_{2}$, sometimes collectively referred by the symbol $\xi$. The natural classical action for a membrane moving in flat space-time is given by the integrated proper volume swept out by the membrane. This action is of the Nambu--Goto form
\begin{eqnarray}
S_{\mathrm{NG}} = - \int d^{3}\xi \sqrt{-h}
\label{ss}
\end{eqnarray}
where $h$ is the determinant of the induced metric
\begin{equation}
h_{ij} = \partial_{i} X^{\mu}\partial_{j}X_{\mu}
\label{ss1}
\end{equation}
The indices $i$ and $j$ run from 0 to 2. Note that like the string we have kept the membrane tension implicit.
The action (\ref{ss}) is again reparametrization invariant for which $X^{\mu}$ should transform as in equation (\ref{113}) with parameters $\left(\tau, \sigma_{1}, \sigma_{2}\right)$. The primary constraints following from the Nambu--Goto action are \cite{rbk}
\begin{eqnarray}
\Omega_{a} = \Pi_{\mu}\partial_{a}X^{\mu} \approx 0, \qquad 
\Omega_{3} = \frac{1}{2}\left(\Pi^2 + \bar{h}\right) \approx 0
\label{I1}
\end{eqnarray}
In the above equations $\bar{h}= det(h_{ab})$, $h_{ab} = \partial_aX^{\mu}\partial_b X_{\mu}.$ The indices $a$, $b$ run from 1 to 2 i.e. a,b label the spatial part of the world volume of the membrane. 

   Since the membrane action (\ref{ss}), like the string case (\ref{111}), possesses reparametrization invariance, the canonical Hamiltonian following from the action vanishes. Thus the total Hamiltonian is only a linear combination of the constraints (\ref{I1})
\begin{equation}
{\cal{H}}_{T} = - \rho_{a}\Pi_{\mu}\partial_{a}X^{\mu} - \frac{\lambda}{2}\left(\Pi^2 + \bar{h}\right) 
\label{I2}
\end{equation}
The corresponding Polyakov form is
\begin{equation}
S_{\mathrm{P}} = -\frac{1}{2}\int d^{3}\xi{\sqrt - g} \left(g^{ij} \partial_{i} X^{\mu}\partial_{j} X_{\mu} - 1\right)
\label{s011}
\end{equation}
where other than $X^{\mu}$ the metric elements $g_{ij}$ are also considered as independent fields. The equivalence of (\ref{s011}) with the N--G form (\ref{ss}) can be established by substituting the solution of $g_{ij}$ in (\ref{s011}). It is instructive to
compare (\ref{s011}) with its counterpart (\ref{1112}). There is an extra `cosmological term' in the action (\ref{s011}). This is necessary because the Polyakov form of the membrane action does not have Weyl invariance. Notably, the Polyakov metric has six independent metric coefficients only three of which can be fixed by using the reparametrization invariances. This distinguishes the Polyakov formalism of the membrane from its string counterpart where the metric can be completely fixed.

    We now come to the the construction of an interpolating action for the membrane N--G action (\ref{ss}). The first step is to consider the Lagrange multipliers as independent fields and write an alternative first order Lagrangian for the membrane similar to (\ref{1113}). The equation of motion for $\Pi_{\mu}$ following from that first order Lagrangean will be
\begin{equation}
\Pi_{\mu} = -\frac{\dot{X_{\mu}}+\rho_a\partial_{a}X^{\mu}}{\lambda}
\label{I4}
\end{equation}
Substituting $\Pi_{\mu}$ from (\ref{I4}) in the first order Lagrangean  
we get, upon simplification, the interpolating Lagrangian or the membrane
\begin{equation}
{\cal{L}}_I =-\frac{1}{2\lambda}\left[\dot{X^{\mu}}\dot{X_{\mu}}
                 + 2\rho_{a}\dot{X}_{\mu}\partial_{a}X^{\mu}
                 +\rho_{a}\rho_{b}\partial_{a}X^{\mu}\partial_{b}X_{\mu}\right]
                 +\frac{\lambda}{2}\bar{h}
\label{I5}
\end{equation}
 We can check our results for the membrane by going over to the string limit. In case of the string which is a 1-brane, $ a,b = 1$. Then $\bar{h} = {\rm{det}}(h_{ab}) = \partial_{\sigma}X^{\mu}\partial_\sigma X_{\mu}= {X}^{\prime 2}_{\mu}$. It is easy to see that with these substitutions the Lagrangean (\ref{I5}) becomes identical with the corresponding Lagrangian (\ref{1116}) of the string. We have anticipated the name interpolating Lagrangean from our experience in the string case. Below, we will establish this by generating both the N--G and the Polyakov forms of the membrane action from (\ref{I5}).
\section{Reduction to the N--G and Polyakov membrane action}
\label{membranereduction}
\pagestyle{myheadings}
\markright{Reduction to the N--G and Polyakov membrane action}
Let us first discuss the passage to the N--G form. From the interpolating Lagrangean it is easy to write down the equations of motion for $\lambda$ and $\rho_a$. The Euler-Lagrange equation for $\lambda$ is
\begin{equation}
\frac{1}{2\lambda^2}\left[\dot{X_{\mu}}\dot{X^{\mu}} + 2\rho_a\dot{X_{\mu}}\partial_aX^{\mu} + \rho_a\rho_b\partial_aX_{\mu}\partial_bX^{\mu}\right] + \frac{1}{2}\bar{h} = 0\label{NG1}
\end{equation}
and that for $\rho_a$ are
\begin{equation}
\partial_{a}X_{\mu}\partial_{b}X^{\mu} \rho_{b} = -\dot{X}_{\mu}\partial_{a}X^{\mu}
\label{NG2}
\end{equation}
From the last equation we can solve $\rho_{a}$
\begin{equation}
\rho_{a} = -h_{0b}\bar h^{ba}
\label{NG3}
\end{equation}
where $h_{ab}$ has been defined below equation (\ref{I1}) and $\bar h^{ab}$ \footnote {Note that $\bar h^{ab}$ is different from the space part of $h^{ij}.$} is the inverse matrix of $h_{ab}$. Using (\ref{NG3}) in (\ref{NG1}) we get after some calculations
\begin{equation}
\lambda = -\frac{\sqrt{-h}}{\bar{h}}
\label{NG4}
\end{equation}
where we take the negative sign due to similar reason as in the string case. Substituting $\rho_a$ and $\lambda$ in (\ref{I5}) we retrieve the N--G action. The reduction is completely analogous to the string case. In fact the solutions to $\rho_a$ and $\lambda$ go to the corresponding solutions of the string case when only one spatial degree of freedom is retained in the brane volume.

    Already in the string case the reduction of the interpolating action to the Polyakov form was non-trivial.
In case of the membrane it is further complicated by a mismatch in the number of degrees of freedom count. The Polyakov action of the membrane contains six independent metric components. Thus there are six more independent fields apart from $X^{\mu}$. In contrast, the interpolating Lagrangean contain only three additional fields ($\rho_1$, $\rho_2$ and $\lambda$). \footnote{This mismatch is important for the choice of gauge fixing conditions in the Polyakov theory \cite{t, rbk}. Here it affects the simulation of the Polyakov Lagrangean from the interpolating Lagrangean. Understandably, this mismatch will be more pronounced for higher branes. Note that there is no such mismatch in the string case.} It will thus be required to introduce $(6 - 3 =)\,3$ arbitrary variables to get the Polyakov Lagrangean from the interpolating one. The interpretation of such variables will then be investigated self-consistently. We, therefore, modify (\ref{I5}) as 
\begin{eqnarray}
{\cal{L}}_I & =& -\frac{1}{2\lambda}\left[\dot{X^{\mu}}\dot{X_{\mu}} + 2\rho_a\dot{X_{\mu}}\partial_aX^{\mu} + \left(\rho_a\rho_b\partial_aX^{\mu}\partial_bX_{\mu} - \lambda^{2} S_{ab}\partial_{a}X_{\mu} \partial_{b} X^{\mu}
\right)\right]\nonumber\\ && - \frac{\lambda}{2}\left(S_{ab} \partial_{a} X_{\mu} \partial_{b} X^{\mu} - \bar{h} - \rm{det S}\right) -\frac{\lambda}{2}\rm{det S}
\label{NG5}
\end{eqnarray}
Here $S_{ab}$ is a $2\times 2$ symmetric matrix whose elements are arbitrary functions of $\xi^{i}$, the parameters labeling the membrane volume. Note that we have introduced as many arbitrary functions which are needed to match the extra number of degrees of freedom as mentioned above. Now exploiting the arbitrariness of the functions $S_{ab}$ we demand that they be chosen to satisfy
\begin{equation}
S_{ab} \partial_{a} X_{\mu} \partial_{b}X^{\mu} - \bar{h} - {\rm{det}}S = 0
\label{NG6}
\end{equation}
The condition (\ref{NG6}) can be written in a suggestive form if we substitute 
\begin{equation}
S_{ab} = \epsilon_{ac} \epsilon_{bd} G_{cd} 
\label{NG61}
\end{equation}
It is easy to check that ${\rm{det}}S = {\rm{det}}G $. Using (\ref{NG61}), the condition (\ref{NG6}) can be cast as
\begin{equation}
\mathrm{det}\left(G_{ab} - h_{ab}\right) = 0
\label{NG66}
\end{equation}
We observe that this condition is reminiscent of a weaker version of the first class constraint $g_{ab} = h_{ab}$ of the Polyakov action. In the following we will find that this coincidence is not accidental. Now substituting $S_{ab}$ by $G_{ab}$ in (\ref{NG5}) we get
\begin{eqnarray}
{\cal{L}}_I & =& -\frac{1}{2\lambda}\left[\dot{X^{\mu}}\dot{X_{\mu}} + 2\rho_a\dot{X_{\mu}}\partial_aX^{\mu} + \left(\rho_a\rho_b\partial_aX^{\mu}\partial_bX_{\mu} - \lambda^{2} \epsilon_{ac} \epsilon_{bd} G_{cd}\partial_{a}
 X_{\mu} \partial_{b} X^{\mu}\right)\right]\nonumber\\
&& - \frac{\lambda}{2}\left(\epsilon_{ac} \epsilon_{bd} G_{cd}\partial_{a} X_{\mu} \partial_{b} X^{\mu} - \bar{h} - \rm{det G}\right) -\frac{\lambda}{2}\rm{det G}
\label{NG51}
\end{eqnarray}
It is now possible to reduce equation (\ref{NG51}) in the form
\begin{eqnarray}
{\cal{L}}_I = -\frac{1}{2}\sqrt{-g} g^{ij}\partial_iX_{\mu}\partial_{j} X^{\mu} - \frac{\lambda}{2} {\rm{det}} G
\label{NG7}\\
{\rm where,} \qquad
g^{ij} = \left(-g\right)^{-\frac{1}{2}}
\left(\begin{array}{cc}
\frac{1}{\lambda} & \frac{\rho_a}{\lambda} \\ 
\frac{\rho_a}{\lambda} & \frac{\rho_{a} \rho_{b} - \lambda^2 \epsilon_{ac}\epsilon_{bd} G_{cd} }{\lambda} \\
\end{array}\right)
\label{idm}
\end{eqnarray}
From the above identification we get after a straightforward calculation that
\begin{equation}
{\rm{det}} g^{ij} = \frac{\lambda {\rm{det}}G }{\left(-g\right)^{\frac{3}{2}}}
\label{NG8}
\end{equation}
But we require ${\rm {det}} g^{ij} $ = $ g^{-1} $. Comparing, we get the condition
\begin{equation}
\lambda {\rm{det}} G = -\sqrt{-g}
\label{NG81}
\end{equation}
Using the above condition in (\ref{NG7}) we find
\begin{equation}
{\cal{L}}_I = -\frac{1}{2}\sqrt{-g}\left(g^{ij}\partial_{i} X_{\mu}
\partial_{j} X^{\mu} - 1 \right)
\label{NG9}
\end{equation}
which is the Polyakov version of the membrane action. Note the automatic appearance of the cosmological term  in this simulation. This is a new feature for the membrane which was not present in the analogous construction
for the string.

    At this point it is appropriate to check the consistency of the above construction as we did in the string case. Referring to the identification (\ref{idm}) we find from the expressions of $g_{00}$ and $g_{0a}$
\begin{eqnarray}
\frac{1}{\lambda} = \sqrt{-g} g^{00}, \qquad
\rho_{a} = \frac{g^{0a}}{g^{00}}
\label{c1a}
\end{eqnarray}
From the above expressions we can solve $\rho_{a}$ and $\lambda$ in terms of the appropriate elements of $g^{ij}$. Since $\rho_{a}$ and $\lambda$ occur in different specific combinations in the space part 
\begin{equation}
g^{ab} = \left(-g\right)^{-\frac{1}{2}}
         \frac{\rho_{a} \rho_{b} - \lambda^2 \epsilon_{ac}\epsilon_{bd}G_{cd} } {\lambda} 
\label{idm1}
\end{equation}
it is necessary to see what happens when the above solutions of $\rho_{a}$ and $\lambda$ are substituted in (\ref{idm1}). In particular, from
\begin{equation}
g ^{11} = \frac{1}{\sqrt {- g}}\frac{\rho_1^2 - \lambda^{2} G_{22}}{\lambda}
\label{c2}
\end{equation}
we get after some manipulations 
\begin{equation}
G_{22} = \frac{g^{00}g^{11} - (g^{01})^2}{g^{-1}} = g_{22}\label{c3}
\end{equation}
Similarly, starting from the remaining terms of (\ref{idm1}) we arrive at
\begin{equation}
G_{ab} = g_{ab}
\label{c4}
\end{equation}
The arbitrary functions $G_{ab}$ introduced earlier are thus identified with the spatial part of $g_{ij}$. Note that this coincidence is due to the special choice of the arbitrary functions (\ref{NG61}).\footnote{From (\ref{c4}) we observe that equation (\ref{NG66}) is really the weaker version of the first class constraint $g_{ab} = h_{ab}$ following from the Polyakov action.} Further, from (\ref{c4}) we get 
\begin{equation}
{\rm{det}}G = \bar g
\label{c5}
\end{equation}
where $\bar g$ is the determinant of $g_{ab}$. The solution of $\lambda$ from (\ref{c1a}) is 
\begin{eqnarray}
\lambda = \frac{1}{\sqrt{-g} g^{00}}
\label{c6}
\end{eqnarray}
Hence we can calculate $\lambda {\rm{det}}G$ as 
\begin{eqnarray}
\lambda {\rm{det}}G = \frac{1}{\sqrt{-g} g^{00}} \bar g
\label{c7}
\end{eqnarray}
But $\bar g = g g^{00}$. Substituting this in (\ref{c7}) we see that the value of $\lambda {\rm{det}}G$ is the same as (\ref{NG81}). Therefore, the identification (\ref{c4}) is consistent with (\ref{NG81}). Finally, one may enquire whether the form of $G_{ab}$ given by (\ref{c4}) is consistent with direct computation of the inverse of (\ref{idm}). It is indeed gratifying to observe that the space part of the inverse matrix coincides with $G_{ab}$. The consistency of the construction (\ref{idm}) is thus completely verified.
\section{Gauge symmetry and Diffeomorphism of the membrane}
\label{membranesym}
\pagestyle{myheadings}
\markright{Gauge symmetry and Diffeomorphism of the membrane}
The investigation of the gauge symmetry of the interpolating membrane can be pursued following essentially the same steps as in the string case discussed earlier in section-$4$. There we argued that the gauge variations of the fields $\rho$ and $\lambda$ can be obtained using the symmetries of the N-G action, where they appear as Lagrange multipliers, by applying the formula (\ref{218}). The same arguments also apply here. So we construct the generator of the gauge transformations as 
\begin{equation}
G = \int d\xi \alpha_{i}\left( \xi \right) \Omega_{i}
\left( \xi \right) ; \left(i= 1,2,3 \right)
\label{GM1}
\end{equation}
where $\Omega_{i}$ are the constraints given in (\ref{I1}), and $ \alpha_{i}\left( \xi \right) $ are the three arbitrary gauge parameters. The algebra of the constraints can be worked out using (\ref{I1}) and the basic PB's of the theory which are similar to (\ref{116}) with the world-volume parameters $\left(\tau, \sigma_{1}, \sigma_{2}\right)$.
\begin{eqnarray}
\left\{ \Omega_{a}\left( \xi \right), \Omega_{b}\left( \xi^{\prime} \right)
\right\} &=& \left[\Omega_{b}\left( \xi \right) \partial_{a}\left(\delta
\left(\xi - \xi^{\prime} \right)\right)
- \Omega_{a}\left( \xi^{\prime} \right)  \partial_{b}^{\prime} \left(\delta
\left(\xi - \xi^{\prime} \right)\right)\right] 
\nonumber
\\
\left\{ \Omega_{a}\left( \xi \right), \Omega_{3}\left( \xi^{\prime} \right)
\right\} &=& \left[\Omega_{3}\left( \xi \right) + \Omega_{3}\left( \xi^{\prime}
\right)\right] \partial_{a}\left(\delta\left(\xi - \xi^{\prime}\right)\right) 
\nonumber
\\
\left\{ \Omega_{3}\left( \xi \right), \Omega_{3}\left( \xi^{\prime} \right)
\right\} &=& 4 \left[ \bar {h}\left( \xi \right)\bar {h}^{ab}
\left( \xi \right)\Omega_{b}\left( \xi \right) + \bar {h}\left( \xi^{\prime} \right)\bar {h}^{ab}\left( \xi^{\prime} \right)
\Omega_{b}\left( \xi^{\prime} \right) \right]
\partial_{a}\left(\delta\left(\xi - \xi^{\prime}\right)\right) \nonumber\\
\label{d1}
\end{eqnarray}
From this algebra we read off the non-zero structure functions\footnote{These cumbersome expressions will be brought into a nice compact form when we shall deal with the $p$-brane case} as defined in (\ref{315}),
\begin{eqnarray}
C_{11}{}^{1}\left( \xi, \xi^{\prime}, \xi^{\prime \prime} \right)
&=& \left[ \delta\left( \xi - \xi^{\prime \prime} \right) +
\delta\left( \xi^{\prime} - \xi^{\prime \prime} \right) \right]
\partial_{1}\left(\delta \left( \xi - \xi^{\prime} \right) \right) \nonumber \\
C_{12}{}^{1}\left( \xi, \xi^{\prime}, \xi^{\prime \prime} \right)
&=& \delta\left( \xi^{\prime} - \xi^{\prime \prime} \right)
\partial_{2}\left(\delta \left( \xi - \xi^{\prime} \right) \right), \qquad 
C_{12}{}^{2}\left( \xi, \xi^{\prime}, \xi^{\prime \prime} \right)
= \delta\left( \xi - \xi^{\prime \prime} \right)
\partial_{1}\left(\delta \left( \xi - \xi^{\prime} \right) \right) \nonumber \\
C_{21}{}^{1}\left( \xi, \xi^{\prime}, \xi^{\prime \prime} \right)
&=& \delta\left( \xi - \xi^{\prime \prime} \right)
\partial_{2}\left(\delta \left( \xi - \xi^{\prime} \right) \right), \qquad 
C_{21}{}^{2}\left( \xi, \xi^{\prime}, \xi^{\prime \prime} \right)
= \delta\left( \xi^{\prime} - \xi^{\prime \prime} \right)
\partial_{1}\left(\delta \left( \xi - \xi^{\prime} \right) \right) \nonumber \\
C_{22}{}^{2}\left( \xi, \xi^{\prime}, \xi^{\prime \prime} \right)
&=& \left[ \delta\left( \xi - \xi^{\prime \prime} \right) +
\delta\left( \xi^{\prime} - \xi^{\prime \prime} \right) \right]
\partial_{2}\left(\delta \left( \xi - \xi^{\prime} \right) \right) \nonumber \\
C_{13}{}^{3}\left( \xi, \xi^{\prime}, \xi^{\prime \prime} \right)
&=& C_{31}{}^{3}\left( \xi, \xi^{\prime}, \xi^{\prime \prime} \right)
= \left[ \delta\left( \xi - \xi^{\prime \prime} \right) +
\delta\left( \xi^{\prime} - \xi^{\prime \prime} \right) \right]
\partial_{1}\left(\delta \left( \xi - \xi^{\prime} \right) \right) \nonumber \\
C_{23}{}^{3}\left( \xi, \xi^{\prime}, \xi^{\prime \prime} \right)
&=& C_{32}{}^{3}\left( \xi, \xi^{\prime}, \xi^{\prime \prime} \right)
= \left[ \delta\left( \xi - \xi^{\prime \prime} \right) +
\delta\left( \xi^{\prime} - \xi^{\prime \prime} \right) \right]
\partial_{2}\left(\delta \left( \xi - \xi^{\prime} \right) \right) \nonumber \\
C_{33}{}^{1}\left( \xi, \xi^{\prime}, \xi^{\prime \prime} \right)
&=& 4 \left[ \bar {h}\left( \xi \right) \bar {h}^{11}\left( \xi \right)
\partial_{1}\left(\delta\left(\xi - \xi^{\prime}\right)\right) + \bar {h}\left( \xi \right) \bar {h}^{21}\left( \xi \right) \partial_{2}\left(\delta\left(\xi - \xi^{\prime}\right)\right) \right]
 \delta \left(\xi -\xi^{\prime \prime} \right)\nonumber \\
&& + 4 \left[ \bar {h}\left( \xi^{\prime} \right)\bar {h}^{11}
\left( \xi^{\prime} \right) \partial_{1}\left(\delta\left(\xi - \xi^{\prime}
\right)\right) + \bar {h}\left( \xi^{\prime} \right)\bar {h}^{21}
\left( \xi^{\prime} \right) \partial_{2}\left(\delta\left(\xi - \xi^{\prime}
\right)\right) \right]\delta \left(\xi^{\prime}  - \xi^{\prime \prime}
\right) \nonumber \\
C_{33}{}^{2}\left( \xi, \xi^{\prime}, \xi^{\prime \prime} \right)
&=& 4 \left[ \bar {h}\left( \xi \right) \bar {h}^{12}\left( \xi \right)
\partial_{1}\left(\delta\left(\xi - \xi^{\prime}\right)\right) + \bar {h}\left( \xi \right) \bar {h}^{22}\left( \xi \right)
\partial_{2}\left(\delta\left(\xi - \xi^{\prime}\right)\right) \right]
 \delta \left(\xi -\xi^{\prime \prime} \right)\nonumber \\
&&+ 4 \left[ \bar {h}\left( \xi^{\prime} \right)\bar {h}^{12}
\left( \xi^{\prime} \right) \partial_{1}\left(\delta\left(\xi - \xi^{\prime}
\right)\right) + \bar {h}\left( \xi^{\prime} \right)\bar {h}^{22}
\left( \xi^{\prime} \right) \partial_{2}\left(\delta\left(\xi - \xi^{\prime}
\right)\right) \right]\delta \left(\xi^{\prime}  - \xi^{\prime \prime}
\right) \nonumber \\
\label{d2}
\end{eqnarray}
Relabeling $\rho_{a}$ and $\lambda$ as
\begin{equation}
\lambda_{a} = \rho_{a} \hspace{1cm} \rm{and} \hspace{1cm}
\lambda_{3} = \frac{\lambda}{2}
\label{d3}
\end{equation}
and using the structure functions (\ref{d2}) we calculate the required gauge variations by applying equation (\ref{314})
\begin{eqnarray}
\delta \lambda_{a} &=& - \dot \alpha_{a}
+ \left(\alpha_{b}\partial_{b}\lambda_{a}
 - \lambda_{b}\partial_{b}\alpha_{a} \right) + 4 \bar{h}\bar{h}^{ab}
\left(\alpha_{3}\partial_{b}\lambda_{3} - \lambda_{3}\partial_{b}
\alpha_{3}\right)\nonumber\\
\delta \lambda_{3} &=& -\dot \alpha_{3}
+\left(\alpha_{3}\partial_{a}\lambda_{a} - \lambda_{a} \partial_{a}\alpha_{3}
\right)
+\left(\alpha_{a}\partial_{a}\lambda_{3} - \lambda_{3} \partial_{a}\alpha_{a}
\right)
\label{GM4}
\end{eqnarray}
Now we use equation (\ref{d3}) to convert $\lambda_{i}$ back to
 $\rho_1$, $\rho_2$ and $\lambda$\footnote{Note that these variations of $\rho_{a}$ and $\lambda$ can be directly obtained from the definitions (\ref{NG3}) and (\ref{NG4}) in complete parallel with the analogous computation for string (see under equation (\ref{GM51})).}
\begin{eqnarray}
\delta \rho_{a} &=& - \dot \alpha_{a}
+ \left(\alpha_{b}\partial_{b}\rho_{a}
 - \rho_{b}\partial_{b}\alpha_{a} \right) + 2 \bar{h}\bar{h}^{ab}
\left(\alpha_{3}\partial_{b}\lambda - \lambda\partial_{b}\alpha_{3}\right)
\nonumber\\
\delta \lambda &=& -2 \dot \alpha_{3}
+2 \left(\alpha_{3}\partial_{a}\rho_{a} - \rho_{a} \partial_{a}\alpha_{3}
\right)
+\left(\alpha_{a}\partial_{a}\lambda - \lambda \partial_{a}\alpha_{a}\right)
\label{GM5}
\end{eqnarray}
At this point it is instructive to study the string limit of equations (\ref{GM5}). In the string limit we put $a = b = 1$. The matrix $h_{ab} = \partial_{a}X^{\mu}\partial_{b}X_{\mu}$ now contains only one term, namely $ h_{11}$. So $\bar h^{ab}$ now also contains a single term $ h^{11} = \frac{1}{h_{11}}$. Hence in the string limit $\bar h \bar h^{ab}$ becomes $h_{11} \times \frac{1}{h_{11}}$ i.e. $1$.  It is now apparent that in the string limit we recover the expressions (\ref{GM51}) from (\ref{GM5}) with the replacement of $\alpha_{3}$ by $\alpha_{2}$.

Now we are in a position to investigate the parallel between gauge symmetry and reparametrization symmetry of the membrane actions using our alternative approaches developed in the string example. We, therefore, require to find a mapping between the two sets of transformation parameters corresponding to gauge and diff. transformations. This is achieved by comparing the variations of $X^{\mu}$ under the two types of transformation. The variations of $X^{\mu}$ in (\ref{I5}) under (\ref{GM1}) is 
\begin{equation}
\delta X_{\mu} = \left\{X_{\mu}, G \right\}
 = \left( \alpha_{a} \partial_{a} X_{\mu} + 2 \alpha_{3} \Pi_{\mu} \right)
\label{GM7}
\end{equation}
Looking at the scenario from the point of view of Polyakov action we find that under reparametrization $ \left( \tau^{\prime} \to \tau - \Lambda_{0}, \sigma_{1}{}^{\prime} \to \sigma_{1} - \Lambda_{1}, \sigma_{2}{}^{\prime} \to \sigma_{2} - \Lambda_{2} \right)$ the variations of $X^{\mu}$ is
\begin{equation}
\delta X^{\mu} = \Lambda^{i} \partial_{i} X^{\mu} = \Lambda^{0} \dot X^{\mu} + \Lambda^{1} \partial_{1} X^{\mu}
+ \Lambda^{2} \partial_{2} X^{\mu}
\label{GM8}
\end{equation}
Using (\ref{I4}) we eliminate $\dot X^{\mu}$ in terms of the momenta $\Pi^{\mu}$ in equation (\ref{GM8}). Now comparing with (\ref{GM7}) we obtain the mapping
\begin{eqnarray}
\Lambda^{0} = - 2 \sqrt{- g } g^{00} \alpha_{3} = - \frac{ 2 \alpha_{3}}{\lambda}, \qquad
\Lambda^{a} = \alpha_{a} - 2 \sqrt{- g } g^{0a} \alpha_{3} = \alpha_{a} - \frac{ 2 \rho_{a} \alpha_{3}}{\lambda}
\label{GM9}
\end{eqnarray}
Under (\ref{GM9}) the transformations on $X^{\mu}$ due to reparametrization become identical with its corresponding gauge variation. The complete equivalence between the transformations can again be established by computing $\delta \rho_{a}$ and $\delta \lambda$ from the alternative approaches. The mapping (\ref{idm}) yields,
\begin{equation}
\rho_{a} = \frac{g^{0a}}{g^{00}}
\label{GM11}
\end{equation}
We require to express these in terms of $g_{ij}$. To this end we start from the identity
\begin{equation}
g^{ij} g_{jk} = \delta^{i}{}_{k}
\label{GM12}
\end{equation}
and obtain the following equations for $\rho_{a}$
\begin{eqnarray}
\rho_{1} g_{11} + \rho_{2} g_{21} = - g_{01}, \qquad 
\rho_{1} g_{12} + \rho_{2} g_{22} = - g_{02}
\label{GM13a}
\end{eqnarray}
Solving the above equations we can express $\rho_{a}$ entirely in terms of $g_{ij}$. The variations of $g_{ij}$ under reparametrization is obtained from (\ref{3122}) where $i,j$ now assume values $0$, $1$ and $2$. So the corresponding variations of $\rho_{a}$ are given by 
\begin{eqnarray}
\delta \rho_{a} = - \partial_{0} \Lambda^{a} +
\rho_{a} \partial_{0} \Lambda^{0} - \rho_{b} \partial_{b} \Lambda^{a} 
 + \rho_{a}\rho_{b} \partial_{b} \Lambda^{0}
+ \Lambda^{k}\partial_{k} \rho_{a} - \lambda^{2}\epsilon_{ab}\epsilon_{cd} g_{bc}
\partial_{d} \Lambda^{0}
\label{GM14a}
\end{eqnarray}
Now introducing the mapping (\ref{GM9}) in (\ref{GM14a}) and using the first class constraints $g_{ab} = h_{ab}$ we find that the variations of $\rho_{a}$ are identical with their gauge variations in (\ref{GM5}). We then compute the variation of $\delta\lambda$. This can be conveniently done by starting from the variation of the ratio
\begin{equation}
\frac{g^{11}}{g^{00}} =\left(\rho_{1}^{2} - \lambda^2 g_{22}\right)
\label{3125A}
\end{equation}
obtained from the identification (\ref{idm}). Converting the l.h.s appropriately in terms of $g_{ij}$, we take the gauge variation to get
\begin{equation}
\delta \left\{\frac{g^{11}}{g^{00}}\right\} = \delta\left\{\frac{g_{00}g_{22} - g_{02}{}^{2}}{g_{11}g_{22} - g_{12}{}^{2}}\right\}= \delta \left(\rho_{1}^{2} - \lambda^2 g_{22}\right)
\label{3125B}
\end{equation}
and using the reparametrisation variations of $g_{ij}$ and $\rho_{a}$ we get the expression of $\delta \lambda$ in terms of the reparametrization parametres $\Lambda_{i}$. Using the mapping (\ref{GM9}) we substitute $\Lambda_{i}$ by $\alpha_{i}$ and  the resulting expression for $\delta \lambda$ agrees with that given in (\ref{GM5}). The complete matching thus obtained illustrates the equivalence of reparametrization symmetry with gauge symmetry for the membrane. What remains now is to generalise the results obtained so far in this chapter to the arbitrary $p-$brane case. 

\section{Generalisation to bosonic $p$-brane }
\label{pbrane}
\pagestyle{myheadings}
\markright{Generalisation to bosonic $p$-brane}
The $p$-brane sweeps out a $p+1$ dimensional world volume in the embedding $D-$dimensional space-time. The dynamics of the brane can be analysed either from the N--G or Polyakov action formalisms. 
Similar to the string and membrane case the equivalence of these two approaches is usually established by solving out the independent metric of the Polyakov action in favour of the space-time coordinates to arrive at the N--G action. We, on the contrary, have been addressing the reverse problem in this chapter by demonstrating how the independent metric can be generated by exploiting the gauge symmetry of the N--G action for the $p$-brane. An intermediate step is the construction of a $p$-brane interpolating action. Though such actions have been already introduced in section \ref{stringreview} and \ref{membranereview} the methods used were specific to the particular choice of $p=1 (strings), p=2 (membranes)$ only, which do not readily admit a generalisation to the arbitrary $p-$case which is essential for the present analysis. In this section we, therefore, construct the $p$-brane interpolating action from the usual N--G version and proceed further to generate the $\left(p+1\right)-$dimensional Polyakov metric in a generalised form applicable for arbitrary spatial dimensions. 

The interpolating action is based on the first class constraints of the N--G theory. We generate the independent Polyakov metric from the corresponding Lagrange multipliers enforcing these constraints. This reveals a deep connection of the metric components with the gauge symmetries of the brane. The mismatch between the number of independent gauge degrees of freedom and the number of independent metric elements brings out the arbitrariness inherent in the Polyakov formulation explicit in our construction. Fixing the arbitrariness in terms of the embedding makes the transition to the Polyakov form complete. Notably, the cosmological term emerges as a logical consequence of our analysis. 

The process of introducing the independent Polyakov metric in the world volume through the interpolating action formalism has a very interesting outcome. First class constraints of the N--G theory generate temporal development and also shifts in the space like directions. The independent metric constructed with the help of the Lagrange multipliers enforcing these constraints naturally emerge with a decomposition of the $(p+1)$-dimensional metric into the $p$-dimensional spatial part plus the multipliers which are the analogues of the lapse and shift variables of general relativity. Indeed, the metric generated in our formalism appears in a canonical form which is shown to be identical with the famous Arnowitt--Deser--Misner (A--D--M) representation in general relativity. In other words our analysis provides a genesis of the A-D-M representation from a string theoretic perspective.  
 
\subsection{Interpolating action for the bosonic $p$-brane}
\label{pbrane1}
\pagestyle{myheadings}
\markright{Interpolating action for the bosonic $p$-brane}
The $p$-brane is a $p-$dimensional object which sweeps out a $\left(p+1\right)-$dimensional world volume parametrised by $\tau$ and $\sigma_{a}$. The index $a$ run from $1$ to $p$. Henceforth these parameters are collectively referred as $\xi_{i}\left(\xi_{0}= \tau, \xi_{a}= \sigma_{a}\right)$. The N--G action of bosonic $p$-brane is the integrated  world volume 
\begin{eqnarray}
S_{\mathrm{NG}} = - \int d^{p+1}\xi \sqrt{-h}
\label{ngaction}
\end{eqnarray}
where $h$ is the determinant of the induced metric
\begin{equation}
h_{ij} = \partial_{i} X^{\mu}\partial_{j}X_{\mu}
\label{indhij}
\end{equation}
Note that, like earlier, we have also kept the $p$-brane tension implicit. The canonical momenta conjugate to $X_{\mu}$ are 
\begin{eqnarray}
\Pi_{\mu} = \frac{\bar{h}}{\sqrt{-h}}\left\{\partial_{0}X_{\mu} - \partial_{a}X_{\mu} {\bar{h}}^{ab} h_{0b}\right\}
\label{momentumX}
\end{eqnarray}
where ${\bar{h}}$ is the determinant of the matrix $h_{ab}$.  Also ${\bar{h}}^{ab}${\footnote{Note that $\bar h^{ab}$ is different from the space part of $h^{ij}.$}} is the inverse of $h_{ab}$. The primary constraints following from (\ref{momentumX}) are,  
\begin{eqnarray}
\Omega_{0} = \frac{1}{2}\left(\Pi^2 + \bar{h}\right) \approx 0; \quad
\Omega_{a} = \Pi_{\mu}\partial_{a}X^{\mu} \approx 0 
\label{constraints}
\end{eqnarray}
We use the nontrivial Poission's bracket of the theory  
\begin{eqnarray}
 \{X^{\mu}\left(\tau,\xi\right),
 \Pi_{\nu}\left(\tau,\xi^{\prime}\right)\} = \eta_{\nu}^{\mu}
 \delta\left(\xi - \xi^{\prime}\right)
\label{XPicommutator}
\end{eqnarray}
to work out the algebra of the constraints
\begin{eqnarray}
\left\{ \Omega_{0}\left( \xi \right), \Omega_{0}\left( \xi^{\prime} \right)
\right\} &=& 4 \left[ \bar {h}\left( \xi \right)\bar {h}^{ab}
\left( \xi \right)\Omega_{b}\left( \xi \right) +
\bar {h}\left( \xi^{\prime} \right)\bar {h}^{ab}\left( \xi^{\prime} \right)
\Omega_{b}\left( \xi^{\prime} \right) \right]
\partial_{a}\left(\delta\left(\xi - \xi^{\prime}\right)\right)\nonumber \\
\left\{ \Omega_{a}\left( \xi \right), \Omega_{0}\left( \xi^{\prime} \right)
\right\} &=& \left[\Omega_{0}\left( \xi \right) + \Omega_{0}\left( \xi^{\prime}
\right)\right] \partial_{a}\left(\delta\left(\xi - \xi^{\prime}\right)\right)
\nonumber\\
\left\{ \Omega_{a}\left( \xi \right), \Omega_{b}\left( \xi^{\prime} \right)
\right\} &=& \left[\Omega_{b}\left( \xi \right) \partial_{a}\left(\delta
\left(\xi - \xi^{\prime} \right)\right)
- \Omega_{a}\left( \xi^{\prime} \right)  \partial_{b}^{\prime} \left(\delta
\left(\xi - \xi^{\prime} \right)\right)\right] 
\label{constraintalgebra}
\end{eqnarray}
which clearly turned out to be weakly involutive. So that the set (\ref{constraints}) is first class. Since the $p$-brane action (\ref{ngaction}) possesses reparametrization invariance, the canonical Hamiltonian vanishes. Thus the total Hamiltonian is only a linear combination of the constraints (\ref{constraints})
\begin{equation}
{\cal{H}}_{T} = - \frac{\lambda_{0}}{2}\left(\Pi^2 + \bar{h}\right)  - \lambda^{a}\Pi_{\mu}\partial_{a}X^{\mu} 
\label{totalHamiltnian}
\end{equation}
where $\lambda_{0}$ and $\lambda^{a}$ are the Lagrange multipliers.

The Polyakov action, on the other hand, is given by
\begin{equation}
S_{\mathrm{P}} = -\frac{1}{2}\int d^{p+1}\xi{\sqrt - g}\left\{g^{ij} \partial_{i} X^{\mu}\partial_{j} X_{\mu} -\left(p - 1\right)\right\}
\label{polyakovaction}
\end{equation}
where metric $g_{ij}$ are considered as independent fields. The equations of motion for $g_{ij}$ are 
\begin{equation}
g_{ij} = h_{ij}
\label{polyeqnmotion}
\end{equation}
Substituting these in (\ref{polyakovaction}) one can retrieve the N--G form (\ref{ngaction}). Note that the cosmological term ${\sqrt - g}\left(p - 1\right)$ in the action vanishes for $p = 1$. We thus observe that the presence of the cosmological term is characteristic of the higher branes as opposed to the strings. The reason for this difference is the Weyl invariance of the string which is not shared by the higher branes. In our action level construction this cosmological term will emerge systematically. 

 To construct the interpolating action for the $p$-brane, we proceed as we did for the string and membrane case. We first consider the Lagrange multipliers as independent fields and write an alternative first order Lagrangian 
\begin{equation}
{\cal{L}}_I = \Pi_{\mu}\dot{X}^{\mu} - {\cal{H}}_{T}
\label{firstorderLagrangean}
\end{equation}
The equation of motion for $\Pi_{\mu}$ following from (\ref{firstorderLagrangean}) is
\begin{equation}
\Pi_{\mu} = -\frac{\dot{X}_{\mu}+\lambda^a\partial_{a}X_{\mu}}{\lambda_{0}}
\label{Pieqnmotion}
\end{equation}
Substituting $\Pi_{\mu}$ from (\ref{Pieqnmotion}) in the first order Lagrangean (\ref{firstorderLagrangean}) we get the interpolating Lagrangian for the $p$-brane
\begin{eqnarray}
{\cal{L}}_I = -\frac{1}{2\lambda_{0}}\left[\dot{X}^{\mu}\dot{X}_{\mu}
                 + 2\lambda^{a}\dot{X}_{\mu}\partial_{a}X^{\mu}
                 +\lambda^{a}\lambda^{b}\partial_{a}X^{\mu}\partial_{b}X_{\mu}\right]
                 +\frac{\lambda_{0}}{2}\bar{h}
\label{interpolatingLagrangean}
\end{eqnarray}
Before we engage ourselves in the construction of the Polyakov metric from the interpolating action which is a rather complicated task, let us briefly describe the trivial passage to the N--G form. From the interpolating Lagrangian it is easy to write down the equations of motion for $\lambda_{0}$ and $\lambda^{a}$
\begin{eqnarray}
\lambda_{0}{}^{2} = \frac{-h}{{\bar{h}}^{2}}; \qquad \lambda^{a} = -h_{0b}\bar h^{ba}
\label{LambdaRhoeqnmotion}
\end{eqnarray}
From the first equation of (\ref{LambdaRhoeqnmotion}) $\lambda_{0}$ is determined modulo a sign. This can be fixed by demanding the consistency of (\ref{momentumX}) with (\ref{Pieqnmotion}), the equation of motion for $\Pi_{\mu}$ following from the first order Lagrangian (\ref{firstorderLagrangean}). Thus we have 
\begin{equation}
\lambda_{0} = -\frac{\sqrt{-h}}{\bar{h}}
\label{Lambda}
\end{equation}
Substituting $\lambda^a$ and $\lambda_{0}$ in (\ref{interpolatingLagrangean}) we retrieve the Nambu--Goto action (\ref{ngaction}). 
\subsection{Derivation of the Polyakov metric}
\label{pbrane2}
\pagestyle{myheadings}
\markright{Derivation of the Polyakov metric}
The reduction of the interpolating Lagrangian to the Polyakov form of the $p$-brane action is highly non-trivial. 
In deriving the interpolating Lagrangean from the N--G theory we have promoted the $\left(p+1\right)$ Lagrange 
multipliers as independent fields. Note that in the Polyakov action the extra degrees of freedom is more than this number. The precise size of the mismatch is $\left(p\right)\left(p+1\right)/2$. We thus observe that the interpolating action is a less redundant description than the Polyakov action. So to make the transition from the interpolating Lagrangean to the Polyakov form we require to introduce just as many independent fields. This can be done by including an arbitrary spatial part $G^{ab}$ in ${\cal{L}}_I$, which has the right number of independent components. We therefore modify the interpolating Lagrangian (\ref{interpolatingLagrangean}) for the $p$-brane in the following way\footnote{This specific choice of the arbitrary part will be convenient in the subsequent calculation.} 
\begin{eqnarray}
{\cal{L}}_I & = & -\frac{1}{2\lambda_{0}}\left[\dot{X}^{\mu}\dot{X}_{\mu}
                 + 2\lambda^a\dot{X}_{\mu}\partial_a X^{\mu} + \left(\lambda^a\lambda^b \partial_a X^{\mu}\partial_b X_{\mu} 
                  -   \lambda_{0}{}^{2}\bar{G} G^{ab}  \partial_{a}X_{\mu} \partial_{b} X^{\mu}\right)\right]\nonumber \\
                  &&- \frac{\lambda_{0}}{2}\left(\bar{G} G^{ab}
                     \partial_{a} X_{\mu} \partial_{b} X^{\mu} - \bar{h}\right) 
\label{modintL}
\end{eqnarray}
where $\bar{G}$ is the determinant of $G_{ab}$ which is the inverse of the arbitrary matrix $G^{ab}$, $\left( a,b = 1, 2,... p\right)$. Observe that (\ref{modintL}) can be cast as 
\begin{eqnarray}
{\cal{L}}_I = -\frac{1}{2}\sqrt{-g} g^{ij}\partial_i X_{\mu}
               \partial_j X^{\mu} - \frac{\lambda_{0}}{2}\left(\bar{G} G^{ab}
              \partial_{a} X_{\mu} \partial_{b} X^{\mu} - \bar{h}\right) \label{almostpoly}\\
{\rm where,} \qquad 
g^{ij} = \left(-g\right)^{-\frac{1}{2}}
\left(\begin{array}{cc}
\frac{1}{\lambda_{0}} & \frac{\lambda^a}{\lambda_{0}} \\
\frac{\lambda^b}{\lambda_{0}} & \frac{\lambda^{a} \lambda^{b} - \lambda_{0}{}^2 \bar{G} G^{ab} }{\lambda_{0}} \\
\end{array}\right)
\label{idmp}
\end{eqnarray}
Since $g = {\mathrm det} g_{ij}$ and $g_{ij}$ is the inverse of $g^{ij}$ this imposes stringent constraints on the 
construction (\ref{idmp}). So its consistency must explicitly be examined. Observe that by exploiting the dynamics of the $p$-brane we are able to generate an independent metric on the world volume of the brane. The arbitrary function $G_{ab}$ signifies a fundamental elasticity in the spatial part of the metric.
The Lagrangean (\ref{almostpoly}) is almost in the required Polyakov form except for the omission of the cosmological constant. Also there is an additional term which is not there in the Polyakov Lagrangean. It is precisely the consistency requirement of the construction (\ref{idmp}) which identifies this extra piece in (\ref{almostpoly}) with the cosmological constant, provided we fix the elasticity in the embedding. The validity of these assertions will be demonstrated now. 

From the identification (\ref{idmp}) we get after a straightforward calculation that
\begin{equation}
{\rm{det}} g^{ij} = \left(-1\right)^{p}\frac{\lambda_{0}{}^{\left(p-1\right)}}{\left(\sqrt{-g}\right)^{\left(p+1\right)}}{\rm{det}}\left(\bar G G^{ab}\right)
\label{detidm}
\end{equation}
But we require ${\mathrm{det}} g^{ij}$ = $ g^{-1} $. Comparing, we get the condition
\begin{equation}
\lambda_{0}{}^{\left(p-1\right)} = \left(-1\right)^{\left(1-p\right)} \left(\frac{\sqrt{-g}}{{\bar G}}\right)^{\left(p-1\right)}
\label{consistancy1}
\end{equation}
Starting from our construction (\ref{idmp}) one can solve for $\lambda_{0}$ and $\lambda^{a}$ as 
\begin{eqnarray}
\frac{1}{\lambda_{0}} = \sqrt{-g} g^{00}, \quad
\lambda^{a}= \frac{g^{0a}}{g^{00}}
\label{c1}
\end{eqnarray}
Using (\ref{c1}) in (\ref{idmp}), we get after a few steps 
\begin{equation}
{G}^{ab} = \frac{g}{\bar G}\left(g^{ab} g^{00} - g^{0a} g^{0b}\right)
\label{spatialidm}
\end{equation}
Inverting ${G}^{ab}$ we arrive at
\begin{equation}
 G_{ab} = \left(\frac{g g^{00}}{{\bar G}}\right)g_{ab}
\label{inversespatialidm}
\end{equation}
From (\ref{spatialidm}) we obtain after some calculations  
\begin{equation}
{\mathrm {det}}{G}^{ab} = \left(\frac{g}{\bar G}\right)^{p}{\mathrm {det}}{g}^{ij} \left(g^{00}\right)^{p-1}  
\label{detinversespatialidm}
\end{equation}
But, by definition, ${\mathrm {det}}{G}^{ab} = 1/\bar{G}$. Using this in (\ref{detinversespatialidm}) we find,
\begin{equation}
{\bar G}^{\left(p-1\right)} = \left(g g^{00}\right)^{\left(p-1\right)}
\label{consistancy2}
\end{equation}
There is an apparent ambiguity of sign in determining $\bar {G}$ from (\ref{consistancy2}) when $p$ is odd. For now we take the positive solution for all $p$. Then from (\ref{inversespatialidm})
\begin{equation}
G_{ab} = g_{ab}
\label{elasticity}
\end{equation}
The consistency requirment thus restricts the arbitrariness of $G_{ab}$ through (\ref{elasticity}).
We use (\ref{consistancy1}) and (\ref{elasticity}) to express the $\sqrt{-g}$ factor in terms of the $\bar{G}$ and $\lambda_{0}$ as{\footnote{ Note that for odd $p$ another sign ambiguity appears here. This is actually related with the corresponding uncertainty about sign stated above. We shall explore the connection subsequently.}}
\begin{equation}
\sqrt{-g} = -\lambda_{0} \bar{G} = - \lambda_{0} {\rm det} g_{ab}
\label{theg}
\end{equation}
Finally, with (\ref{theg}) we are ready to reduce the interpolating action to the Polyakov form. Note that the spatial part of the metric $g^{ij}$ is still arbitrary. Also no attention has so far been paid to the background space time in which the brane is embedded. We now propose the rigid structure 
\vskip -0.50cm
\begin{equation}
g_{ab} = h_{ab}
\label{rigidity}
\end{equation}
\vskip -0.25cm
Note that this is just the spatial part of (\ref{polyeqnmotion}) which is required to demonstrate the equivalence of the Polyakov form with the N--G. Now equation (\ref{rigidity}), along with (\ref{elasticity}), imposes 
\begin{equation}
G_{ab} = h_{ab}
\label{result}
\end{equation}
Plugging it in the Lagrangean (\ref{almostpoly}) and using (\ref{theg}) we find that the last term of (\ref{almostpoly}) is precisely equal to the cosmological constant occurring in the Polyakov action (\ref{polyakovaction}). This completes the reduction of the interpolating Lagrangian to the Polyakov form. The connection (\ref{rigidity}) fixes the brane in its embedding. A couple of interesting observations also follow from this. First, we can understand now the nature of the ambiguities of sign encountered above for odd $p$ more clearly. If we chose the opposite sign in (\ref{elasticity}) then we would have $G_{ab} = - h_{ab}$ and $\lambda_{0}$ should then be expressed from (\ref{consistancy1}) as $\lambda_{0} = \frac{\sqrt{-g}}{\bar{G}}$. Otherwise there would be contradiction with (\ref{Lambda}). Next, for $p = 1$ we find that the imposed rigidity admits a residual scale transformation. This is the well known Weyl invariance of the string. 
\subsection{Emergence of Arnowitt--Deser--Misner decomposition from the brane dynamics}
\label{pbrane3}
\pagestyle{myheadings}
\markright{Emergence of Arnowitt--Deser--Misner decomposition from the brane dynamics}
The interpolating action formalism enables us to introduce an independent metric in the world volume swept out by the N--G brane. The process depends crucially on the first class constraints of the theory. We have also clearly identified the arbitrariness in the spatial part of the metric in (\ref{idmp}). Our method thus introduces the metric in a very special way such that the world volume is decomposed into the $p$ dimensional spatial part along with the multipliers which generate temporal evolution  with shifts in the space-like directions. This decomposition of the metric is reminiscent of the A--D--M decomposition in geometrodynamics. 
We now show how the A--D--M decomposition emarges from the dynamics of the generic $p$-brane.
To see this we have to use (\ref{theg}) to first express the metric in terms of its arbitrary spatial part and the Lagrange multipliers only. The construction (\ref{idmp}) then reduces to
\begin{equation}
g^{ij} = 
\left(\begin{array}{cc}
-\frac{1}{\lambda_{0}{}^{2}{\rm det} g_{ab}} & -\frac{\lambda^a}{\lambda_{0}{}^{2}{\rm det} g_{ab}} \\
-\frac{\lambda^b}{\lambda_{0}{}^{2}{\rm det} g_{ab}} & \left(\bar{g}^{ab} -\frac{\lambda^{a} \lambda^{b}}{\lambda_{0}{}^{2}{\rm det} g_{ab}}\right) \\
\end{array}\right)
\label{idmm}
\end{equation}
where $\bar{g}^{ab}$ is the inverse of the spatial metric $g_{ab}$. Note that as in the case of $\bar{h}^{ab}$, $\bar{g}^{ab}$ is also different from the spatial part of the identification matrix $g^{ij}$. In the A--D--M construction the metric $g^{ij}$ of physical space time is represented as 
\begin{equation}
g^{ij} = 
\left(\begin{array}{cc}
-\frac{1}{\left(N\right)^{2}} & \frac{N^a}{\left(N\right)^{2}} \\
\frac{N^b}{\left(N\right)^{2}} & \left(\bar{g}^{ab} -\frac{N^{a} N^{b}}{\left(N\right)^{2}}\right) \\
\end{array}\right)
\label{idmm1}
\end{equation}
where $N$ and $N^{a}$ are respectively the lapse and shift variables and $\bar{g}^{ab}$ is the inverse of the `metric' $g_{ab}$ on the spatial hypersurface. Using the correspondence 
\begin{eqnarray}
\left(N^{a}\right)\mapsto -\lambda^{a},\quad \mathrm {and}\quad N \mapsto \lambda_{0} \sqrt{{\rm det} g_{ab}}
\label{ADM5}
\end{eqnarray}
it is easy to convince oneself that the A--D--M decomposition of the brane volume emerges from our analysis. Note that in the correspondences (\ref{ADM5}) apart from the Lagrange multipliers only the  space part of the metric $g_{ij}$ is involved. The flexibility in $g_{ab}$ is apparent in our equation (\ref{elasticity}). Modulo this arbitrariness the lapse and shift variables are the fields $\lambda_{0}$ and $\lambda^{a}$ in our interpolating Lagrangean (\ref{almostpoly}). They in turn owe their existence to the constraints (\ref{constraints}) which are nothing but the superhamiltonian and supermomentum of the theory. Our interpolating Lagrangean (\ref{interpolatingLagrangean}) can thus be considered as the brane analog of the A--D--M formulation of geometrodynamics.
\section{Constraint structure and Gauge symmetry of the $p$-brane}
\label{pbrane4}
\pagestyle{myheadings}
\markright{Constraint structure and Gauge symmetry of the $p$-brane}
As is demonstrated in the earlier sections of this chapter the interpolating action formalism is based on the gauge symmetries of the N--G theory which in turns correspond to the invariance under reparametrisation of the brane world volume. Considering the pivotal role played by the gauge symmetries in our analysis, it is only natural to undertake a thorough investigation of their connection with the reparametrisation symmetries. In the present section we shall address this issue by employing the Hamiltonian method \cite{brr} summarised in section \ref{Algo}. We shall see that this analysis will provide results structurally similar to those obtained in context of membrane. Nevertheless, it will provide an elegant generalisation in a much more compact form, as we shall see shortly.
 
Let us begin with an analysis of the constraint structure of interpolating action (\ref{interpolatingLagrangean}). The independent fields are $X_{\mu}$, $\lambda_{0}$ and $\lambda^{a}$. By definition, the corresponding momenta $\Pi_{\mu}$, $\Pi_{\lambda_{0}}$ and $\Pi_{\lambda^{a}}$ are given by 
\begin{eqnarray}
\Pi_{\mu} & = & -\frac{1}{\lambda_{0}}\left(\dot{X}_{\mu} + \lambda^{a} \partial_{a}X_{\mu}\right) \label{intprimaryconstraints1} \\
\Pi_{\lambda_{0}} & = & 0, \qquad \Pi_{\lambda^{a}} = 0 
\label{intprimaryconstraints}
\end{eqnarray}
respectively. We can easily identify the set (\ref{intprimaryconstraints}) as the primary constraints of the theory. In addition to the Poisson brackets (\ref{XPicommutator}) we now also have
\begin{eqnarray}
\{\lambda_{0}\left(\tau,\xi\right),\Pi_{\lambda_{0}}\left(\tau,\xi^{\prime}\right)\} = \delta\left(\xi - \xi^{\prime}\right), \qquad 
\{\lambda^{a}\left(\tau,\xi\right), \Pi_{\lambda^{b}}\left(\tau,\xi^{\prime}\right)\} = \delta\left(\xi - \xi^{\prime}\right) \delta^{a}_{b} 
\label{intcommutator}
\end{eqnarray}
The canonical Hamiltonian following from (\ref{interpolatingLagrangean}) is
\begin{equation}
{\cal{H}}_c = -\lambda^{a}\Pi_{\mu}\partial_{a}X^{\mu} - \frac{\lambda_{0}}{2}\left(\Pi^2 +X^{\prime 2}\right) 
\label{intcanonicalH}
\end{equation}
which reproduces the total Hamiltonian of the N--G action. Conserving the primary constraints (\ref{intprimaryconstraints}), two new secondary constraints emerge which are, as expected from our previous experience, the primary constraints of the N--G action (\ref{constraints}). No more secondary constraints are obtained. The system of constraints for the Interpolating Lagrangean thus comprises of the set (\ref{constraints}) and (\ref{intprimaryconstraints}). These constraints from an involutive algebra. Hence they are first class and therefore generate gauge transformations on ${\cal{L}}_I$. The number of independent gauge parameters is equal to the number of independent primary first class constraints.  As before, we can abstract the most general local symmetry transformations of the Lagrangean by applying the Hamiltonian procedure of \cite{brr}.

In the interpolating action (\ref{interpolatingLagrangean}) the fields are $X^{\mu}, \lambda_{0}$ and $\lambda^{a}$. The set of first-class constraints are given by (\ref{intprimaryconstraints}) along with (\ref{constraints}). Denoting these by the set $\left\{\Psi_{k}\right\}$ we write the generator of the gauge transformations of (\ref{interpolatingLagrangean}) as 
\begin{equation}
G =\int d\xi \alpha_{k}\Psi_{k}
\label{gaugegenerator}
\end{equation}
where $\alpha_{k}$ are the gauge parameters. Instead of constructing the gauge transformation generators from (\ref{gaugegenerator}) by including the whole set of first class constraints (\ref{constraints}, \ref{intprimaryconstraints}) and then eliminating the dependent gauge parameters using (\ref{219}), we take the easier alternative like we did earlier in sections \ref{stringsym} and \ref{membranesym}. We understand that the gauge variations of the fields $\lambda_{0}$ and $\lambda^{a}$ appearing in the intermediate first order form (\ref{firstorderLagrangean}) can be calculated alternatively, ( using (\ref{218})) from the N--G theory where they appear as Lagrange multipliers. The generator of gauge transformations has already been given in (\ref{gaugegenerator}) where $\Omega_{i}$ now stands for the first-class constraints of the N--G theory, i.e. (\ref{constraints}) only. The variations of $\lambda_{i}$ are obtained from (\ref{314}) whereas the structure functions $C_{kj}{}^{i}\left(\xi^{\prime}, \xi^{\prime \prime}, \xi\right)$ are given by (\ref{315}) \footnote{Note that the variables $\sigma, \sigma^{\prime}$ etc in equations (\ref{314}) and (\ref{315}) are to be replaced by the collective variables $\xi, \xi^{\prime}$ etc in the present case.}. The nontrivial structure functions $C_{\alpha \beta}{}^{\gamma}\left(\xi, \xi^{\prime}, \xi^{\prime \prime}\right)$ are obtained from the N--G constraint algebra (\ref{constraintalgebra}).
\begin{eqnarray}
C_{00}^{b} & = & 4\left[ \bar{h}\bar{h}^{ab} \delta\left(\xi - \xi^{\prime}\right)+ \bar{h}\bar{h}^{ab} \delta\left(\xi^{\prime} - \xi^{\prime \prime}\right)\right] \partial_{a}\left\{\delta\left(\xi - \xi^{\prime}\right)\right\}  \\
C_{a0}^{0} & = & \left[\delta\left(\xi - \xi^{\prime \prime}\right)+ \delta\left(\xi^{\prime} - \xi^{\prime \prime}\right)\right] \partial_{a}\left\{\delta\left(\xi - \xi^{\prime}\right)\right\} \\
C_{ab}^{c} & = & \left[\delta\left(\xi - \xi^{\prime \prime}\right)\partial_{a}\left\{\delta\left(\xi - \xi^{\prime}\right)\right\} \delta^{c}{}_{b} + \delta\left(\xi^{\prime} - \xi^{\prime \prime}\right)\partial_{b}\left\{\delta\left(\xi - \xi^{\prime}\right)\right\} \delta^{c}{}_{a}\right] 
\label{abc}
\end{eqnarray}
and using the structure functions (\ref{abc}) we calculate the required gauge variations by applying equation (\ref{314})
\begin{eqnarray}
\delta \lambda^{a} &=& - \dot \alpha_{a} + \left(\alpha_{b}\partial_{b}\lambda^{a} - \lambda^{b}\partial_{b}\alpha_{a} \right) + 4 \bar{h}\bar{h}^{ab} \left(\alpha_{0}\partial_{b}\lambda_{0} - \lambda_{0}\partial_{b} \alpha_{0}\right)\nonumber\\
\delta \lambda_{0} &=& -\dot \alpha_{0} +\left(\alpha_{0}\partial_{a}\lambda^{a} - \lambda^{a} \partial_{a}\alpha_{0}
\right) +\left(\alpha_{a}\partial_{a}\lambda_{0} - \lambda_{0} \partial_{a}\alpha_{a} \right)
\label{deltalambda}
\end{eqnarray}
Now we can systematically investigate and explicitly establish the parallel between gauge symmetry and reparametrization symmetry of the $p-$brane actions. The Polyakov action offers the most appropriate platform to test this proposition. The reparametrisation of the Polyakov metric componants are first worked out. The identification (\ref{idmp}) immediately allows us to find  corresponding transformations of $\lambda_{0}$ and $\lambda^{a}$. We already have their gauge variations in (\ref{deltalambda}). The equivalence between the two concepts (i.e. gauge variation and reparametrisation) can be demonstrated by comparing the changes of $\lambda_{0}$ and $\lambda^{a}$ from the alternative approaches. But for this comparison we first need to devise an exact mapping between the reparametrization parameters and gauge parameters. This is again done, as in the string and membrane case, by demanding that the symmetry transformations on $X^{\mu}$ agree from both the approaches. 
 
Under infinitesimal reparametrisation of the world volume coordinates 
\begin{eqnarray}
\xi^{\prime}{}_{i} =\xi_{i} - \Lambda_{i} 
\label{reparametrisation}
\end{eqnarray}    
where $\Lambda_{i}$
are arbitrary functions of $\xi_{i}$, the variations of the fields $X_{\mu}$ and $g_{ij}$ are
\begin{eqnarray}
\delta X^{\mu} & = & \Lambda^{i} \partial_{i} X^{\mu}
= \Lambda^{0} \dot X^{\mu} + \Lambda^{a}\partial_{a}X^{\mu} \label{deltaX}\\
\delta g_{ij} & = & D_{i}\Lambda_{j} + D_{j}\Lambda_{i} \label{deltag} \\
{\rm where } \qquad
D_{i}\Lambda_{j} & = & \partial_{i} \Lambda_{j} - \Gamma_{ij}{}^{k} \Lambda_{k} \label{3123p}
\end{eqnarray}
with $\Gamma_{ij}{}^{k}$ being the Christoffel symbols \cite{jvh}. 
From the Lagrangean corresponding to (\ref{polyakovaction}) we find
\begin{equation}
\Pi^{\mu} = - \sqrt{- g } g^{00} \dot X^{\mu} - \sqrt{- g } g^{0a}
 \partial_{a}X^{\mu}
\label{3118p}
\end{equation}
Substituting $\dot X^{\mu}$ from (\ref{3118p}) in (\ref{deltaX}) we get after some calculation
\begin{equation}
\delta X^{\mu} = \Lambda^{0} \frac{\sqrt{- g } }{g g^{00}} \Pi^{\mu}  + \left(\Lambda^{a} - \frac{g^{0a} }{g^{00}}  \Lambda^{0}\right) \partial_{a}X^{\mu}
\label{3119p}
\end{equation}
Now the variation of $X^{\mu}$ in (\ref{interpolatingLagrangean}) under (\ref{gaugegenerator}) is
\begin{equation}
\delta X_{\mu} = \left\{X_{\mu}, G \right\}
 = \left( \alpha_{a} \partial_{a} X_{\mu} + 2 \alpha_{0} \Pi_{\mu} \right)
\label{GM7p}
\end{equation}
Comparing the above expression of $\delta X^{\mu}$ with that of (\ref{3119p}) we find the mapping
\begin{eqnarray}
\Lambda^{0} =- \frac{\alpha_{0}}{\lambda_{0}}, \qquad \Lambda^{a} = \alpha_{a} -\frac{\lambda^{a}\alpha_{0}}{\lambda_{0}}
\label{3120}
\end{eqnarray}
With this mapping the gauge transformation on $X_{\mu}$ in both the formalism agree. The complete equivalence between the transformations can now be demonstrated by computing $\delta \lambda^{a}$ and $\delta \lambda_{0}$ from the alternative approaches. The identification (\ref{idmp}) yields
\begin{equation}
\lambda^{a} = \frac{g^{0a}}{g^{00}}
\label{GM11p}
\end{equation}
We require to express these in terms of $g_{ij}$. To this end we start from the identity
\begin{equation}
g^{ij} g_{jk} = \delta^{i}{}_{k}
\label{GM12p}
\end{equation}
and obtain the following equations for $\lambda^{a}$
\begin{eqnarray}
\lambda^{a} g_{ab} & = & - g_{0b} \label{GM13}\\
{\rm which \ gives } \qquad
\lambda^{a} & = & - \bar{g}^{ab} g_{0b} \label{GM14}
\end{eqnarray}
Taking variation on both sides of (\ref{GM13})we get 
\begin{equation}
\delta \lambda^{a} g_{ab} = -\delta g_{0b} - \lambda^{a} \delta g_{ab} 
\label{1stvariation}
\end{equation}
Rearranging the terms conveniently we write
\begin{equation}
\delta \lambda^{a} = -\delta g_{0b} \bar{g}^{ab} - \lambda^{c} \delta g_{cb} \bar{g}^{ba}
\label{mastervariation}
\end{equation}
Now using (\ref{deltag}) we compute $\delta g_{0b}$ and $\delta g_{ab}$ and simplify the relation (\ref{mastervariation}) keeping in mind the fact that $\bar{g}^{ab}$ is the inverse of the spatial metric $g_{ab}$. This gives the variations of $\lambda^{a}$ under reparametrization as
\begin{eqnarray}
\delta \lambda^{a} &=& - \partial_{0} \Lambda^{a} + \lambda^{a} \partial_{0} \Lambda^{0} - \lambda^{b} \partial_{b} \Lambda^{a} + \lambda^{a}\lambda^{b} \partial_{b} \Lambda^{0} + \Lambda^{k}\partial_{k} \lambda^{a} + \left(g_{00} - g_{c0}\bar{g}^{cb}g_{0b}\right)\bar{g}^{ad}\partial_{d} \Lambda^{0}\nonumber\\
\label{GM14p}
\end{eqnarray}
We further simplify the last term using a relation\footnote{This relation is obtained by substituting for $\lambda_{0}$ and $\lambda^{a}$ from (\ref{LambdaRhoeqnmotion}) in the interpolating Lagrangean (\ref{interpolatingLagrangean})}
\begin{equation}
\frac{1}{\lambda^{2}}\left[h_{00} - h_{0c}\bar{h}^{cb}h_{0b}\right] + \bar{h} = 0
\label{therelation}
\end{equation}
and also (\ref{polyeqnmotion}) to get
\begin{eqnarray}
\delta \lambda^{a} &=& - \partial_{0} \Lambda^{a} + \lambda^{a} \partial_{0} \Lambda^{0} - \lambda^{b} \partial_{b} \Lambda^{a} + \lambda^{a}\lambda^{b} \partial_{b} \Lambda^{0} + \Lambda^{k}\partial_{k} \lambda^{a} + \lambda_{0}^{2}\bar{h}\bar{h}^{ad}\partial_{d} \Lambda^{0}
\label{GM15}
\end{eqnarray}
Now introducing the mapping (\ref{3120}) in (\ref{GM15}) we find that the variations of $\lambda^{a}$ are identical with their gauge variations in (\ref{deltalambda}). Finally we compute the variation of $\lambda_{0}$. This can be conveniently done by taking the expression of $\lambda_{0}$ in (\ref{theg}) and using the variations (\ref{deltag}). We get the expression of $\delta \lambda_{0}$ in terms of the reparametrization parametres $\Lambda_{i}$ as 
\begin{equation}
\delta \lambda_{0} = \lambda_{0} \partial_{i}\Lambda^{i} + \Lambda^{k} \partial_{k} \lambda_{0} - 2\lambda_{0} \partial_{a}\Lambda^{a} + 2 \lambda_{0} \lambda^{a} \partial_{a} \Lambda^{0}
\label{GM16}
\end{equation}
Again using the mapping (\ref{3120}) we substitute $\Lambda_{i}$ by $\alpha_{i}$ and  the resulting expression for $\delta \lambda_{0}$ agrees with that given in (\ref{deltalambda}). The complete matching, thus obtained, illustrates the equivalence of reparametrization symmetry with gauge symmetry for the generic $p$-brane. 

With this we conclude our analysis of internal symmetries of free stringy models. In the next section we shall consider an interacting string and impose string boundary conditions to examine whether any modification is required in our interpolating formalism.

\section{Noncommutativity in Interpolating string}
\label{ncstring}
\pagestyle{myheadings}
\markright{Noncommutativity in Interpolating string}
So far in this chapter we have described the string, membrane and generic $p-$brane in a new Lagrangean formalism which well-suited the study of their gauge invariance and diffeomorphism. Interestingly the study of open strings propagating in the presence of a background Neveu--Schwarz two form field $B_{\mu\nu}$ exhibits a manifest noncommutative
structure among the space-time coordinates of the D-branes \cite{SW}. Several approaches have been taken to obtain such results, for example a Dirac approach \cite{dir} is employed with the string boundary conditions (BC(s)) imposed as second class constraints in \cite{ar, br}. Alternatively, in a series of recent papers \cite{rbbcsg, rbk, bcsgaghs} it has been shown explicitly that noncommutativity can be obtained in a more transparent way by modifying the canonical Poisson bracket (PB) structure, so that it is compatible with the BC(s). 

Acknowledging the above facts, we formulate an interpolating action for interacting bosonic strings in the present section. We modify the basic PB structure to make them compatible with BC(s) following the approach in \cite{rbbcsg, rbk, bcsgaghs}. This leads to the emergence of the noncommutativity among the stringy coordinates in case of both free and interacting strings \cite{sgaghas}. For simplicity, in this thesis we discuss the free string case only. Our results go over smoothly to the Polyakov version once proper identifications are made. Interestingly, we observe that a gauge fixing is necessary to give an exact NC solution between the string coordinates. This gauge fixing condition restrict us to a reduced phase space of the interpolating theory
which in turn minimizes the gauge redundancy of the theory by identifying a particular combination of the constraints (that occurs in the full gauge independent theory) leading to a new involutive constraint algebra which is markedly different from that given in \cite{rbbcsg}.  

With the new involutive constraint algebra we revisit the study of gauge symmetry to find surprising changes in the structure functions of the theory. We proceed to compute the gauge variations of the fields and show the underlying unity of diffeomorphism with the gauge symmetry in this new framework.
\subsection{Interacting string in Nambu--Goto formalism}
\label{ncstring1}  
\pagestyle{myheadings}
\markright{Interacting string in Nambu--Goto formalism}
We start with the N--G action for a bosonic string moving in the presence of a constant background Neveu-Schwarz two-form field $B_{\mu\nu}$
\begin{eqnarray}
S_{NG}=\int^{\infty}_{- \infty} d\tau \int^{\pi}_{0}d\sigma 
\left[{\mathcal L}_{0} + eB_{\mu\nu}\dot{X}^\mu
X^{\prime \nu}\right]
\label{NGaction} 
\end{eqnarray}
where ${\mathcal L}_{0}$ is the free N--G Lagrangian density given in (\ref{111}). The string tension is again kept implicit for convenience. The Euler-Lagrange (EL) equations and BC obtained by varying the action are
\begin{eqnarray}
\dot{\Pi}^{\mu} + K^{\prime \mu} & = & 0 \qquad {\rm and} \qquad
K^{\mu}\vert_{\sigma = 0, \pi} = 0 \label{EL} \\
{\rm where,} \qquad \qquad
\Pi_{\mu} &=& \frac{\partial {{\mathcal L}}}{\partial \dot{X}^{\mu}}
= {\mathcal L}_{0}^{-1}\left(-X^{\prime 2}\dot{X}_{\mu}
+(\dot{X}.X^{\prime})X^{\prime}_{\mu}\right) + eB_{\mu\nu}X^{\prime \nu}
\nonumber \\
K_{\mu} &=& \frac{\partial {{\mathcal L}}}{\partial X^{\prime \mu}}
= {\mathcal L}_{0}^{-1}\left(- \dot{X}^2 X^{\prime}_{\mu}
+ (\dot{X} \cdot X^{\prime})\dot{X}_{\mu}\right) -e B_{\mu\nu}\dot{X}^{\nu}.
\label{NGmomentum} 
\end{eqnarray}
Note that $\Pi_{\mu}$ is the canonically conjugate momentum to $X^{\mu}$. The nontrivial PB(s) of the theory are same as in (\ref{116}). The primary constraints of the theory are
\begin{eqnarray}
\Omega_{1} = \Pi_{\mu}X^{\prime \mu}\approx 0, \qquad
\Omega_{2} = \left(\Pi_{\mu} - e B_{\mu \nu} X^{\prime \nu}\right)^2 + X^{\prime 2} \approx 0
\label{NGconstraints} 
\end{eqnarray}
Using the PB structure (\ref{116}), it is easy to check that these constraints generate the same first class (involutive) algebra as the free N--G string given by equations (\ref{117}), (\ref{118}) and (\ref{119}). Also, the canonical Hamiltonian density vanishes
and the total Hamiltonian density is thus given by a linear combination of the first class constraints (\ref{NGconstraints})
\begin{eqnarray}
{\mathcal H}_{T}=-\rho\Omega_{1}-\frac{\lambda}{2}\Omega_{2}
\label{totalH} 
\end{eqnarray}
where $\rho$ and $\lambda$ are Lagrange multipliers. It is easy to check that time preserving the primary constraints yields no new secondary constraints. Hence the total set of constraints  of the interacting N--G theory is given by the first-class system (\ref{NGconstraints}).

Now we enlarge the domain of definition of the bosonic field $X^{\mu}$ from $[0,\pi]$ to $[-\pi,\pi]$ by defining \cite{bcsgaghs}
\begin{eqnarray}
X^{\mu}(\tau , -\sigma) = X^{\mu}(\tau , \sigma)\ \ ; \ \  
B_{\mu \nu} \to - B_{\mu \nu} \ \mathrm{under}\ \sigma \to -\sigma.
\label{37}
\end{eqnarray}
The second condition implies that $B_{\mu \nu}$, albeit a constant, transforms as a pseudo scalar under $\sigma \to -\sigma$ in the extended interval. This ensures that the interaction term $eB_{\mu\nu}\dot{X}^{\mu}X^{\prime \nu}$ in (\ref{NGaction}) remains invariant under $\sigma \to - \sigma$ like the free N--G Lagrangian density ${\cal{L}}_{0}$ (\ref{111}). Consistent with this, we have
\begin{eqnarray}
\Pi^{\mu}(\tau , -\sigma) = \Pi^{\mu}(\tau , \sigma),
\quad 
X^{\prime \mu}(\tau , -\sigma) = - X^{\prime \mu}(\tau , \sigma) 
\label{37h}
\end{eqnarray}
Now, from  (\ref{NGconstraints}), (\ref{37}) we note that the constraints $\Omega_1(\sigma) \approx 0$ and $\Omega_2(\sigma) \approx 0$ are odd and even respectively under $\sigma \rightarrow -\sigma$. Now demanding the total Hamiltonian density ${\cal{H}}_T$ (\ref{totalH}) also remains invariant under $\sigma \to - \sigma$, one finds that $\rho$ and $\lambda$ must be odd and even respectively under $\sigma \to - \sigma$. With the salient features of N--G interacting string mentioned we now move on to construct the interpolating action of the interacting string.

\subsection{Interpolating Lagrangean, its boundary conditions and additional constraints}
\label{ncstring2}
\pagestyle{myheadings}
\markright{Interpolating Lagrangean, its boundary conditions and additional constraints}
To construct the interpolating action of the interacting string we exactly follow the procedure of the earlier sections. We write down the Lagrangean of the interacting N--G action (\ref{NGaction}) in the first-order form
\begin{eqnarray}
{\mathcal L}_{I}=\Pi_{\mu}\dot{X}^{\mu}-{\mathcal H}_{T}
\label{1stL} 
\end{eqnarray}
which upon substituting (\ref{totalH}) becomes
\begin{eqnarray}
{\mathcal L}_{I}=\Pi_{\mu}\dot{X}^{\mu}+\rho\Pi_{\mu}X^{\prime\mu}
+\frac{\lambda}{2}\left[(\Pi^2 + X^{\prime 2})-2eB_{\mu\nu}
\Pi^{\mu}X^{\prime\nu}+e^2B_{\mu\nu}B^{\mu}_{\ \rho}X^{\prime\nu}
X^{\prime\rho}\right]
\label{subL} 
\end{eqnarray}
In this Lagrangian $\lambda$ and  $\rho$ (originally introduced as Lagrange multipliers in the N--G action (\ref{NGaction})) are treated as independent fields, which behave as scalar and pseudo-scalar fields respectively in the extended world-sheet, as we have already discussed. The EL equation for the auxiliary field $\Pi_{\mu}$ is given by 
\begin{eqnarray}
\dot{X}^{\mu}+\rho X^{\prime\mu}+\lambda\Pi^{\mu}-
e\lambda B^{\mu\nu}X^{\prime}_{\nu}=0\,.
\label{momentumeqn} 
\end{eqnarray}
We eliminate the auxiliary field obtained from (\ref{subL}) by substituting $\Pi_{\mu}$ from (\ref{momentumeqn}) back in (\ref{subL}) which yields the interpolating Lagrangian of the interacting string
\begin{eqnarray}
{\mathcal L}_{I}=-\frac{1}{2\lambda}\left[\dot{X}^{2}+
2\rho(\dot{X}.X^{\prime})+(\rho^2-\lambda^2)X^{\prime 2}-2\lambda eB_{\mu\nu}
\dot{X}^{\mu}X^{\prime\nu}\right]\,.
\label{interpolatingL} 
\end{eqnarray}

The reproduction of the N--G action (\ref{NGaction}) from the interpolating action of the interacting string is trivial and can be done by eliminating $\rho$ and $\lambda$ using their respective EL equations of motion (\ref{1117}) and (\ref{1118b}) which can be shown to follow from (\ref{interpolatingL}).

If, on the other hand, we identify $\rho$ and $\lambda$ with the components of the world-sheet metric as given in (\ref{1119}), then the above Lagrangian (\ref{interpolatingL}) reduces to the Polyakov form,
\begin{eqnarray}
{\mathcal L}_P= -{1\over 2}\left( {\sqrt {-g}}g^{ab}{\partial}_a
X^{\mu}{\partial}_bX_{\mu}
-e\epsilon^{ab}B_{\mu \nu}\partial_{a} X^{\mu}\partial_{b} X^{\nu}\right)
\ \ ; \  \ \left(a,b = \tau,\sigma\right).
\label{Polyaction} 
\end{eqnarray}
We can now, likewise construct the interpolating BC from the interpolating Lagrangian (\ref{interpolatingL}), 
\begin{eqnarray}
K^{\mu} = \left[\partial {{\cal L}}_{I}\over \partial 
X^{\prime}_{\mu}\right]_{\sigma =0,\pi} = \left({\rho \over \lambda}
{\dot X}^{\mu}+{\rho^2 - 
\lambda^2 \over \lambda}X'^{\mu} + eB^{\mu}_{\ \nu}\dot{X}^{\nu}
\right)_{\sigma = 0, \pi} = 0. 
\label{EL1} 
\end{eqnarray}
That this can be interpreted as interpolating BC, can be easily seen by using the expressions (\ref{1117}) and (\ref{1118b}) for $\rho$ and $\lambda$ in (\ref{EL1}) to yield
\begin{eqnarray}
\left[{\mathcal L}^{- 1}_{0}\left( - \dot{X}^{2} X^{\prime \mu}
+ \left(\dot{X} X^{\prime}\right)\dot{X}^{\mu}\right) - 
e B^{\mu \nu}\dot{X}_{\nu}\right]_{\sigma = 0, \pi} = 0
\label{17} 
\end{eqnarray}
This is the BC of the interacting N--G string (\ref{EL}). 

Alternatively, we can identify $\rho$ and $\lambda$ with the metric components as in (\ref{1119}) to recast (\ref{EL1}) as
\begin{eqnarray}
\left(g^{1a}\partial_{a}X^{\mu}(\sigma) +
\frac{1}{\sqrt{-g}}e B^{\mu}_{\ \nu}\partial_{0}
X^{\nu}(\sigma) 
\right)_{\sigma = 0, \pi} = 0.
\label{EL1a} 
\end{eqnarray}
which is easily identifiable with Polyakov form of BC \cite{rbbcsg} following from the action (\ref{Polyaction}).

Using phase space variables $X^{\mu}$ and $\Pi_{\mu}$, (\ref{EL1}) can be rewritten as
\begin{eqnarray}
K^{\mu} = \left[\left(\rho \Pi^{\mu} + \lambda X^{\prime \mu}\right)
+ e B^{\mu}_{\ \nu}\left(\Pi^{\nu} - eB^{\nu}_{\ \rho}
X^{\prime \rho}\right)\right]_{\sigma = 0, \pi} = 0.
\label{EL1ab} 
\end{eqnarray}
Hence it is possible to interpret either of (\ref{EL1}) or (\ref{EL1ab}) as an interpolating BC.

To get the constraint structure of the interpolating interacting string we write the canonically conjugate momenta of the independent fields in (\ref{interpolatingL}), i.e., $X^{\mu}$, $\rho$ and $\lambda$. They are denoted by $\Pi_{\mu}$, $\pi_{\rho}$ and $\pi_{\lambda}$ respectively and given as
\begin{eqnarray}
\Pi_{\mu} = -\frac{1}{\lambda}\left(\dot{X}_{\mu} +
\rho X^{\prime}_{\mu}\right) + eB_{\mu\nu}X^{\prime \nu}, \qquad
\pi_{\rho} =  0, \qquad 
\pi_{\lambda} = 0
\label{211a}
\end{eqnarray}
So in addition to the PB(s) (\ref{116}), we now also have the PB(s) (\ref{212}). The canonical Hamiltonian following from (\ref{interpolatingL}) reads
\begin{equation}
{\cal{H}}_c = -\rho \Pi_{\mu}X^{\prime\mu} - \frac{\lambda}{2}\left\{\left(\Pi_{\mu} - e B_{\mu \nu} X^{\prime \nu}\right)^2 + X^{\prime 2}\right\} 
\label{canonicalH2}
\end{equation}
which reproduces the total Hamiltonian (\ref{totalH}) of the N--G action. From the definition of the canonical momenta  (\ref{211a}) we can easily identify the primary constraints
\begin{eqnarray}
\Omega_{3} = \pi_{\rho} \approx 0, \qquad 
\Omega_{4} = \pi_{\lambda} \approx 0\,
\label{int_primary}
\end{eqnarray}
The conservation of the above primary constraints leads to secondary constraints which are nothing but the primary constraints of the N--G theory $\Omega_1$ and $\Omega_2$ of (\ref{NGconstraints}). No more secondary constraints are obtained. The system of constraints for the Interpolating Lagrangian thus comprises of the set 
(\ref{int_primary}) and (\ref{NGconstraints}). The PB(s) of the constraints of (\ref{int_primary}) vanish within themselves. Also the PB of these with (\ref{NGconstraints}) vanish. Equipped with the constraint structure and the BC's of the interacting theory we now proceed to check the compatibility of the BC's with the PB's.
\subsection{Modified PBs and new Constraint structure for free interpolating string}
\label{ncstring3}
\pagestyle{myheadings}
\markright{Modified PBs and new Constraint structure for free interpolating string}
The boundary condition for free interpolating string is obtained by setting $B_{\mu \nu} = 0$ in (\ref{EL1ab}) which gives 
\begin{eqnarray}
K^{\mu} = \left[\left(\rho \Pi^{\mu} + \lambda X^{\prime \mu}\right) \right]_{\sigma = 0, \pi} = 0.
\label{20} 
\end{eqnarray}
The above BC is not compatible with the basic PB (\ref{116}). To incorporate this, appropriate modifications in the basic PBs are in order. In \cite{hrt, rbbcsg, rbk, bcsgaghs}, the equal time brackets were given in terms of certain combinations ($\Delta_{+}(\sigma , \sigma^{\prime})$) of periodic delta function\footnote{The form of the periodic delta function is given by $\delta_{P}(x-y) = \delta_{P}(x-y+ 2\pi) =\frac{1}{2\pi} \sum_{n\in Z}e^{in(x-y)}$ and is related to the usual
Dirac $\delta$-function as $\delta_P(x-y) = \sum_{n\in Z}\delta(x - y + 2\pi n)$.}
\begin{eqnarray}
\{X^{\mu}(\tau , \sigma) ,  \Pi_{\nu}(\tau , \sigma^{\prime})\} & = & \delta^{\mu}_{\nu} \Delta_{+}(\sigma, \sigma^{\prime})
\label{23} \\
{\rm where,} \quad 
\Delta_{+}\left(\sigma , \sigma^{\prime}\right) & = & \delta_{P}(\sigma - \sigma^{\prime}) + \delta_{P}(\sigma + \sigma^{\prime}) = \frac{1}{\pi} + \frac{1}{\pi}\sum_{n\neq 0} \mathrm{cos}(n\sigma^{\prime})\mathrm{cos}(n\sigma) \nonumber \\
\Delta_{-}\left(\sigma , \sigma^{\prime}\right) & = & \delta_{P}(\sigma - \sigma^{\prime}) - \delta_{P}(\sigma + \sigma^{\prime}) =  \frac{1}{\pi}\sum_{n\neq 0} \mathrm{sin}(n\sigma^{\prime})\mathrm{sin}(n\sigma) 
\label{24}
\end{eqnarray}
rather than an ordinary delta function to ensure compatibility with Neumann BC 
\begin{eqnarray}
\partial_{\sigma}X^{\mu}(\sigma)\vert_{\sigma = 0, \pi} = 0\,, 
\label{16b}
\end{eqnarray}
in the bosonic sector. Observe that the other brackets 
\begin{eqnarray}
\{X^{\mu}\left(\sigma\right), X^{\nu}\left(\sigma^{\prime}\right)\}  & = & 0 \label{24q}\\
\{\Pi^{\mu}\left(\sigma\right), \Pi^{\nu} \left(\sigma^{\prime}\right)\} & = & 0 \label{24p}
\end{eqnarray}
are consistent with the Neumann boundary condition (\ref{16b}). Now a simple inspection shows that the BC (\ref{20}) is also compatible with (\ref{23})\footnote{Note that there is no inconsistency in (\ref{16b}) as $\partial_{\sigma}\Delta_{+}\left(\sigma, \sigma^{\prime}\right) \vert_{\sigma = 0, \pi}= 0$.} and (\ref{24p}), but not with (\ref{212}) and (\ref{24q}). Hence the brackets (\ref{212}) and (\ref{24q}) need to be altered suitably. Since $\rho$ and $\lambda$ are odd and even functions of $\sigma$ respectively, we propose
\begin{eqnarray}
\{\rho(\tau,\sigma ) , \pi_{\rho}(\tau, \sigma^{\prime} )\} = \Delta_{-}(\sigma , \sigma^{\prime}), \qquad
\{\lambda(\tau,\sigma ) , \pi_{\lambda}(\tau, \sigma^{\prime} )\} = \Delta_{+}(\sigma , \sigma^{\prime})
\label{32j}
\end{eqnarray}
and also make the following ansatz for the bracket among the coordinates (\ref{24q})
\begin{eqnarray}
\{X^{\mu}(\tau , \sigma) ,  X^{\nu}(\tau , \sigma^{\prime})\} = C^{\mu \nu}(\sigma, \sigma^{\prime})\ \ ;\ \ 
\mathrm{where} \ \ \ C^{\mu \nu}(\sigma, \sigma^{\prime}) = -\  C^{\nu \mu}(\sigma^{\prime}, \sigma)\,.
\label{25}
\end{eqnarray}
One can easily check that the brackets (\ref{32j}) are indeed compatible with the BC (\ref{20}). Now imposing the BC (\ref{20}) on the above equation (\ref{25}), we obtain the following condition
\begin{eqnarray}
\partial_{\sigma}C^{\mu \nu}\left(\sigma, 
\sigma^{\prime}\right)|_{\sigma = 0, \pi} 
= \frac{\rho}{\lambda}\eta^{\mu \nu} \Delta_{+}\left(\sigma, 
\sigma^{\prime}\right)|_{\sigma = 0, \pi}\,.
\label{26}
\end{eqnarray}
Now to find a solution for $C^{\mu\nu}(\sigma ,\sigma^{\prime})$, we choose{\footnote {The condition (\ref{27}) reduces to a restricted class of metric for Polyakov formalism that satisfy $\partial_{\sigma}g_{01}=0$. Such conditions also follow from a standard treatment of the light-cone gauge \cite{pol}.}} 
\begin{eqnarray}
\partial_{\sigma}\left(\frac{\rho}{\lambda}\right) = 0
\label{27}
\end{eqnarray}
which gives a solution of $C^{\mu\nu}(\sigma ,\sigma^{\prime})$ as
\begin{eqnarray}
C^{\mu\nu}(\sigma ,\sigma^{\prime}) = \eta^{\mu \nu}\left[
\kappa(\sigma)\Theta(\sigma ,\sigma^{\prime}) - 
\kappa(\sigma^{\prime})\Theta(\sigma^{\prime} ,\sigma )\right] 
\label{28}
\end{eqnarray}
where the generalised step function $\Theta (\sigma ,\sigma^{\prime})$ satisfies,
\begin{eqnarray}
\partial_{\sigma }\Theta (\sigma ,\sigma^{\prime}) = \Delta_{+}(\sigma ,\sigma^{\prime})\,
\label{29}
\end{eqnarray}
Here, $\kappa(\sigma) = \frac{\rho}{\lambda}(\sigma)$ is a pseudo-scalar. The $\sigma$ in the parenthesis has been included deliberately to remind the reader that it transforms as a pseudo-scalar under $\sigma \to -\sigma$ and should not be read as a functional dependence. The pseudo-scalar property of $\kappa(\sigma)$ is necessary for 
$C^{\mu\nu}(\sigma ,\sigma^{\prime})$ to be an even function of $\sigma$ as $X(\sigma)$ is also an even function of $\sigma$
in the extended interval $[-\pi, \pi]$ of the string (\ref{37}).
\noindent
An explicit form of $\Theta(\sigma ,\sigma^{\prime})$ is given in \cite{hrt}
\begin{eqnarray}
\Theta (\sigma ,\sigma^{\prime}) & = & {\sigma \over \pi} + {1 \over \pi }
\sum_{n\neq 0}{1\over n}\mathrm{sin}(n\sigma)\mathrm{cos}(n\sigma^{\prime})
\label{30} \\
\mathrm{where} \quad
\Theta (\sigma ,\sigma^{\prime}) & = & 1~~~ \mathrm{for} ~~\sigma >\sigma^{\prime}
\quad \mathrm{and} \quad
\Theta (\sigma ,\sigma^{\prime}) = 0 ~~~\mathrm{for}
~~ \sigma <\sigma^{\prime}
\label{31jj}
\end{eqnarray}
Using the above relations, the simplified structure of (\ref{24q}) reads,
\begin{eqnarray}
\{X^\mu (\tau,\sigma ),X^{\nu}(\tau, \sigma^{\prime} )\}|_{\sigma = \sigma^{\prime}}  =  0 \, & \mathrm{or,} & \,
\{X^\mu (\tau,\sigma ),X^{\nu}(\tau, \sigma^{\prime} )\}|_{\sigma >\sigma^{\prime}} = \kappa(\sigma)\,\eta^{\mu \nu} 
\nonumber \\
& \mathrm{or,} & \, \{X^\mu (\tau,\sigma ),X^{\nu}(\tau, \sigma^{\prime} )\}|_{\sigma <\sigma^{\prime}}  =  -\kappa(\sigma^\prime)\,\eta^{\mu \nu}
\label{32}
\end{eqnarray}
We therefore propose the brackets (\ref{23}) and (\ref{32}) as the basic PB(s) of the theory. Using these one can easily obtain the following involutive algebra between the constraints
\begin{eqnarray}
\{\Omega_1(\sigma) , \Omega_1(\sigma^{\prime})\} &=&  
\Omega_1(\sigma^{\prime})\partial_{\sigma}\Delta_{+}
\left(\sigma , \sigma^{\prime}\right) + \Omega_1(\sigma)
\partial_{\sigma}\Delta_{-}\left(\sigma , \sigma^{\prime}\right) \,
\nonumber \\ 
\{\Omega_1(\sigma) , \Omega_2(\sigma^{\prime})\} &=&  \left(
\Omega_2(\sigma) + \Omega_2(\sigma^{\prime})\right)
\partial_{\sigma}\Delta_{+}\left(\sigma , \sigma^{\prime}\right)\,
\nonumber \\ 
\{\Omega_2(\sigma) , \Omega_2(\sigma^{\prime})\} &=& 4 \left(
\Omega_1(\sigma)\partial_{\sigma}\Delta_{+}
\left(\sigma , \sigma^{\prime}\right) + \Omega_1(\sigma^{\prime})
\partial_{\sigma}\Delta_{-}\left(\sigma , \sigma^{\prime}\right)\right).
\label{33}
\end{eqnarray}
Note that a crucial intermediate step in the above derivation is to use the relation 
\begin{eqnarray}
\{X^{\prime\mu}(\sigma), X^{\prime\nu}(\sigma^{\prime})\} = 0 
\label{importantstep}
\end{eqnarray}
which follows from the basic bracket (\ref{32}) \cite{rbbcsg}. 
\subsection{The reduced interpolating Lagrangean}
\label{ncstring4}
\pagestyle{myheadings}
\markright{The reduced interpolating Lagrangean}
It is interesting to observe that the condition (\ref{27}) (which is necessary for giving an exact NC solution (\ref{28})) reduces the gauge redundancy of the interpolating formalism as $\rho$ and $\lambda$ are no more independent.
Consequently, one should look for only a particular combination of the constraints (\ref{NGconstraints}) which gives a involutive algebra. To this end we go back to the interpolating Lagrangean (\ref{interpolatingL}) and study the effect of (\ref{27}) on it. Earlier (\ref{interpolatingL}) contained two additional fields $\rho$ and $\lambda$. However the interpolating Lagrangian depends only on one of these fields $\lambda$ (say) once the condition (\ref{27}) is imposed and one gets the following reduced form of the Lagrangian
\begin{eqnarray}
{\mathcal{L}}_{\rm{red}} = -\frac{1}{2 \lambda}\dot{X}^{2} - \kappa(\sigma) \dot{X}\cdot X^{\prime}
\label{mod}
\end{eqnarray}
Owing to the condition (\ref{27}), the free canonical Hamiltonian reduces to
\begin{equation}
{\cal{H}}_c = -\kappa\left(\sigma\right) \lambda \Pi \cdot X^{\prime} - \frac{\lambda}{2}\left\{\Pi^{2} + X^{\prime 2}\right\} 
\label{n1}
\end{equation}
having only one primary constraint
\begin{equation}
\pi_{\lambda} \approx 0.
\label{n2}
\end{equation}
Conserving (\ref{n2}) with the canonical Hamiltonian (\ref{n1}) we get the secondary constraint
\begin{equation}
\Omega(\sigma) = \frac{1}{2}\left[\Pi^{2} + X^{\prime 2} + 
2\kappa\left(\sigma\right)
 \Pi\cdot X^{\prime}\right] \approx 0
\label{n3}
\end{equation}
which generates the first class algebra 
\begin{equation}
\left\{\Omega(\sigma),\Omega(\sigma^{\prime})\right\}  = 
2\left[\kappa\left(\sigma\right)
\Omega(\sigma) 
\partial_{\sigma}\Delta_{+}(\sigma,\sigma^{\prime}) - 
\kappa\left(\sigma^{\prime}\right) 
\Omega(\sigma^{\prime}) \partial_{\sigma^{\prime}}
\Delta_{+}(\sigma,\sigma^{\prime})\right] 
\label{n4a}
\end{equation}
in the reduced framework. We shall study the consequences of the earlier constraint algebra (\ref{33}) and also the reduced one (\ref{n4a}) when we make an exhaustive analysis of gauge symmetry. 
\subsection{Analysis of gauge symmetry}
\label{ncstring5}
\pagestyle{myheadings}
\markright{Analysis of gauge symmetry}
In this section we will discuss the gauge symmetries of the different actions of the present string model and investigate their correspondence with the reparametrisation invariances. For simplicity we shall stick to the free string case. Though this has been done earlier in section \ref{stringsym}, the canonical symplectic structure for the open string considered therein were not compatible with the general BC(s) of the theory. Here we shall investigate the gauge symmetry with the new modified PB structures  (\ref{23}), (\ref{32j}), (\ref{32}) and (\ref{importantstep}) which correctly take into account the BC(s) of the theory. Importantly, the modified PB structure reveals a NC behavior among the string coordinates (\ref{25}, \ref{28}). As we have seen in section \ref{ncstring3}, an explicit account of noncommutativity requires a gauge fixing (\ref{27}), thereby reducing the gauge redundancy of the interpolating picture. Note that in the reduced phase space there is only one generator of gauge transformation, i.e. (\ref{n3}).

Our discussion will be centered on the reduced interpolating Lagrangian (\ref{mod}) as it provides an easy access to the analysis of gauge symmetry. The constraint structure (\ref{n2}), (\ref{n3}) and constraint algebra (\ref{n4a}) of the reduced interpolating Lagrangian has already been discussed. All the constraints are first class and therefore generate gauge transformations on ${\cal{L}}_{\rm{red}}$ but the number of independent gauge parameters is equal to the number of independent primary first class constraints, i.e. one.  Once again we shall apply the systematic algorithm of abstracting the most general local symmetry transformations of the Lagrangian \cite{brr}.

The full constraint structure of the interpolating theory (before we enforce the gauge fixing) comprises of the constraints (\ref{int_primary}) along with (\ref{NGconstraints}). We could proceed from these and construct the generator of gauge transformations. The generator of the gauge transformations of (\ref{interpolatingL}) is obtained by including the whole set of first class constraints $\Omega_{i}$ given by (\ref{int_primary}) and (\ref{NGconstraints}) as
\begin{equation}
G =\int d\sigma \alpha_{i}\Omega_{i}, \, \left(i = 1,...4\right)
\label{3112i}
\end{equation}
where only two of the $\alpha_{i}$'s are the independent gauge parameters. Using (\ref{219}) the dependent gauge parameters could be eliminated. After finding the gauge generator in terms of the independent gauge parameters, the variations of the fields $X^{\mu}$, $\rho$ and $\lambda$ can be worked out. But the number of independent gauge parameters are same in both N--G (\ref{NGaction}) and interpolating (\ref{interpolatingL}) version. So the gauge generator\footnote{Note that the gauge parameters $\alpha_{1}$ and $\alpha_{2}$ are odd and even respectively under $\sigma \to -\sigma$.} is the same for both the cases, namely
\begin{equation}
G =\int d\sigma\left( \alpha_{1}\Omega_{1} +  \alpha_{2}\Omega_{2}\right)
\label{3112aaa}
\end{equation}
Also, looking at the intermediate first order form (\ref{subL}) it appears that the fields $X^{\mu}$ were already there in the N--G action (\ref{NGaction}). The other two fields of the interpolating Lagrangian are $\rho$ and $\lambda$ which are nothing but the Lagrange multipliers enforcing the first class constraints (\ref{NGconstraints}) of the N--G theory. Hence their gauge variation can be worked out from (\ref{314}). We prefer to take this alternative route. For convenience we relabel $\rho$ and $\lambda$ by $\lambda_{1}$ and $\lambda_{2}$
\begin{equation}
\lambda_{1} = \rho \hspace{1cm} \rm{and} \hspace{1cm}
\lambda_{2} = \frac{\lambda}{2}
\label{313i}
\end{equation}
The nontrivial structure functions $C_{\alpha \beta}{}^{\gamma} \left(\sigma, \sigma^{\prime}, \sigma^{\prime \prime}\right)$ are worked out using the constraint algebra (\ref{33}) in the definition (\ref{315}) as
\begin{eqnarray}
C_{1 1}{}^{1}\left(\sigma ,\sigma^{\prime},\sigma^{\prime \prime}\right) &=&
\left(\partial_{\sigma}\Delta_{+}\left(\sigma , \sigma^{\prime}
\right)\right)\Delta_{-}\left(\sigma^{\prime} , \sigma^{\prime \prime}\right)
+ \left(\partial_{\sigma}\Delta_{-}\left(\sigma , \sigma^{\prime}
\right)\right)\Delta_{-}\left(\sigma , \sigma^{\prime \prime}\right)
\nonumber \\
C_{2 2}{}^{1}\left(\sigma, \sigma^{\prime}, \sigma^{\prime \prime}\right) &=&
4\left(\partial_{\sigma}\Delta_{+}\left(\sigma , \sigma^{\prime}
\right)\right)\Delta_{-}\left(\sigma , \sigma^{\prime \prime}\right)
+ 4\left(\partial_{\sigma}\Delta_{-}\left(\sigma , \sigma^{\prime}
\right)\right)\Delta_{-}\left(\sigma^{\prime} , \sigma^{\prime \prime}\right)
\nonumber \\
C_{1 2}{}^{2}\left(\sigma, \sigma^{\prime}, \sigma^{\prime \prime}\right) &=&
\partial_{\sigma}\Delta_{+}\left(\sigma , \sigma^{\prime}
\right)
\left[\Delta_{+}\left(\sigma , \sigma^{\prime \prime}\right) 
+ \Delta_{+}\left(\sigma^{\prime} , \sigma^{\prime \prime}\right) 
\right]
\nonumber \\
C_{2 1}{}^{2}\left(\sigma, \sigma^{\prime}, \sigma^{\prime \prime}\right) &=&
\partial_{\sigma}\Delta_{-}\left(\sigma , \sigma^{\prime}
\right)
\left[\Delta_{+}\left(\sigma , \sigma^{\prime \prime}\right) 
+ \Delta_{+}\left(\sigma^{\prime} , \sigma^{\prime \prime}\right) 
\right]
\label{319i}
\end{eqnarray}
All other $ C_{\alpha b}{}^{\gamma}$'s are zero. Note that these structure functions are potentially different from those appearing in section \ref{stringsym} in the sense that here periodic delta functions are introduced to make the basic brackets compatible with the nontrivial BC. Using the expressions of the structure functions (\ref{319i}) in equation (\ref{314}) we can easily derive
\begin{eqnarray}
\delta \lambda_{1} &=& - \dot \alpha_{1}
+ \left(\alpha_{1}\partial_{1}\lambda_{1}
 - \lambda_{1}\partial_{1}\alpha_{1} \right) 
+ 4 \left(\alpha_{2}\partial_{1}\lambda_{2} - \lambda_{2}\partial_{1}
\alpha_{2}\right)\nonumber\\
\delta \lambda_{2} &=& -\dot \alpha_{2}
+\left(\alpha_{2}\partial_{1}\lambda_{1} - \lambda_{1} \partial_{1}\alpha_{2}
\right)
+\left(\alpha_{1}\partial_{1}\lambda_{2} - \lambda_{2} \partial_{1}\alpha_{1}
\right)
\label{GM41i}
\end{eqnarray}
From the relabeling (\ref{313i}), we get the variations of $\rho$ and $\lambda$ as
\begin{eqnarray}
\delta \rho &=& - \dot \alpha_{1}
+ \left(\alpha_{1}\partial_{1}\rho
 - \rho\partial_{1}\alpha_{1} \right) + 2 \left(\alpha_{2}\partial_{1}\lambda - \lambda \partial_{1}\alpha_{2}\right)
\nonumber\\
\delta \lambda &=& -2 \dot \alpha_{2}
+2 \left(\alpha_{2}\partial_{1}\rho - \rho \partial_{1}\alpha_{2}
\right)
+\left(\alpha_{1}\partial_{1}\lambda - \lambda \partial_{1}\alpha_{1}\right)
\label{GM51i}
\end{eqnarray}
In the above we have found out the symmetry transformations of the fields $\rho$ and $\lambda$ in the usual interpolating Lagrangian (\ref{interpolatingL}). Interestingly, they agree with (\ref{GM51}) which were obtained using the structure functions (\ref{319}), markedly different from the ones used here (\ref{319i}). To work out the gauge variation in the reduced framework we choose 
\begin{eqnarray}
&& \alpha_{1}(\sigma) = 2\kappa(\sigma)\alpha_{2}(\sigma)
\label{choice}\\
{\rm and \ write \ (\ref{3112aaa}) \ as,}\,
&& G  = \int d\sigma \alpha_{2}(\sigma)\left[\Pi^{2} + X^{\prime 2} + 
2\kappa\left(\sigma\right) \Pi\cdot X^{\prime}\right] 
\label{gen}
\end{eqnarray}
which is nothing but the generator of gauge transformation in the reduced interpolating framework (\ref{n3}). 

The nontrivial structure functions $C\left(\sigma, \sigma^{\prime}, \sigma^{\prime \prime}\right)$ obtained from
(\ref{n4a}) using (\ref{315}) are
\begin{eqnarray}
C\left(\sigma, \sigma^{\prime}, \sigma^{\prime \prime}\right) = 4 \left[1_{\sigma} \partial_{\sigma}\left( \Delta_{+}\left(\sigma , \sigma^{\prime}\right)\right)
\Delta_{+}\left(\sigma , \sigma^{\prime \prime}\right) 
- 1_{\sigma^{\prime}}\partial_{\sigma^{\prime}}
\left(\Delta_{+}\left(\sigma , \sigma^{\prime}\right)\right)
\Delta_{+}\left(\sigma^{\prime} , \sigma^{\prime \prime}\right) 
\right].
\label{rev4}
\end{eqnarray}
Substituting the structure functions in equation (\ref{314}) yields the variation in $\lambda$ to be
\begin{eqnarray}
\delta \lambda = - 2 \dot{\alpha_{2}}
+ 4\, 1_{\sigma}\left(\alpha_{2}\partial_{\sigma}\lambda
 - \lambda\partial_{\sigma}\alpha_{2} \right).
\label{rev5}
\end{eqnarray}
which can also be obtained by substituting (\ref{choice}) in the second equation of (\ref{GM51i}). 

We are still to find out to what extent the exact correspondence between gauge symmetry and reparametrisation holds, now that we have modified our PB structure and also have performed a partial gauge fixing. This can be done very easily if we stick to the method we have already successfully applied for string, membrane and $p-$brane earlier. We have calculated the gauge variation of the extra fields $\rho$ and $\lambda$. To explicitly show that they are connected to the reparametrization we need to device a map between the gauge parameters and the diffeomorphism parameters. We obtained these maps by demanding the consistency of the variations $\delta X^{\mu}$ due to gauge transformation and reparametrization. They were found to be identical with (\ref{3120a}).

To get the variation of $\rho$ and $\lambda$ induced by the reparametrisation (\ref{31143}) we use the identification (\ref{1119}) in the diffeomorphism variation of $g_{ab}$ (\ref{3122}). The reparametrisation variations of $\rho$ and $\lambda$ reproduce (\ref{GM51i}) once we substitute the reparametrisation parameters $(\Lambda_{0}, \Lambda_{1})$ by the gauge parameters $(\alpha_{1}, \alpha_{2})$ using the map (\ref{3120a}). This establishes complete equivalence of the gauge transformations with the diffeomorphisms of the string.

Again in the reduced picture using the condition (\ref{choice}) in (\ref{3120a}) leads to the map
\begin{eqnarray}
\Lambda^{0}  =  - \frac{1}{\lambda}\,\alpha \quad ; \quad \Lambda^{1} = 0
\label{tm11}
\end{eqnarray}
Using this map in the reparametrisation variation of $\lambda$ (which is again calculated using (\ref{1119}) and (\ref{3122})) reproduces (\ref{rev5}). Thus we establish complete equivalence of gauge symmetry and diffeomorphism in the reduced case as well.
\section{Conclusion}
\label{stringconc}
\pagestyle{myheadings}
\markright{Conclusion}
In this chapter we have developed new action formalisms which interpolate between the Nambu--Goto and the Polyakov type of actions of bosonic membranes, $p$-branes and interacting strings. Such Lagrangeans are based on the first-class constraints, or in other words, the local gauge symmetries of the theories. For such stringy models the interpolating Lagrangeans are shown to provide highly convenient platforms to analyse the interconnection of gauge symmetry and diffeomorphism.

This interpolating formulation was first introduced in \cite{rbbcsg} for free strings. Our analysis for the membrane \cite{bms1}, however, revealed some interesting new aspects. The difference originated from the mismatch of the number of independent metric components with the number of independent reparametrizations in the membrane problem. A definite number of arbitrary variables (that properly accounted for the mismatch) were required to be introduced in the interpolating Lagrangean to reduce it to the Polyakov form. 
A remarkable feature of this analysis was the natural emergence of the cosmological term in the Polyakov action from the internal consistency conditions. A thorough analysis of the gauge symmetries of interpolating actions for strings and membranes was performed using a general method \cite{brr} based on Dirac's theory of constrained Hamiltonian analysis \cite{dir}. Specifically we have demonstrated the equivalence of the reparametrization invariances of different string and membrane actions with the gauge invariances generated by the first class constraints. The appearance (or otherwise) of the Weyl invariance was shown to be a logical consequence of our construction.

We have generalised the above results to the bosonic $p$-brane case in \cite{bms2} where we have shown how an independent metric can be generically introduced in the world volume of the brane. This, again, has been done with the help of the interpolating action based on the first-class constraints. The specific method adopted here leads to the introduction of the metric in a very special way, namely we have achieved a segregation of the $\left(p + 1\right)$ dimensional world volume in the $p-$dimensional spatial part and the Lagrange multipliers analogous to the lapse and shift variables of classical gravity. Using this correspondence we have shown that the Arnowitt--Deser--Misner like decomposition of the brane world volume emerges from our analysis. A comprehensive analysis of the gauge symmetries of the $p-$brane interpolating action has been elaborated and equivalence of gauge and reparametrisation invariances has been established. 

We also discussed the interpolating formulation for an open string propagating in presence of a background Neveu--Schwarz two-form field. It was shown that imposing string boundary conditions demands proper modifications of the Poission bracket (PB) structure of the theory. This in turns generates noncommutative behaviour among the string coordinates. In the gauge-independent settings such results persists even for the free string case. Since modification of the PBs altered the constraint structure of the theory considerably, we reanalysed the underlying unity of gauge symmetry and reparametrisation with the new constraint algebra for the free string case.

\clearpage
\label{sec:gravity}
\chapter{Gauge Symmetry and Diffeomorphism Invariance in second order metric Gravity: A Hamiltonian approach}
\label{gravity}
\pagestyle{myheadings}
\markright{Gauge Symmetry and Diffeomorphism Invariance in second order metric Gravity: A Hamiltonian approach}
\section{Introduction}
\label{gravity1}
\pagestyle{myheadings}
\markright{Introduction}
In the previous chapter we have discussed the application of a Hamiltonian method of \cite{brr}, described in section \ref{Algo}, to the symmetry analysis of strings and higher branes. Considered from the point of view of world-volume coordinates, these systems are generally covariant systems. The most famous example of generally covariant theory is Einstein's General theory of relativity (GTR). We will now apply the same method to investigate the symmetry aspects of GTR.

Einstein's General theory of relativity stands as a successful theory of classical gravity which is also unique in the sense that here space-time manifold itself acquires dynamics. 
The metric tensor $g_{\mu\nu}$ which is a measure of invariant distance between space-time points constitute the dynamical fields of the theory.
As is well known, this feature presents great difficulties in the quantization of gravity. Many variants and extensions of GTR have been proposed which have been argued to be more suitable from one point of view or another. However, a successful theory of Quantum Gravity still eludes us \cite{QG, Carlip}. It is therefore all the more relevant to understand the classical foundations of the theories of Gravitations from different angles.

  The theories of gravitation are distinguished by a common feature which is  general covariance. From the active point of view this is the invariance of the space-time manifold labeled by the coordinates $x^{\mu}$ under the transformations 
\begin{eqnarray}
 x^{\mu} \to x^{\prime}{}^{\mu} = x^{\mu} - \Lambda^{\mu}\left(x \right)
\label{Diff}
\end{eqnarray}
where $\Lambda^{\mu}\left(x \right)$ are arbitrary infinitesimal functions of $x^{\mu}$. This is an automorphism $M \to M$ that moves points within the manifold. Consequently there arises a certain arbitrariness of description of the gravitational field by the metric tensor 
$g_{\mu\nu}$ which can be obtained from their transformations under (\ref{Diff}). Looking from the Hamiltonian (canonical) point of view this arbitrariness is reflected in the transformations generated by the first class constraints of the theory i.e. the gauge transformations.  Stated otherwise, there should exist the right number of gauge invariances corresponding to the invariances (\ref{Diff}). The connection is however non-trivial and therefore has been a topic of continuing interest in the literature \cite{Komar, Teitelboim, Castelani,Pons1, Pons2}. 

The equivalence between the diffeomorphism (diff.) and gauge invariances is completely established when one can prescribe an exact mapping between the two sets of independent transformation parameters. While on the diff. side the independent parameters are intuitively clear, the same can not be said about the gauge parameters. Thus different works related to the subject vary not only in their interpretation of gauge transformation but also in their approach of abstracting the independent gauge parameters. As a concrete example we may consider the problem in connection with the second order metric gravity theory. In \cite{Castelani} the gauge transformations are viewed as mapping solutions to solutions and independent gauge generators are obtained following a``more Lagrangean" approach of \cite{Sudarshan} which makes use of the Lagrange equations of motion. Gauge transformations can, on the other hand, be considered as mapping field configurations to field configurations. In fact this is the essence of Dirac's point of view. In \cite{Pons1, Pons2} this point of view is adopted. They find the connection between the diffeomorphism group and the gauge group by a certain projection technique from the configuration-velocity space to the phase space. Though the approaches in these works differ, they share the following common features:
\begin{enumerate}
\item All these works utilise a combination of Lagrangean and Hamiltonian methods. They can not be identified as strict Hamiltonian approaches.
\item In one way or another these works make use of the Lagrange's equations of motion. 
\end{enumerate}
 These aspects are precisely the points of departure in our work \cite{diff} which we shall discuss in the present chapter. Our primary goals are the following:  
\begin{enumerate}
\item The construction of a dedicated Hamiltonian approach {\it a la} Dirac which will lead to the equivalence between the diffeomorphism and gauge transformations. 
\item To derive the most general gauge transformation generator without taking recourse to the velocity-space approach. 

\end{enumerate}
As concrete example we will consider the second order metric gravity theory here, though our approach will be easily applicable to other theories of gravitation as well. To summarise the principal results and fix the notations used in the rest of this chapter we provide a short introduction of the canonical theory of second order metric gravity in $\left(3+1\right)$ dimensions. 
\section{Second order canonical formalism of metric gravity}
\label{gravity2}
\pagestyle{myheadings}
\markright{Second order canonical formalism of metric gravity}
We begin with the Einstein--Hilbert action on a manifold $M$
\begin{eqnarray}
S = \int \left(-{}^{\left(4\right)}g\right)^{1/2} {}^{\left( 4\right) }R\left(x\right) d^{4}x
\label{S} 
\end{eqnarray}
where ${}^{\left( 4\right) }R\left(x\right)$ is the Ricci scalar and ${}^{\left(4\right)}g$ is the determinant of the metric ${}^{\left(4\right)}g_{\mu\nu}$. The pre-superscript ${}^{\left(4\right)}$ indicates that the corresponding quantities are defined on the four-dimensional manifold $M$. This is required to distinguish these quantities from their analogue defined on the three-hypersurface which are written without any such pre-superscript. 

   By adding suitable divergence to the action (\ref{S}) we write an equivalent Lagrangean \cite{hrt, Sunder} 
\begin{eqnarray}
\int d^{3}x {\cal{L}} =  \int d^{3}x N^{\perp} \left(g\right) ^{1/2} \left(K_{ij}K^{ij} - K^{2} + R \right) 
\label{L} 
\end{eqnarray}
where $ K = K^{i}{}_{i} = g_{ij} K^{ij}$ and $R$ is the Ricci scalar on the three surface. The lapse variable $N^{\bot}$ and shift variables $N^{i}$ represent arbitrary variation, respectively normal to and along the three-surface on which the state of the system are defined 
\begin{eqnarray}
N^{j} &=& g^{ij}g_{0i} \label{Nj} \\
N^{\perp} &=& \left(- g^{00}\right)^{-1/2} \label{N}
\end{eqnarray}
Note that $N^{i}$ is contained in the Lagrangean (\ref{L}) through the definition of $K_{ij}$ given by
\begin{eqnarray}
K_{ij} &=& \frac{1}{2 N^{\bot}} \left(- \dot{g}_{ij} + N_{i \mid j } + N_{j \mid i} \right)
\label{K} 
\end{eqnarray}
where the ${\mid}$ indicates covariant derivative on the three-surface. Since the lapse and shift variables represent arbitrary deformations of the hypersurface one can expect them not to be restricted by the Hamiltonian equations. 
Hence the Lagrangean (\ref{L}) is suitable for canonical analysis because it does not contain time derivatives of $N^{\mu} \left(N^{\bot},N^{i}\right)$. One can immediately write down the primary constraints following from the definition of the conjugate momenta of $N^{\mu}$
\begin{eqnarray}
\pi_{\mu} &=& \frac{\partial {\cal{L}}}{\partial{\dot{N}}^{\mu}} = 0
\label{M1} 
\end{eqnarray} 
The second fundamental form of the three-surface $K_{ij}, \left(i,j = 1,2,3\right)$ contains the velocities $\dot{g}_{ij}$ and therefore related to the momenta canonical to $g_{ij}$ by 
\begin{eqnarray}
\pi^{ij} &=& \frac{\partial {\cal{L}}}{\partial{\dot{g}}_{ij}} = - \left(g\right) ^{1/2}\left(K^{ij} - K g^{ij} \right)
\label{M2} 
\end{eqnarray} 
The inverse relation expresses $K_{ij}$ in terms of the dynamical variables of the theory
\begin{eqnarray}
K^{ij} =  - \left(g\right) ^{-1/2}\left(\pi^{ij} - \frac{1}{2}\pi g^{ij} \right); \qquad {\rm {where}} \qquad \pi = g_{ij} \pi^{ij}
\label{K1} 
\end{eqnarray} 
The non-trivial Poission Brackets (PB) between the pair of conjugate variables of the theory are 
\begin{eqnarray}
\left\lbrace g_{ij} \left( x\right), \pi^{kl} \left( x^{\prime} \right)\right\rbrace &=& \frac{1}{2}\left( \delta^{i}{}_{k}\delta^{j}{}_{l} + \delta^{j}{}_{k}\delta^{i}{}_{l}\right) \delta^{\left(3 \right) }\left( x - x^{\prime}\right) \nonumber \\
\left\lbrace N^{\mu} \left( x\right), \pi_{\nu} \left( x^{\prime} \right)\right\rbrace &=& \delta^{\mu}{}_{\nu} \delta^{\left(3 \right) }\left( x - x^{\prime}\right) 
\label{PB}
\end{eqnarray} 
Using equations (\ref{L}), (\ref{K}), (\ref{M1}) and (\ref{K1}) the canonical Hamiltonian can be worked out as
\begin{eqnarray}
H_{c} & = & \int d^{3}x \left(\pi_{\mu} \dot{N}^{\mu} + \pi^{ij} \dot{g}_{ij} - {\cal{L}}\right) = \int d^{3}x \left( N^{\perp}{\cal{H}}_{\perp} + N^{i}{\cal{H}}_{i} \right) 
\label{gH} \\
{\rm where,} \qquad
{\cal{H}}_{\perp} & = &  g ^{-1/2} \left( \pi_{ij}\pi^{ij} - \frac{1}{2} \pi^{2}\right) - \left(g\right) ^{1/2}R, \qquad
{\cal{H}}_{i}  =  - 2 \pi_{i}{}^{j}{}_{\mid j }
\label{H2} 
\end{eqnarray}
The basic brackets (\ref{PB}) are used to conserve the primary constraints 
\begin{eqnarray}
\Omega_{\mu} = \pi_{\mu}  \approx 0
\label{PC} 
\end{eqnarray}
with the Hamiltonian (\ref{gH}) to generate the secondary constraints 
\begin{eqnarray}
\Omega_{4} =  {\cal{H}}_{\perp} \approx 0, \qquad 
\Omega_{4+i} =  {\cal{H}}_{i} \approx 0
\label{SC} 
\end{eqnarray}
Using the basic PBs the constraint algebra becomes \cite{dir}
\begin{eqnarray}
\left\lbrace \Omega_{4}\left( x\right), \Omega_{4}\left( x^{\prime}\right) \right\rbrace   
&=& g^{ri} \left[ \Omega_{4+i} \left( x\right) +  \Omega_{4+i} \left( x^{\prime}\right)\right] 
\delta_{,i}\left(x - x^{\prime} \right) \nonumber\\
\left\lbrace\Omega_{4+i}\left( x\right), \Omega_{4}\left( x^{\prime}\right) \right\rbrace  
&=& \Omega_{4} \delta_{,i}\left(x - x^{\prime} \right) \nonumber\\
\left\lbrace\Omega_{4+i}\left( x\right), \Omega_{4+j}\left( x^{\prime}\right) \right\rbrace 
&=&  \Omega_{4+i}\left( x^{\prime}\right)\delta_{,j}\left(x - x^{\prime} \right)
 + \Omega_{4+j}\left( x\right)\delta_{,i}\left(x - x^{\prime} \right) 
\label{CA} 
\end{eqnarray}
This weakly involutive algebra  signifies that the set (\ref{PC}) - (\ref{SC}) are first-class constraints. 
This concludes our review of the canonical formulation of metric gravity. In the next section we will discuss our methodology of analyzing the gauge symmetry and establishing its underlying unity with the reparametrization invariance of the theory.
\section{Our methodology}
\label{gravity3}
\pagestyle{myheadings}
\markright{Our methodology}
In the Canonical approach to the metric gravity a time parameter needs to be identified. This is attained by dividing space-time in to a collection of space-like three-surfaces with a time-like direction of evolution. This is the famous Arnowitt--Deser--Misner (A--D--M) decomposition \cite{adm} where the arbitrariness of the foliation is reflected by one `lapse' and three `shift' variables in equations (\ref{Nj}, \ref{N}). One can cast the original Einstein--Hilbert action modulo boundary terms in the form (\ref{L}) where no time derivative of these variables appear. As a consequence their corresponding conjugate momenta (\ref{M1}) vanish imposing four primary constraints in (\ref{PC}). Conservation of these constraints gives rise to four secondary constraints in (\ref{SC}). All these constraints are first-class. Since the Hamiltonian is a linear combination of these constraints no further constraints appear. According to the Dirac conjecture the gauge generator is a linear combination of all these first-class constraints. There are thus eight gauge parameters appearing in the generator. However, only four of them are independent since the number must be equal to the number of primary first-class constraints (\ref{PC}). As has been pointed out in the introduction, the crucial first step in establishing a one-to-one correspondence between the diffeomorphisms and the gauge variations is to identify the independent gauge parameters.  This is where the Hamiltonian technique of \cite{brr} which we have summarised in the previous chapter comes int play. This method has been applied to analyze the gauge invariances in various field and string theoretic models in the literature \cite{brr, ban, ex1, bms1, bms2, sgaghas}. This approach of analyzing the gauge invariances can be contrasted with the approach of \cite{Castelani} where a more Lagrangean approach of \cite{Sudarshan} was adopted and also with the approach of \cite{Pons1} where the gauge transformations are obtained as Legendre map from the coordinate-velocity space to the phase space. Also this algorithm is a ``dedicated" Hamiltonian algorithm in the sense that it requires only the Hamiltonian and the first-class constraints of the theory and no reference to the associated action is necessary. 

 Once the independent gauge parameters are identified we require to find a connection through which the gauge variations and diff. variations may be related. Again the lapse and shift variables provide this connection. Their gauge variations can be immediately written down. Since they are related to the 0i-th components of the metric their variation due to reparametrization (\ref{Diff}) can be independently worked out. This will be used to establish the exact mapping between the independent gauge and diff. parameters. The mapping obtained by this connection will then be tested on the other variables to verify the consistency of the procedure. This will explicitly demonstrate the unity of the different symmetries involved. Also this mapping will enable us to compare our results with those available in the literature \cite{Castelani, Pons1, Pons2}. 
\section{Gauge invariance vs. Diffeomorphism in second order metric Gravity}
\label{gravity4}
\pagestyle{myheadings}
\markright{Gauge invariance vs. Diffeomorphism in second order metric Gravity}
We begin our analysis of gauge symmetry of the metric gravity by writing the gauge generator as a linear combination of all the first-class constraints of the theory
\begin{eqnarray}
G = \int d^{3}x \left(\epsilon^{0}\Omega_{0} + \epsilon^{i}\Omega_{i} + \epsilon^{4}\Omega_{4} + \epsilon^{4+i}\Omega_{4+i}\right)
\label{GG} 
\end{eqnarray}
which is obtained from (\ref{217}) in the continum limit. The set of constraints $\Omega$ is given by (\ref{PC})-(\ref{SC}). To find the independent gauge parameters from the set $\left(\epsilon^{0}, \epsilon^{i}, \epsilon^{4}, \epsilon^{4+i}\right)$ we require to solve the analogue of (\ref{219}), with the indicated parameters. For this we require to compute the structure functions of the involutive algebra (\ref{CA}).

The structure functions $C_{ab}{}^{c}$ are obtained from equation (\ref{2110}) of which only $C_{b_{1}a}^{a_{2}}$ will be required in our analysis. The details of these structure functions are given in \cite{Teitelboim}. However, the later coefficients vanish in the present case since the primary first-class constraints $\Omega_{\mu}$ in (\ref{PC}) gives strictly zero brackets with all the constraints of the theory. The non-trivial structure factors $V_{\alpha}^{\beta}\left( x, x^{\prime}\right) $ are obtained from equation (\ref{2110a}) written in the continum limit as
\begin{equation}
\left\lbrace H_{c}, \Omega_{\alpha}\left( x\right) \right\rbrace   = \int d^{3}x^{\prime} V_{\alpha}^{\beta}\left( x, x^{\prime}\right) \Omega_{\beta}\left( x^{\prime}\right)
\end{equation}
Using the constraint algebra (\ref{CA}) we get 
\begin{eqnarray}
V_{4}{}^{4+s}\left(x, x^{\prime} \right) &=&  N^{\bot}\left(x^{\prime} \right) g^{rs}\left(x^{\prime} \right) \partial^{\prime}{}_{r}\delta\left(x - x^{\prime} \right) 
- \partial_{r} N^{\bot} g^{rs}\delta\left(x - x^{\prime} \right) \nonumber \\
V_{4}{}^{4}\left(x, x^{\prime} \right) &=&  N^{i}\left(x^{\prime} \right)\partial^{\prime}_{i}\delta\left(x - x^{\prime} \right), \qquad 
V_{4+s}^{4}\left(x, x^{\prime} \right) = -\partial_{s}N^{\bot}\left(x \right)\delta\left(x - x^{\prime} \right) \nonumber \\
V_{4+s}^{4+i}\left(x, x^{\prime} \right) &=& -\partial_{s}N^{i}\delta\left(x - x^{\prime} \right)
+ N^{l}\left(x^{\prime} \right)\partial^{\prime}_{l}\delta\left(x - x^{\prime} \right)\delta^{i}{}_{s}  \nonumber \\
V_{\mu}^{4}\left(x, x^{\prime} \right) & = &\delta^{0}{}_{\mu}\delta\left(x - x^{\prime} \right), \qquad 
V_{\mu}^{4+i}\left(x, x^{\prime} \right) = \delta^{i}{}_{\mu}\delta\left(x - x^{\prime} \right)
\label{V} 
\end{eqnarray}

The basic equations connecting the gauge parameters (i.e. (\ref{219})) now become 
\begin{equation}
  0 = \frac{d\epsilon^{a_2}\left(x \right)} {dt}
 - \int d^{3}x^{\prime} \epsilon^{a}\left(x^{\prime} \right) V_{a}^{a_2}\left(x^{\prime}, x\right) 
\label{219a}
\end{equation} 
Using (\ref{V}) in (\ref{219a}) four equations involving the eight gauge parameters are obtained which can be written as
\begin{eqnarray}
\epsilon^{0} \left(x\right) &=&\left[  \dot{\epsilon}^{4} + \epsilon^{4+s} \partial_{s}N^{\bot} - N^{i} \partial_{i}\epsilon^{4}\right] \left(x\right)\\
\epsilon^{i} \left(x\right)&=&\left[  \dot{\epsilon}^{4+i} + \epsilon^{4+s} \partial_{s}N^{i} - N^{l}\partial_{l} \epsilon^{4+i}
- N^{\bot}g^{ri}\partial_{r}\epsilon^{4} + \epsilon^{4}g^{ri}\partial_{r}N^{\bot}\right] \left(x\right)
\label{IGP} 
\end{eqnarray}
The equations (\ref{IGP}) suggest that the set $\left( \epsilon^{0}, \epsilon^{i}\right) $ will be the appropriate choice of the dependent gauge parameters. 
Substituting the above expressions in (\ref{GG}) we obtain the gauge generator solely in terms of the independent gauge parameters the number of which matches with the number of independent primary first-class constraints, as it should be \cite{HTZ, brr}. Also note that the most general form of the gauge generator contains time derivatives of the independent gauge parameters. It is remarkable that in our approach this feature follows naturally from the formalism and needs no special treatment. 

Before proceeding further let us note that the assumption on which (\ref{219}) is based only involves the relation between the velocities and the canonical momenta and the arbitrary Lagrange multipliers, i.e. the first of Hamiltons equations in \cite{brr}
\begin{eqnarray}
\dot{q} = \left[q, H_{c}\right] + \lambda^{a_{1}}\left[q, \Omega_{a_{1}}\right]
\label{q} 
\end{eqnarray}
So the full dynamics is not required to impose restrictions on the gauge parameters. Since this is the only input in our method of abstraction of the independent gauge parameters in the context of second order metric gravity we find that our analysis will be valid off-shell to this extent. Of course dynamics will be needed to establish the equivalence of the full phase-space variables under the two types of transformations \cite{Carlip}.

After identifying the most general gauge generator of the theory we now proceed to derive the desired mapping between the gauge and the reparametrization parameters. This is conveniently obtained from the gauge variations of $N^{i}$, comparing them with the corresponding variations due to reparametrization (\ref{Diff}). The gauge variations of the shift variables are 
\begin{eqnarray}
\delta N^{i}\left(x \right) & = & \left\{N^{i}\left(x\right), G\right\}
=
\left[\dot{\epsilon}^{4+i} + \epsilon^{4+s} \partial_{s}N^{i} - N^{l}\partial_{l} \epsilon^{4+i}
 - N^{\bot}g^{ri}\partial_{r}\epsilon^{4} + \epsilon^{4}g^{ri}\partial_{r}N^{\bot}\right] 
\label{GV2}
\end{eqnarray}
To find the corresponding variations due to reparametrization we have to use the variations of the four-metric ${}^{\left( 4\right) }g_{\mu \nu}$ under the infinitesimal transformation (\ref{Diff})
\begin{eqnarray}
\delta  {}^{\left( 4\right) }g_{\mu \nu}  = {}^{\left( 4\right) }g_{\gamma \nu} \partial_{\mu}\Lambda^{\gamma} + {}^{\left( 4\right) }g_{\gamma \mu} \partial_{\nu}\Lambda^{\gamma} + \Lambda^{\gamma}\partial_{\gamma}  {}^{\left( 4\right) }g_{\mu \nu}
\label{gDiff}
\end{eqnarray}
Using (\ref{gDiff}) and (\ref{Nj}) we can compute the desired variations under the reparametrization (\ref{Diff})
\begin{eqnarray}
\delta N^{i}\left(x \right) & = & \left(\frac{d}{dt} - N^{k}\partial_{k} \right) \left(  \Lambda^{i} +  \Lambda^{0}N^{i}\right) 
+ \left(  \Lambda^{k} +  \Lambda^{0}N^{k}\right) \partial_{k}N^{i} 
 - \left( N^{\bot} \right)^{2} g^{ij} \partial_{i} \Lambda^{0} 
\label{R2}
\end{eqnarray}
where we have also used the inverse of the relations (\ref{Nj}) and (\ref{N}), namely 
\begin{eqnarray}
g_{ij}N^{j}  =  N^{i}, \qquad g_{ij}N^{i}N^{j} - \left(N^{\bot} \right)^{2} = g_{00}
\label{IN}
\end{eqnarray}
Comparing the variations of the shift variable $N^{i}$ from (\ref{GV2}) and (\ref{R2}) we obtain the sought-for mapping between the reparametrization parameters and the independent gauge parameters
\begin{eqnarray}
\epsilon^{4+i} & = & \Lambda^{i} +  \Lambda^{0}N^{i}, \qquad \epsilon^{4}  =  N^{\bot}\Lambda^{0}
\label{M}
\end{eqnarray}
Note that similar mapping between the different sets of parameters were obtained earlier in \cite{Castelani} and also in \cite{Pons1}. Note however that in comparison to these earlier works we follow a strictly Hamiltonian approach. Also our analysis requires only the first set of the Hamiltons equations (\ref{q}). Thus the connection we derive is valid off-shell which is a completely new result. Moreover, we provide a structured algorithm for metric gravity where the occurrence of time derivative of the gauge parameter need not be addressed separately. Though discussed in connection with the second order metric gravity it is apparent that this algorithm is applicable in the same general form to other theories of gravitation as well.

    A through consistency check of the whole formalism is now in order. The mapping (\ref{M}) when used in the gauge variation of the lapse variable $N^{\bot}$
\begin{eqnarray}
\delta N^{\bot}\left(x \right) & = &\left[  \dot{\epsilon}^{4} + \epsilon^{4+s} \partial_{s}N^{\bot} - N^{i} \partial_{i}\epsilon^{4}\right] \left(x\right)
\label{GV1}
\end{eqnarray}
gives its variation in terms of the diff. parameters 
\begin{eqnarray}
\delta N^{\bot}\left(x \right) & = & \left(\frac{d}{dt} - N^{i}\partial_{i}\right)\Lambda^{0}N^{\bot} 
+ \Lambda^{0}N^{i}\partial_{i}N^{\bot} + \Lambda^{i}\partial_{i}N^{\bot}
\label{R1}
\end{eqnarray}
which is identical with the variation calculated from (\ref{gDiff}).
Similarly, we work out the gauge variation of $g_{ij}$ generated by $G$ (\ref{GG}) which gives 
\begin{eqnarray}
\delta g_{ij} \left( x \right)   =   \left\lbrace g_{ij} \left( x \right), G \right\rbrace 
=  -2 \epsilon^{4} K_{ij} +  \epsilon^{4+k} \partial_{k}g_{ij}
 + g_{ki}\partial_{j} \epsilon^{4+k} + g_{kj}\partial_{i} \epsilon^{4+k} 
\label{gGV}
\end{eqnarray}
and use the mapping (\ref{M}) in it. The resulting expression can be identified with the reparametrization variation of $g_{ij}$ given by 
\begin{eqnarray}
\delta g_{ij} \left( x \right)  = \left( \Lambda^{0} \frac{d}{dt}  - \Lambda^{k} \partial_{k} \right) g_{ij}  + N_{i} \partial_{j} \Lambda^{0} 
+ N_{j} \partial_{i} \Lambda^{0} + g_{ki}\partial_{j} \Lambda^{k} + g_{kj}\partial_{i} \Lambda^{k}
\label{gR}
\end{eqnarray}
This completes the explicit identification of the gauge invariance and diffeomorphism in second order metric gravity theory. 
\section{Conclusion}
\label{gravity5}
\pagestyle{myheadings}
\markright{Conclusion}
We discussed a novel approach of obtaining the most general gauge invariances of the second order metric gravity theory following the general Hamiltonian method of \cite{brr} and used this analysis to establish a one-to-one mapping between the gauge and reparametrization parameters. We have performed explicit computation to check the consistency of our method. Though we re-derive already available results \cite{Castelani, Pons1} our method is completely new in the following senses: 
\begin{enumerate}
\item This is a new dedicated Hamiltonian approach to the problem and does not require to refer to the velocity space at any stage in the calculational algorithm. As far as we know this is the first time such a calculational scheme is advanced in canonical gravity. 
\item This approach reveals properly to what extent the mapping between diffeomorphisms and gauge invariances can be considered valid off-shell. Our Hamiltonian method clearly reveals that it is dependent only on the first set of Hamilton's equations which connects the velocities, momenta and the Lagrange multipliers. In other words the specific phase space structure is only important but not the full dynamics. Note however dynamics must be invoked in establishing the equivalence of transformations of the full set of phase space variables as we have already mentioned.
\end{enumerate}
In addition to these attractive features our method has the advantage of providing a structured algorithm which can easily be applied to other theories of gravitation.

\clearpage
\label{sec:nc}
\chapter{Aspects of Noncommutative gauge symmetry and their application to the study of Noncommutative Gauge Theory}
\label{nc}
\pagestyle{myheadings}
\markright{Aspects of Noncommutative gauge symmetry and their application to the study of Noncommutative Gauge Theory}
\section{Introduction}
\label{ncintro}
\pagestyle{myheadings}
\markright{Introduction}
 Our investigation of symmetries in open string brings us in connection with noncommutativity among the space-time coordinates. We have already given a brief review of the origin and various motivations of noncommutative structure of space-time in the introduction of this thesis. In view of the recent string theoretic results the idea of a noncommutative space time has resurfaced in the late nineties and field theories defined over this NC space are currently the subject of very intense research \cite{szabo}. The idea is to introduce a space time where the coordinates $x^{\mu}$ satisfy the noncommutative (NC) algebra  
\begin{equation}
\left[x^{\mu}, x^{\nu}\right] = i \theta^{\mu \nu}
\label{ncgometry}
\end{equation}
where the anti-symmetric tensor $\theta^{\mu \nu}$ may be constant, or space-time dependent having Lie-algebraic or quantum group structure. In the remainder of this thesis we will assume constant (canonical) noncommutativity and exploit the NC gauge symmetry to analyse some field theoretic and quantum mechanical models defined over such NC space-time.

Mainly, two different approach have been devised to analyse these NC field theories. One approach (we call it the Operator approach in this thesis) is to work in terms of the operators defined in a certain Hilbert space which carries a representation of the basic NC algebra. The fields are defined as operators in this Hilbert space by the Weyl--Wigner correspondence \cite{szabo}. Alternatively one can work in the deformed phase space where the ordinary product is replaced by the star product and the fields are defined as functions of the phase space variables with the product of two fields \footnote{Note that a hat sign over the field denotes that it is defined as a function of the NC space-time} $\hat \phi(x)$ and $\hat \psi(x)$ given by the star product
\begin{equation}
\hat \phi(x) \star \hat \psi(x) = \left(\hat \phi \star \hat \psi \right)(x) = e^{\frac{i}{2}
\theta^{\alpha\beta}\partial_{\alpha}\partial^{'}_{\beta}} 
  \hat \phi (x) \hat \psi(x^{'})\big{|}_{x^{'}=x.} 
\label{star}
\end{equation}
Once in the deformed phase space one no longer has to worry about the ordering of the NC fields. These NC fields can then be expressed in terms of ordinary fields which are defined over the commutative space-time using the celebrated Seiberg-Witten (SW) maps \cite{SW}. The SW-type maps, implemented individually on the fields, thus enable us to work out a commutative equivalent description of different NC gauge theories. 

This alternative commutative equivalent approach thus relies heavily on the important correspondence between the ordinary and the NC gauge symmetry which was first perceived by Seiberg and Witten during their study of the Dirac--Born--Infeld (DBI) action of open string dynamics on a D-Brane \cite{SW, frad}. They observed that both ordinary and noncommutative Yang--Mills (YM) fields arise from the same two-dimensional field theory depending on the regularization scheme. Hence there must be a transformation from ordinary to NC YM fields which maps the standard YM gauge symmetry to the NC YM gauge symmetry. Thus a space-time redefinition between the ordinary and noncommutative gauge fields was indicated which led to the Seiberg-Witten (SW) maps \cite{SW}. Before proceeding any further we shall briefly describe the salient features of the SW map.
\section{The Seiberg--Witten Map}
\label{SeibergWitten}
\pagestyle{myheadings}
\markright{The Seiberg--Witten Map}
The SW maps are explicit local transformations connecting a given NC gauge theory with a conventional gauge theory. This means that to any finite order in the NC parameter $\theta$ the NC gauge fields and gauge parameters are to be given by local differential expressions of ordinary gauge fields and gauge parameters.

Let us consider the case in which the noncommutative gauge theory is governed by a Yang-Mills (YM) Lagrangian for the gauge potential $\hat A_\mu$, transforming under $\star-$gauge transformations according to
\begin{eqnarray}
\hat \delta_{\hat \lambda} \hat{A}_\mu(x) = \hat A'_\mu(x) - \hat A_\mu(x) = D_\mu[\hat A] \star \hat \lambda(x)~.
\label{seiberg1}
\end{eqnarray}
where $D_\mu[\hat A] \star$ denotes the star covariant derivative and the hatted fields and parameters denote that they are defined on the NC space-time. The SW map connects the noncommutative YM Lagrangian to some ordinary Lagrangian on the commutative space-time. In this latter ordinary Lagrangean, apart from the fact that fields are multiplied with the ordinary product, the transformation law for the gauge field $A_\mu$ is now governed by the ordinary covariant derivative
\begin{eqnarray}
\delta_{\lambda} A_\mu(x) = A'_\mu(x) - A_\mu(x) =  D_\mu[A] \lambda(x).
\label{seiberg2}
\end{eqnarray}
Hence, the SW map should include, a connection between $\hat A_\mu$ and $A_\mu$ and also another between $\hat \lambda$ and $\lambda$. It turns out that the equivalence holds at the level of orbit space, i.e., the physical configuration space of gauge theories. This means that if two gauge fields $\hat A_\mu$ and $\hat A'_\mu$ belonging to the same orbit can be connected by a noncommutative gauge transformation $\exp_{*}(i \hat \lambda)$, then $A'_\mu$ and $A_\mu$, the corresponding mapped gauge fields will also be gauge equivalent by an ordinary gauge transformation $\exp(i \lambda)$. Importantly, the mapping between $\hat \lambda$ and $\lambda$ must depends on both $\lambda$ as well as $A_\mu$. Indeed, if $\hat \lambda$ were a function of $\lambda$ only, the ordinary and the noncommutative gauge groups would be isomorphic. But even for the simple rank 1 case, the ordinary gauge group is abelian and acts by 
\begin{equation}
\delta A_{\mu} = \partial_{\mu} \lambda
\label{seiberg2a}
\end{equation}
whereas the NC gauge group is nonabelian and acts by 
\begin{equation}
\delta A_{\mu} = \partial_{\mu} \lambda + i \lambda \star A_{\mu} - i A_{\mu}\star\lambda 
\label{seiberg2b}
\end{equation}
Since an abelian group can not be isomorphic to a nonabelian group, no redefinition of the ordinary gauge parameter can map it to the NC gauge parameter while interwining with the gauge invariances.

Then, deriving the SW maps is to find relations like 
\begin{eqnarray}
\hat A = \hat A [A;\theta], \qquad  \hat \lambda = \hat \lambda [\lambda, A;\theta]
\label{seiberg3}
\end{eqnarray}
so that the equivalence between orbits holds
\begin{eqnarray}
\hat A [A] + \hat\delta_{\hat\lambda}\hat A [A]= \hat A[A + \delta_\lambda A].
\label{seiberg4}
\end{eqnarray}
Using the explicit form of gauge transformations and expanding to first order in $\theta$, the solution of
eq (\ref{seiberg4}) is
\begin{eqnarray}
\hat A_\mu[A] = A_\mu -\frac{1}{4} \theta^{\rho \sigma} \{A_\rho, \partial_\sigma A_\mu + F_{\sigma \mu}\} + {\cal O}(\theta^2); \quad 
 \hat\lambda [\lambda,A] = \lambda + \frac{1}{4} \theta^{\rho \sigma}\{\partial_\rho \lambda, A_\sigma\} + {\cal O}(\theta^2)
\label{seibergsolsw}
\end{eqnarray}
where the products on the right hand side, such as $\{A_\rho, \partial_\sigma A_\mu\}=A_\rho.\partial_\sigma A_\mu
+ \partial_\sigma A_\mu.A_\rho$ are ordinary matrix products \footnote{Note that solution to S--W equation (\ref{seiberg4}) are not unique. In general homogeneous terms can be added to the solution (\ref{seibergsolsw}) \cite{stern1}. }. Concerning the field strength, the connection is given by
\begin{eqnarray}
\hat F_{\mu\nu}[A] = F_{\mu \nu} +\frac{1}{4}\delta\theta^{\alpha \beta}\left(2\{F_{\mu \alpha},F_{\nu \beta}\} - \{A_\alpha, D_\beta F_{\mu\nu} + \partial_\beta F_{\mu\nu}\} \right) + O(\delta\theta^2)
\label{seibergswf}
\end{eqnarray}
Similarly, SW maps for the matter fields either in the adjoint or the fundamental representation can be worked out. Note that in the next two sections we shall focus on NC abelian gauge theory where the gauge fields are simple functions rather than matrix-valued ones. Latter in section \ref{ncgr} we shall deal with non-abelian gauge theory in the context of NC general relativity. The necessary details of the particular gauge theories dealt with and the corresponding SW maps for the gauge and matter fields are given in appropriate places.


\section{Application to the abelian Chern--Simons and Maxwell theory in NC space-time: Search for NC solitons with a smooth commutative limit}
\label{chernmax}
\pagestyle{myheadings}
\markright{Application to the abelian Chern--Simons and Maxwell theory in NC space-time: ...}
    A particularly interesting scenario in the context of NC gauge theories is the occurrence of classical stable soliton solutions in odd dimensional scalar NC field theories with self interaction only \cite{gms} and also in a NC $U(1)$ gauge theory coupled with matter in the adjoint representation of the gauge group \cite{poly}. 
The NC $U(1)$ gauge theories coupled with adjoint matter is important for their possible application in constructing D-branes as solitons of the tachyon field in NC open string theory \cite{hkl, as}. It is indeed worthwhile to analyse such theories from different points of view. A solution generating technique  based on the operator approach has been used in \cite{prnjp} which was developed in \cite{hkl}. Here different solutions have been obtained in various limits of the action, 
\begin{equation}
S = \int dt d^{2}x \left( -\frac{1}{4} \left(F_{\mu \nu}\right)^{2} + \frac{1}{2}D^{\mu}\phi D_{\mu}\phi - V\left(\phi - \phi_{\star}\right)\right)
\label{model}
\end {equation}
where $\phi$ is the scalar field in the adjoint representation of the NC U(1) group. The potential $V$ has a local minima at $\phi = \phi_{\star}$ with $V(0)= 0$ and a local maximum at $\phi = 0$. The static soliton solution of the theory has energy
\begin{equation}
E = 2 \pi \theta n \left(\frac{1}{2 \theta^{2}} + V\left( - \phi_{\star}\right)\right)
\end {equation} 
which diverges as $\theta \to 0$. These soliton solutions obtained by using the operator approach thus essentially belong to the singular sector. It is not clear what happens in the $\theta \to 0$ limit.
Specifically, the question is whether there is some non-trivial non-perturbative solutions depending on the NC parameter and vanishing continuously along with it. Clearly, this question can not be answered by the operator approach. It is however very much desirable to explore all sectors of solutions of the theory. At this point our commutative equivalent method which is regular in the $\theta \to 0$ limit becomes instrumental. 
Specifically, it will be very much desirable if this commutative equivalent analysis can be done in a closed form such that results exact to all orders in $\theta$ are obtained. Naturally, the possibility of this rests on the availability of the SW maps in a closed form. 

 A method of obtaining SW maps for certain models has been devised recently which is exact in the NC parameter \cite{hsk, rbhsk}. This is based on the change of variables between open and closed string parameters and connection of the approach with the deformation quantization technique \cite{Kont} has been demonstrated \cite{rbhsk}. Specifically, an exact map for an adjoint scalar field has been found \cite{rbhsk}, consistent with that deduced from RR couplings of unstable non-BPS D-branes \cite{mukhi}. Note that this closed-form S--W map will provide a exact commutative equivalent method valid non-perturbatively which compares favourably with the operator approach. In the following two sections we shall construct exact commutative equivalent descriptions of adjoint scalar matter coupled to NC Chern--Simons (C--S) and Maxwell-type of gauge fields. 

\subsection{An NC Chern--Simons Theory through exact Seiberg--Witten maps}
\label{chern}
\pagestyle{myheadings}
\markright{An NC Chern--Simons Theory thorugh exact Seiberg--Witten maps}
In this section we shall analyze a $U(1)_{\star}$ C--S coupled scalar field theory in $2+1$ dimensional flat space time where the scalar field is in the adjoint representation of the gauge group. Models with the NC scalar field in the adjoint representation have been considered earlier from the operator approach with the gauge field dynamics governed solely by the Maxwell term \cite{poly} and also by a combination of the Maxwell and the C--S term \cite{ prnjp}. Our selection of the model here as well as in the next section is motivated by the fact that in the commutative limit the adjoint scalar field decouples from the gauge interaction. In other words any non-trivial result of our analysis comes from the NC features only.

The action of our theory is given by
\begin{eqnarray}
\hat S &=& \int d^{3}x\left[\frac{1}{2} \left(\hat D_{\mu} \star \hat \phi \right)\star \left(\hat D^{\mu} \star \hat \phi\right) + \frac{k}{2}\epsilon^{\mu \nu \lambda}\left(\hat A_{\mu} \star \partial_{\nu}\hat A_{\lambda} - \frac{2i}{3} \hat A_{\mu} \star \hat A_{\nu} \star \hat A_{\lambda}\right)\right]
\label{ncaction}
\end{eqnarray}
where $\hat \phi$ is the scalar field and $\hat A_{\mu}$ is the NC C--S gauge field. We adopt the Minkowski metric $\eta_{\mu \nu} = {\rm diag} \left( +,-,-,-\right)$. The covariant derivative $\hat D_{\mu}\star \hat \phi$ is in the adjoint representation and is defined as
\begin{equation}
\hat D_{\mu}\star \hat \phi = \partial_{\mu} \hat \phi - i \left[\hat A_{\mu}, \phi\right]_{\star}
\label{covder}
\end{equation}
The action (\ref{ncaction}) is invariant under the $\star$-gauge transformation
\begin{equation}
\hat \delta_{\hat \lambda} \hat A_{\mu} = \hat D_{\mu} \star \hat \lambda, \qquad \hat \delta_{\hat \lambda} \hat \phi = -i \left[\hat \phi, \hat \lambda  \right]_{\star}
\label{starg}
\end{equation}

  The commutative equivalent of (\ref{ncaction}) is obtained by using the exact SW map for $\hat D_{\mu}\star\hat \phi(x)$ \cite{rbhsk} and noting that the C--S action retains its form under SW map \cite{gs}
\begin{eqnarray}
\hat S &\stackrel{\rm{SW \; map}}{=}&  \int d^{3}x \left[\frac{1}{2}\sqrt{{\mathrm {det}} \left( 1 + F \theta \right)}\left(\frac{1}{1 + F \theta }\frac{1}{1 + \theta F}\right)^{\mu \nu} \partial_{\mu} \phi\partial_{\nu} \phi + \frac{k}{2}\epsilon^{\mu \nu \lambda} A_{\mu} \partial_{\nu} A_{\lambda}\right] 
\label{caction}
\end{eqnarray}
In (\ref{caction}) we have used the matrix notation
\begin{eqnarray}
\left(AB\right)^{\mu \nu} = A^{\mu \lambda}B_{\lambda}{}^{\nu} 
\label{matnot}
\end{eqnarray}
Also $\left( 1 + F \theta \right)$ is to be interpreted as a mixed tensor in calculating the determinant. Note that the quartic term in the C--S action vanishes in the commutative equivalent version. The scalar field part of the action (\ref{caction}) can be written as an ordinary scalar field theory coupled with a gravitational field induced by the dynamical gauge field. However, the dynamics of the gauge field, being dictated by the Chern--Simons three-form, is unaffected by the induced gravity. In the next section we shall consider Maxwell coupling where this induced gravity should equally affect the gauge field dynamics also \cite{rbhsk}.

From (\ref{caction}) we readily observe that in the commutative limit ($\theta_{\mu \nu} \to 0$) the gauge field decouples, leading to the well known fact that there is no non-trivial gauge coupling of the neutral scalar field in the corresponding commutative field theory. Clearly, the action (\ref{caction}) is manifestly invariant under the corresponding commutative equivalent to the transformations (\ref{starg}), i.e. under
\begin{equation}
\delta_{\lambda} A_{\mu} =\partial_{\mu}\lambda, \qquad \delta_{\lambda} \phi = 0
\label{g}
\end{equation}

It is now straightforward to write down the equations of motion for  the scalar field $\phi$ and the gauge field $A_{\mu}$ from (\ref{caction}) respectively as
\begin{eqnarray}
\partial_{\alpha}\left\{\sqrt{{\mathrm {det}} \left( 1 + F \theta \right)}\left(\frac{1}{1 + F \theta }\frac{1}{1 + \theta F}\right)^{\alpha \nu} \partial_{\nu} \phi\right\} = 0
\label{eqmphi}
\end{eqnarray}
 and
\begin{eqnarray}
k\epsilon^{\alpha \nu \lambda} \partial_{\nu} A_{\lambda} = j^{\alpha}
\label{eqmA}
\end{eqnarray}
\begin{eqnarray}
{\rm where, \quad}  j^{\alpha} &=& \partial_{\xi}\left[\sqrt{{\mathrm {det}} \left( 1 + F \theta \right)} \left\{\frac{1}{4} \left(\theta\frac{1}{1 + F \theta } + \frac{1}{1 + \theta F} \theta\right)^{\alpha \xi} \left(\frac{1}{1 + F \theta }\frac{1}{1 + \theta F}\right)^{\mu \nu} \right.\right.\nonumber\\ 
&&+
\left(\frac{1}{1 + F \theta }\frac{1}{1 + \theta F} \theta \right)^{\mu \alpha} \left(\frac{1}{1 + \theta F}\right)^{\xi \nu} \nonumber \\
&&+ \left. \left. \left(\frac{1}{1 + F \theta}\right)^{\mu \alpha} \left(\theta \frac{1}{1 + F \theta }\frac{1}{1 + \theta F} \right)^{\xi \nu}
\right\}
\partial_{\mu} \phi\partial_{\nu} \phi\right]
\label{defJ}
\end{eqnarray}
Certain observations about the above equations are in order. In the commutative limit or (and) vanishing gauge field
\begin{equation}
\left(\frac{1}{1 + F \theta }\frac{1}{1 + \theta F}\right)^{\mu \nu} \to \eta^{\mu \nu} 
\label{mat}
\end{equation}
Thus the equation of motion for $\phi$ in (\ref{eqmphi}) reduces to the expected form $\partial_{\mu}\partial^{\mu}\phi = 0$ in these limits. Again going to the commutative limit we find that the gauge field equation becomes trivial, which is also a characteristic feature of C--S theories without any matter coupling. By direct computation, we get from (\ref{defJ}) 
\begin{equation}
\partial_{\alpha} j^{\alpha}= 0
\label{cont}
\end{equation}
This exhibits the consistency of (\ref{eqmA}). Naturally, $j^{\alpha}$ is interpreted as the matter current.

     At this point it can be noted that the usual approach of obtaining the commutative equivalent of (\ref{ncaction}) is to expand the star products and use separate maps for the gauge fields and matter fields in the form of perturbative expansions in the NC parameter $\theta$ \cite{all}. To the lowest order in $\theta$ the explicit forms of the SW maps are known as \cite{SW,bichl, vic}
\begin{eqnarray}
\hat \psi = \psi - \theta^{mj}A_{m}\partial_{j}\psi \qquad {\rm and} \qquad  
\hat A_{i} = A_{i} - \frac{1}{2}\theta^{mj}A_{m}\left(\partial_{j}A_{i} + F_{ji}\right)
\label{1stordmp}
\end{eqnarray}
Using these expressions and the star product (\ref{star}) to order $\theta$ in (\ref{ncaction}) we get
\begin{eqnarray}
\hat S \stackrel{\rm{SW \; map}}{=}  \int d^{3}x && \left[\left\{\frac{1}{2} \partial^{\mu} \phi \partial_{\mu} \phi - \theta ^{\alpha \beta} F^{\mu}{}_{\alpha} \partial_{\beta} \phi \partial_{\mu} \phi - \theta ^{\alpha \beta} A_{\alpha}\partial_{\mu} \partial_{\beta} \phi \partial^{\mu} \phi\right\} \right. \nonumber\\
&& + \left.\frac{k}{2}\epsilon^{\mu \nu \lambda} A_{\mu} \partial_{\nu} A_{\lambda}\right] 
\label{1stordac}
\end{eqnarray}
We can show explicitly that the first order approximation of (\ref{caction}) matches exactly with (\ref{1stordac}). Naturally, the equations of motion (\ref{eqmphi}) and (\ref{eqmA}) should agree upto the first order with those following from the conventional first order action (\ref{1stordac}). Expanding (\ref{eqmphi}) and (\ref{eqmA}) to first order in $\theta$ we get
\begin{eqnarray}
\partial_{\alpha}\left[\left\{1 + {\mathrm Tr} \left( F \theta \right)\right\} \partial^{\alpha} \phi - \left(F\theta + \theta F\right)^{\alpha \nu} \partial_{\nu} \phi\right] & = & 0 
\label{eqmphi1} 
\end{eqnarray}
and
\begin{eqnarray}
\partial_{\xi} [ \frac{1}{2} \theta^{\alpha \xi} \partial_{\mu} \phi \partial^{\mu} \phi + \theta^{\mu \alpha} \partial_{\mu} \phi \partial^{\xi} \phi + \theta^{\xi \mu} \partial^{\alpha} \phi \partial_{\mu} \phi ] & = & k\epsilon^{\alpha \nu \lambda} \partial_{\nu} A_{\lambda} 
\label{eqma1}
\end{eqnarray}
respectively. One can verify easily that the same equations follow as Euler--Lagrange equations from (\ref{1stordac}).

   We now turn to the construction of an energy momentum (EM) tensor of our model (\ref{caction}). The issue of energy momentum tensor for a noncommutative gauge theory involves many subtle points as evidenced in the literature \cite{rbyl}. It is thus instructive to address it from different approaches, which in the context of commutative models are known to lead to equivalent conclusions but may not be assumed {\it apriori} for NC gauge theories. Indeed, the commutative equivalent model offers an appropriate platform to discuss these aspects.

    We begin with the construction of the Noether EM tensor. Consider the infinitesimal space time translation $x^{\mu}  \to \left(x^{\mu}+ a^{\mu}\right)$ under which the fields $\phi$ and $A_{\mu}$ transform as 
\begin{eqnarray}
\delta \phi = a^{\mu}\partial_{\mu}\phi, \qquad \delta A_{\mu} = a^{\nu}\partial_{\nu} A_{\mu}
\label{trans}
\end{eqnarray}
From the invariance of the theory we get the following form of the EM tensor in the usual way, 
\begin{eqnarray}
\Theta^{c}_{\rho \sigma} &=& \sqrt{{\mathrm det} \left( 1 + F \theta \right)}\left[\left(\frac{1}{1 + F \theta }\frac{1}{1 + \theta F}\right)_{\rho}{}^{\nu}\partial_{\nu} \phi \partial_{\sigma} \phi\right.\nonumber\\
&&+ \left\{\frac{1}{4} \left(\theta\frac{1}{1 +F \theta} + \frac{1}{1 + \theta F} \theta\right)^{\alpha}{}_{\rho}\left(\frac{1}{1 + F \theta }\frac{1}{1 + \theta F}\right)^{\mu \nu}\right.\nonumber\\
&&+
\left(\frac{1}{1 + F \theta }\frac{1}{1 + \theta F} \theta \right)^{\mu \alpha} \left(\frac{1}{1 + \theta F}\right)_{\rho}{}^{\nu} \nonumber\\
&&+  \left. \left(\frac{1}{1 + F \theta}\right)^{\mu \alpha} \left(\theta \frac{1}{1 + F \theta }\frac{1}{1 + \theta F} \right)_{\rho}{}^{\nu}
\right\} \left( \partial_{\mu} \phi\partial_{\nu} \phi\right)\left( \partial_{\sigma} A_{\alpha} \right) \nonumber\\
&&- \left. \frac{1}{2} \eta_{\rho \sigma} \left(\frac{1}{1 + F \theta }\frac{1}{1 + \theta F}\right)^{\mu \nu} \partial_{\mu} \phi\partial_{\nu} \phi\right]\nonumber\\
&&- \frac{k}{2}\left(\epsilon_{\rho}{}^{\mu \alpha} A_{\mu} \partial_{\sigma} A_{\alpha} + \eta_{\rho \sigma} \epsilon^{\mu \nu \alpha} A_{\mu} \partial_{\nu} A_{\alpha}\right)
\label{noetheremt}
\end{eqnarray}

The Noether E--M tensor, although is useful to construct the generators of space time transformations, is neither gauge invariant nor symmetric. It can be improved to get a gauge invariant EM tensor using Belinfante's method. A better alternative is to consider a subsequent gauge transformation with the spatial translation (\ref{trans}) so that the gauge field transform covariantly and obtain an further improved EM tensor by Noether's method \cite{jackiw} using the modified transformation
\begin{eqnarray}
\delta A_{\mu} = a^{\nu}F_{\nu\mu}
\label{trans1}
\end{eqnarray}
which leads to 
\begin{eqnarray}
{\mathcal T}_{\rho \sigma} &=& \frac{\partial {\mathcal L}}{\partial\left(\partial^{\rho}\phi\right)}\partial_{\sigma}\phi +  \frac{\partial {\mathcal L}}{\partial\left(\partial^{\rho}A_{\alpha}\right)}F_{\sigma\alpha} - \eta_{\rho\sigma}{\mathcal L}\nonumber\\
&=&\sqrt{{\mathrm det} \left( 1 + F \theta \right)}\left[\left(\frac{1}{1 + F \theta }\frac{1}{1 + \theta F}\right)_{\rho}{}^{\nu}\partial_{\nu} \phi \partial_{\sigma} \phi\right.\nonumber\\
&&+ \left\{\frac{1}{4} \left(\theta\frac{1}{1 +F \theta} + \frac{1}{1 + \theta F} \theta\right)^{\alpha}{}_{\rho}\left(\frac{1}{1 + F \theta }\frac{1}{1 + \theta F}\right)^{\mu \nu}\right.\nonumber\\
&&+
\left(\frac{1}{1 + F \theta }\frac{1}{1 + \theta F} \theta \right)^{\mu \alpha} \left(\frac{1}{1 + \theta F}\right)_{\rho}{}^{\nu} \nonumber\\
&&+  \left. \left(\frac{1}{1 + F \theta}\right)^{\mu \alpha} \left(\theta \frac{1}{1 + F \theta }\frac{1}{1 + \theta F} \right)_{\rho}{}^{\nu}
\right\} \left( \partial_{\mu} \phi\partial_{\nu} \phi\right)\left(F_{\sigma\alpha} \right) \nonumber\\
&&- \left. \frac{1}{2} \eta_{\rho \sigma} \left(\frac{1}{1 + F \theta }\frac{1}{1 + \theta F}\right)^{\mu \nu} \partial_{\mu} \phi\partial_{\nu} \phi\right]\nonumber\\
&&- \frac{k}{2}\left(\epsilon_{\rho}{}^{\mu \alpha} A_{\mu} F_{\sigma\alpha} + \eta_{\rho \sigma} \epsilon^{\mu \nu \alpha} A_{\mu} \partial_{\nu} A_{\alpha}\right)
\label{jackiwEMT}
\end{eqnarray}
Apart from the contribution from the C--S part this expression is gauge invariant but not symmetric. In the commutative theories this part of the improved EM tensor becomes symmetric as well. This exception in context of NC gauge theories has already been mentioned and is due to the fact that Lorentz and classical conformal invariance are broken in such theories \cite{rbyl}.

We have observed that the canonical procedures do not lead to a satisfactory EM tensor. An alternative procedure is to vary the action (\ref{caction}) with respect to a background metric and finally keeping the metric flat \cite{nallu}. We thus extend the action (\ref{caction}) as
\begin{eqnarray}
S = \int d^{3}x \sqrt{- g} \mathcal{L}
\label{lag}
\end{eqnarray}
where $g = \mathrm{det} g_{\mu \nu}$ and $g_{\mu \nu}$ is the background metric. The pure C--S part of (\ref{caction}) is generally covariant irrespective of any metric. Thus the Lagrangean $\mathcal{L}$ in (\ref{lag}) is taken to be the Lagrangean of (\ref{caction}) without the C--S kinetic term. The EM tensor is obtained from 
\begin{eqnarray}
\Theta^{\left(s\right)}_{\alpha\beta} = 2 \frac{\partial \mathcal{L}}{\partial g^{\alpha \beta}} - {\mathcal{L}} g_{\alpha \beta}
\label{SEMT}
\end{eqnarray}
in the limit $g_{\mu \nu} \to \eta_{\mu \nu}$. Computing explicitly we found 
\begin{eqnarray}
\Theta^{\left(s\right)}_{\alpha\beta} = \frac{1}{2} \sqrt{{\mathrm {det}} \left( 1 + F \theta \right)} && \left[\frac{1}{2}\left(\theta F \frac{1}{1 + \theta F} + \frac{1}{1 + F \theta} F \theta\right)_{\alpha \beta} \left(\frac{1}{1 + F \theta }\frac{1}{1 + \theta F}\right)^{\mu \nu} \partial_{\mu} \phi \partial_{\nu} \phi\right. \nonumber\\
&&+ \left.\left(\frac{1}{1 + F \theta }\frac{1}{1 + \theta F}\right)_{\alpha}{}^{\nu} \partial_{\beta} \phi \partial_{\nu} \phi+ \left(\frac{1}{1 + F \theta }\frac{1}{1 + \theta F}\right)_{\beta}{}^{\nu}\partial_{\alpha} \phi \partial_{\nu} \phi\right] \nonumber \\
&& - \eta_{\alpha \beta} \mathcal{L} 
\label{symemt}
\end{eqnarray}
Note that by construction this EM tensor is both symmetric and gauge invariant. We can conclude that of the various expressions given above this form is the most satisfactory and can be identified as the physical EM tensor.

The equations (\ref{eqmphi}) and (\ref{eqmA}) are a set of coupled nonlinear equations. It will thus be instructive to investigate whether they admit any solitary wave solution by systematically looking for the Bogomolnyi bounds of the equations. To this end we require the energy functional which can appropriately be constructed from the physical EM tensor (\ref{symemt}). Note that until now our approach was completely general in the sense that the vanishing of time-space noncommutativity i.e. $\theta^{0i} = 0$ is not assumed. The issue of non zero time-space noncommutativity is an involved subject. It has been argued that $\theta^{0i} \neq 0$ spoils unitarity \cite{gomis, gaume} and causality \cite{sei} but there also exists counter examples \cite{bala, bala1, Daiy, bert2}. However, assuming $\theta^{0i} = 0$ is almost conventional in the study of NC solitons and in the context of odd dimensional theories it is always possible to do so. {\it A la} this tradition we now assume the same i.e. $\theta^{0i} = 0$. In this limit the energy functional becomes
\begin{eqnarray}
E = \int d^{2}x \Theta^{\left(s\right)}_{00} = \int  d^{2}x \frac{1}{2} \sqrt{{\mathrm{det}} \left( 1 + F \theta \right)} && \left\{  2\left(\frac{1}{1 + F \theta }\frac{1}{1 + \theta F}\right)_{0}{}^{\nu} \left( \partial_{0} \phi \partial_{\nu} \phi\right)\right.\nonumber\\ 
&& - \left.\left(\frac{1}{1 + F \theta }\frac{1}{1 + \theta F}\right)^{\mu \nu} \left( \partial_{\mu} \phi \partial_{\nu} \phi\right)\right\}
\label{statenrgy}
\end{eqnarray}
With the stated assumptions about NC tensor $\theta^{\mu \nu}$ the form of the matrices appearing in the above equation can be easily worked out. Explicitly, the matrix for $\left(\frac{1}{1 + F \theta }\frac{1}{1 + \theta F}\right)^{\mu \nu}$ can be written as
\begin{eqnarray}
\left(\begin{array}{ccc}
\left\{1 - \frac{\theta^{2}\left(E_{1}^{2} + E_{2}^{2}\right)}{\left(1 - \theta B\right)^{2}}\right\} & \frac{\theta E_{2}}{\left(1 - \theta B\right)^{2}}  & \frac{-\theta E_{1}}{\left(1 - \theta B\right)^{2}}  \\
\frac{\theta E_{2}}{\left(1 - \theta B\right)^{2}}  & \frac{-1}{\left(1 - \theta B\right)^{2}}  & 0 \\
\frac{-\theta E_{1}}{\left(1 - \theta B\right)^{2}} & 0 & \frac{-1}{\left(1 - \theta B\right)^{2}}\end{array}\right)
\label{thematrix}
\end{eqnarray}
Also ${\rm det}\left(1 + F\theta \right) = \left(1 - \theta B\right)^{2}$. While extracting the square root of this determinant one has to take positive value only. So for $\theta B < 1$, $\sqrt{{\rm det}\left(1 + F\theta \right)} = \left(1 - \theta B\right)$ whereas for $\theta B > 1 $ it is to be replaced by $\left(\theta B - 1 \right)$. The critical point $\theta B = 1$ is known to be a general feature of the NC models, the origin of which can be traced back to noncommutativity in planar quantum mechanics \cite{Nair, ban1}.

 To work out the static limit of the energy functional we first observe from (\ref{defJ}) that for $\theta_{0i} = 0$, $j^{0}$ vanishes in the static limit. This leads to vanishing $B$-field, as can be seen from (\ref{eqmA}), making the coupling trivial. The expression of the energy functional (\ref{statenrgy}) becomes
\begin{equation}
E = \int d^{2}x  \left\{ \left( \partial_{1} \phi \right)^{2} + \left( \partial_{2} \phi \right)^{2} \right\}\label{statenrgyexpli}
\end{equation}
Clearly, the energy functional is positive definite and trivially minimized and there is no non-trivial solutions. We thus observe that there is no BPS soliton of the model. Note that nontrivial soliton solutions has been found in NC adjoint scalar field theories \cite{poly, prnjp} with Maxwell coupling. However, these soliton solutions become singular in the $\theta \mapsto 0$ limit. Since our approach has a smooth commutative limit, based as it is on the SW map, such singular solutions (if any) are not accounted for in our model.
\subsection{Extension to a noncommutative Maxwell--Chern--Simons Theory}
\label{max}
\pagestyle{myheadings}
\markright{Extension to a noncommutative Maxwell--Chern--Simons Theory}
In this section we make a generalisation of the model (\ref{ncaction}) of section \ref{chern} in the sense that along with the NC C--S term we now also consider the NC Maxwell gauge fields. The present model thus reads
\begin{equation}
\hat S = \int d^{3}x\left[-\frac{1}{4 g^{2}}\hat F_{\mu\nu}\star \hat F^{\mu\nu} + \frac{1}{2} \left(\hat D_{\mu} \star \hat \phi \right)\star \left(\hat D^{\mu} \star \hat \phi\right) + \frac{k}{2}\epsilon^{\mu \nu \lambda}\left(\hat A_{\mu} \star \partial_{\nu}\hat A_{\lambda} - \frac{2i}{3}\hat A_{\mu} \star \hat A_{\nu} \star \hat A_{\lambda}\right)\right]
\label{ncactionM}
\end{equation}
 where symbols retain the same meanings as in the previous section. The dynamics of the gauge field is assumed to be governed by a combination of the Maxwell and the C--S term and thus more general than (\ref{model}). Note that we have not included any potential term in (\ref{ncactionM}). Our motivation is to find whether any non-trivial coupling is possible in the smooth $\theta \to 0$ sector. In the event of such coupling an appropriate potential term can be devised to saturate the BPS limits.

 The exact commutative equivalent of (\ref{ncactionM}) is obtained by using the closed-form S--W maps for $D_{\mu}\star\hat \phi(x)$ and $-\frac{1}{4 g^{2}}\hat F_{\mu\nu}\star \hat F^{\mu\nu}$ given in \cite{rbhsk, hsk}\footnote{Note that although the only existing exact S--W map are true for $F = constant$, here what we are using is an exact commutative equivalent form of the action (\ref{ncactionM}) as a whole and not the exact S--W maps for the individual fields. } whereas C--S action retains its form under SW map as usual \cite{gs}
\begin{eqnarray}
\hat S \stackrel{\rm{SW \; map}}{=}  \int d^{3}x \left[\sqrt{{\mathrm {det}} \left( 1 + F \theta \right)}\left\{\frac{1}{4 g^{2}}\left(\frac{1}{1+ F \theta}\right)^{\mu \alpha} F_{\alpha\beta}\left(\frac{1}{1+ F \theta}\right)^{\beta\nu} F_{\nu\mu}\right.\right.\nonumber\\
\qquad \qquad \qquad \qquad + \left.\left. \frac{1}{2}\left(\frac{1}{1 + F \theta }\frac{1}{1 + \theta F}\right)^{\mu \nu} \partial_{\mu} \phi\partial_{\nu} \phi\right\} + \frac{k}{2}\epsilon^{\mu \nu \lambda} A_{\mu} \partial_{\nu} A_{\lambda}\right]
\label{cactionM}
\end{eqnarray}
 Note that the invariance of the theory under $U(1)$ gauge transformation (\ref{g}) is manifest in (\ref{cactionM}). It is now straightforward to write down the equations of motion for the scalar field $\phi$ and the gauge field $A_{\mu}$ from (\ref{cactionM}) respectively as
\begin{eqnarray}
&& \partial_{\alpha}\left\{\sqrt{{\mathrm {det}} \left( 1 + F \theta \right)}\left(\frac{1}{1 + F \theta }\frac{1}{1 + \theta F}\right)^{\alpha \nu} \partial_{\nu} \phi\right\} = 0 
\label{eqmphiM}
\end{eqnarray}
and
\begin{eqnarray}
&& k \epsilon^{\alpha \nu \lambda} \partial_{\nu} A_{\lambda} - \frac{1}{4g^{2}}\partial_{\xi}\left[ \sqrt{{\mathrm {det}} \left( 1 + F \theta \right)}\left\{\frac{1}{2}\left(\theta\frac{1}{1+ F\theta}+\frac{1}{1+ \theta F}\theta\right)^{\alpha\xi}\left(\frac{1}{1+ F\theta} F \frac{1}{1+ F\theta} F\right) \right.\right.\nonumber \\
&& + \left.\left.2\left(\frac{1}{1+ F\theta} F \frac{1}{1+ F\theta}+ \frac{1}{1+ \theta F} F \frac{1}{1+ \theta F} \right)^{\alpha\xi}\right.\right.\nonumber \\ 
&& -\left.\left. 2\left(\theta \frac{1}{1+ F\theta} F \frac{1}{1+ F\theta} F \frac{1}{1+ F\theta}+ \frac{1}{1+ \theta F} F \frac{1}{1+ \theta F} F \frac{1}{1+ \theta F}\theta \right)^{\alpha\xi} \right\}\right] = j^{\alpha} 
\label{eqmAM}
\end{eqnarray} 
where $j^{\alpha}$ is given by 
\begin{eqnarray}
j^{\alpha} = \partial_{\xi} &&  \left[\sqrt{{\mathrm {det}} \left( 1 + F \theta \right)} \left\{\frac{1}{4} \left(\theta \frac{1}{1 + F \theta } + \frac{1}{1 + \theta F} \theta\right)^{\alpha \xi}\left(\frac{1}{1 + F \theta }\frac{1}{1 +\theta F}\right)^{\mu \nu} \right.\right.\nonumber\\
&& + \left(\frac{1}{1 + F \theta }\frac{1}{1 + \theta F} \theta \right)^{\mu \alpha} \left(\frac{1}{1 + \theta F}\right)^ {\xi \nu} 
\nonumber\\  &&
+ \left. \left. \left(\frac{1}{1 + F \theta}\right)^{\mu \alpha} \left(\theta \frac{1}{1 + F \theta }\frac{1}{1 + \theta F} \right)^{\xi \nu} \right\} \partial_{\mu} \phi\partial_{\nu} \phi\right] 
\label{defJM}
\end{eqnarray}
By direct computation from (\ref{defJM}) we get 
\begin{eqnarray}
\partial_{\alpha} j^{\alpha} = 0
\label{continuityM}
\end{eqnarray}
So $j^{\alpha}$ in (\ref{defJM}) can be consistently interpreted as the matter current. If we go to the limit $g \to \infty$, the Maxwell term is eliminated from the action. In this limit we can compare the equations of motion with the results derived in section \ref{chern}. There, we have discussed elaborately, the issues of commutative limit and first order approximations. The observations in this context apply equally for the present model.

 To investigate whether the set of coupled nonlinear equations (\ref{eqmphiM}) and (\ref{eqmAM}) admit any solitary wave solution in a systematic way we look for the Bogomolnyi bounds of these equations. So we again turn to the construction of the energy functional for which we require the physical energy momentum (EM) tensor corresponding to the model (\ref{cactionM}). We have already shown that the canonical (Noether) procedure or the extended method of Jackiw \cite{jackiw} do not give both symmetric and gauge invariant EM tensor. The best way is to find a symmetric and gauge-invariant EM tensor by varying the action (\ref{caction}) with respect to a background metric and finally keeping the metric flat \cite{nallu}. Following this prescription (\ref{SEMT}) the symmetric and gauge-invariant EM tensor is computed
\begin{eqnarray}
\Theta^{\left(s\right)}_{\rho\lambda} &=& \sqrt{{\mathrm {det}}\left( 1 + F \theta \right)}\left[\frac{1}{4g^{2}}\left(\frac{1}{1 + \theta F}F\theta\right)_{\rho\lambda}\left(\frac{1}{1 + \theta F} F \frac{1}{1 + \theta F}F\right) 
+ \frac{1}{g^{2}}\left(\frac{1}{1 + \theta F} F \frac{1}{1 + \theta F}F\right)_{\rho\lambda} \right.\nonumber\\
&& - \left.\frac{1}{4g^{2}}\left(\frac{1}{1 + \theta F} F \frac{1}{1 + \theta F}F\right)g_{\lambda\rho} + \frac{1}{2}\left(\frac{1}{1 + \theta F}F\theta\right)_{\rho\lambda}\left(\frac{1}{1 + F \theta }\frac{1}{1 + \theta F}\right)^{\mu\nu}\partial_{\mu}\phi \partial_{\nu}\phi \right.\nonumber\\
&& \left.+ \left(\frac{1}{1 + F \theta }\frac{1}{1 + \theta F}\right)_{\rho}{}^{\nu}\partial_{\lambda}\phi \partial_{\nu}\phi - \frac{1}{2}\left(\frac{1}{1 + F \theta }\frac{1}{1 + \theta F}\right)^{\mu\nu}\partial_{\mu}\phi \partial_{\nu}\phi g_{\rho\lambda}
\right]
\label{symemtM}
\end{eqnarray}
Naturally this EM tensor is physically significant and we can work out the energy density of the field system from the time-time component of this EM tensor. At this point we again make the assumption conventional in the study of NC solitons, i.e. vanishing time-space noncommutativity ($\theta^{0i} = 0$) since the calculations of non-perturbative solutions in NC $U(1)$ gauge theory reported in the literature assume noncommutativity only in the spatial directions. Going over to this limit, the energy functional becomes
\begin{eqnarray}
E & = & \int d^{3}x \Theta^{\left(s\right)}_{00} = \int d^{3}x \frac{1}{2} \sqrt{{\mathrm{det}} \left( 1 + F \theta \right)}\left\{-\frac{1}{2g^{2}}\left(\frac{1}{1 + F \theta } F \frac{1}{1 + F \theta }F\right) \right.\nonumber\\
&&+\left. \frac{2}{g^{2}}\left(\frac{1}{1 + F \theta } F \frac{1}{1 + F \theta }F\right)_{00}
+2\left(\frac{1}{1 + F \theta }\frac{1}{1 + \theta F}\right)_{0}{}^{\nu} \left( \partial_{0} \phi \partial_{\nu} \phi\right) \right.\nonumber\\
&&-\left.\left(\frac{1}{1 + F \theta }\frac{1}{1 + \theta F}\right)^{\mu \nu} \left( \partial_{\mu} \phi \partial_{\nu} \phi\right)\right\}
\label{statenrgyM}
\end{eqnarray}
The form of the matrices appearing in the above equation are worked out and in the static limit the energy functional becomes
\begin{equation}
E = \int d^{2}x \frac{\sqrt{{\mathrm{det}} \left( 1 + F \theta \right)}}{2\left(1- \theta B\right)^{2}}\left[\frac{1}{g^{2}}\left(E_{i}{}^{2} + B^{2}\right) + \left(\partial_{i}\phi\right)^{2}\right]
\label{statenergyM}
\end{equation}
Note that the square root term in the first factor is taken positive by convention\footnote{ Note that this square-root term can not be  written explicitly since depending on $\theta B < 1$ or $\theta B > 1$ one has to take $\sqrt{{\rm det}\left(1 + F\theta \right)} = \left(1 - \theta B\right)$ or $\left(\theta B - 1 \right)$, as we have mentioned in section \ref{chern}.}.
Evidently, from (\ref{statenergyM}), minimum configuration of the static energy functional corresponds to the trivial solution where $E_{i}$, $B$ and $\partial_{i}\phi$ vanishes everywhere. These solutions also trivially satisfy the equations of motion. Clearly in the static limit there is no coupling between the matter and the gauge field. Hence there is {\it no} BPS soliton in the model.

This result, along with that in section \ref{chern}, fills a gap in the existing literature regarding the possible B--P--S soliton solutions of the model and is consistent with the fact that there is no coupling of the matter and the gauge field in the corresponding commutative theory. Note however, that the solution space for the Bogomolony equations form a subspace of solutions of the equations of motion (\ref{eqmphiM}, \ref{eqmAM}) and our analysis does not rule out the existence of non-trivial non-B--P--S solutions of the equations of motion.
\section{The introduction of time-space noncommutativity:\\ $\left(1+1\right)$-dimensional deformed bosonized Schwinger model}
\label{ncsch}
\pagestyle{myheadings}
\markright{The introduction of time-space noncommutativity: $\left(1+1\right)$-dimensional deformed bosonized...}
In the previous sections we have briefly mentioned a contentious issue in the present scenario of NC field theories, i.e., the aspect of noncommutativity in the time-space sector. It was argued that introduction of space-time noncommutativity spoils unitarity \cite{gomis, gaume} or even causality \cite{sei}. Much attention has been devoted in recent times to circumvent these difficulties in formulating theories with $\theta^{0i} \neq 0$ \cite{bala, bala1, bert2, Daiy}. The $\left(1+1\right)$ dimensional field theoretic models are particularly important in this context because any noncommutative extension of such models essentially contains fuzziness in the time-space sector. Such $\left(1+1\right)$ dimensional field theories in the commutative space-time have long been recognized as the laboratory where important ideas can be tested in a simple setting. However, not much analysis of the corresponding NC theories is available in the literature. In the present sections we will therefore consider a $\left(1+1\right)$ dimensional Schwinger model \cite{SCH} on a NC setting. Apart from the specific NC aspect such studies are also motivated by the inherent interest of the model as $\left(1+1\right)$ dimensional electrodynamics.
\subsection{The ordinary bosonised (vector) Schwinger model}
\label{ncsch1}
\pagestyle{myheadings}
\markright{The ordinary bosonised (vector) Schwinger model}
Historically, the first two dimensional model was proposed  by Thirring \cite{TH} describing a pure fermionic current-current interaction. The interest increased considerably when Schwinger was able to obtain an exact solution of two dimensional electrodynamics with massless spinor \cite{SCH}. The model is defined by the Lagrangian density
\begin{equation}
{\cal L}_F = \bar\psi (i\partial\!\!\!\!/ -eA\!\!\!\!/)\psi - {1\over 4}F^{\mu\nu}F_{\mu\nu}
\label{1} 
\end{equation}
where the  Lorentz  indices  run  over  the  two  values  $0,1$ and the rest of the notation is standard. \footnote{Notice that the coupling constant $e$ has unit mass dimension in this  situation.}
 The Schwinger model, i.e., the theory of mass less fermion interacting with an Abelian gauge field in $(1+1)$ dimensional space time is an exactly solvable field theoretical model. It has been extensively studied over the years \cite{old1, old2, old3, old4,  old5, JR, rabin, GB, PM, ARP, kij, AR, dur, sad, san, siy} mainly due to the emergence of phenomena such as mass generation and confinement of fermions (quarks). 

In $(1+1)$ dimensions an exact mapping can be established between the bosonic and fermionic theories. The singularities of the Schwinger model can be accommodated by regularizing the fermionic current. An equivalent approach is to obtain the effective action by integrating out the fermions. Commonly it is known as bosonization of the Schwinger model. For extension to NC scenario we adopt the more economical bosonised version. Bosonizing the fermion field in (\ref{1}) we get the Lagrangian density ${\cal L}_{B}$ involving a scalar field $\phi$ instead of the Dirac field $\psi$
\begin{equation}
{\cal  L}_B ={1\over 2}\partial_\mu\phi\partial^\mu\phi
+\frac{e}{2}\epsilon_{\mu\nu}F^{\mu\nu}\phi + a e^2 A_\mu A^\mu +{\alpha\over 4}F_{\mu\nu}F^{\mu\nu}
\label{2sch}
\end{equation}
The first piece is the kinetic energy term for the  scalar field whereas the second one describes the interaction between the matter field and gauge field. The last two terms involve something new, {\it  viz.}, two undetermined parameters $a$ and $\alpha$.\footnote{Note that the usual kinetic energy term is absorbed within $\alpha$.} These are fallouts of the regularization process. To be more specific, if the left handed and the right handed component of $\psi$  are integrated out one by one the regularization of the determinant contains such parameters \cite{JR, PM, AR}. Setting $a$ to be zero and $\alpha$ to be $-1$ we get the bosonized version of the usual gauge invariant vector Schwinger model. We will use this version for the NC extension in the reminder of this section.
\subsection{The NC Schwinger Model}
\label{ncsch2}
\pagestyle{myheadings}
\markright{The NC Schwinger model}
The NC version of the bosonised action which we consider is 
\begin{equation}
\int d^2x {\cal  L}_{NCB} = \int d^2x \left[{1\over 2}\left(\hat{D}_\mu\star \hat\phi\right)
\star \left(\hat{D}^\mu\star\hat{\phi} \right)
+\frac{1}{2}e\epsilon^{\mu\nu}\hat{\phi}\star\hat{F}_{\mu\nu}
 - {1\over 4}\hat{F}_{\mu\nu} \star \hat{F}^{\mu\nu}\right].
\label{NCBS}
\end{equation}
where $\hat \phi$ is the NC adjoint scalar field and $\hat A_{\mu}$ is the $\star$-gauge field. We again adopt the Minkowski metric $\eta_{\mu \nu} = {\rm diag} \left( +,-\right)$. The covariant derivative $\hat D_{\mu}\star \hat \phi$ is already defined in (\ref{covder}) and
$\star$ denotes that the ordinary multiplication is replaced by the star multiplication defined by (\ref{star}).
The action (\ref{NCBS}) is invariant under the $\star$-gauge transformation (\ref{starg}).

The physics behind the NC theory (\ref{NCBS}) can be explored by several approaches which sometimes compliment each other \cite{hkl, poly, prnjp, pmas1, pmas2}. In this thesis we will particularly employ the Seiberg--Witten (SW) type transformations \cite{SW, bichl, vic} to construct a commutative equivalent model of the actual NC theory (\ref{NCBS}) in a perturbative framework. Note, however, that even if the fields and the coordinates in the commutative equivalent model are commuting it is not obvious that the usual Hamiltonian procedure could produce dynamics with respect to noncommuting time. This issue has been addressed by Dayi \cite{Daiy} where noncommutativity in time-space sector emerges from a theory with only spatial noncommutativity, due to a duality transformation. Specifically a Hamiltonian formulation was obtained with commutating time which was shown to be identical to order $\theta$ for both the original theory ( with noncommutativity in the spatial sector only ) and its dual containing space time noncommutativity. Following this we propose to carry out our analysis to first order in $\theta$ and assume the applicability of the usual Hamiltonian dynamics for the commutative equivalent model. Since our motivation is to investigate what new features emerge from the presence of noncommutativity in the Schwinger model, introduction of minimal noncommutativity will be sufficient.

Using the SW maps given by (\ref{1stordmp}) \cite{SW, bichl, vic} and the star product (\ref{star}) to the lowest order in $\theta$ in (\ref{NCBS}) we get
\begin{eqnarray}
\hat S &\stackrel{\rm{SW \; map}}{=}& \int d^{2}x \left[\left\{ 1 + \frac{1}{2} {\rm{Tr}}\left(F \theta\right)\right\}{\mathcal{L}}_{c} - \left(F \theta\right)_{\mu}{}^{\beta}\partial_{\beta}\phi \partial^{\mu}\phi - \frac{e}{2} \epsilon^{\mu \nu}\left(F \theta F\right)_{\mu \nu} \phi \right.\nonumber\\
&& \qquad\qquad \left. + \frac{1}{2}\left(F \theta F\right)^{\mu \nu}F_{\mu \nu}\right]
\label{1storderacS}
\end{eqnarray}
where ${\mathcal {L}}_{c}$ stands for the commutative Lagrangean (\ref{2sch}) with $\alpha = -1$ and $a = 0$
\begin{eqnarray}
{\mathcal {L}}_{c} =  \frac{1}{2}\partial_{\mu}\phi \partial^{\mu}\phi + \frac{e}{2}\epsilon_{\mu\nu}F^{\mu\nu}\phi - {1\over 4}F_{\mu\nu}F^{\mu\nu} 
\label{cL}
\end{eqnarray}
Note that we can write the action (\ref{1storderacS}) in a form which, modulo total derivative terms, does not contain second or higher order time derivatives. This happens because we are considering a perturbative calculation to first order in $\theta$. If we would calculate to second order or beyond, higher derivatives of time would appear in the Lagrangean from star product expansion which brings complication in the Hamiltonian formulation \cite{Git, EW}. 

 The Eular--Lagrange equations following from the action (\ref{1storderacS}) are
\begin{eqnarray}
&&\partial_{\xi}\left[\left\{ 1 + \frac{1}{2} {\rm{Tr}}\left(F \theta\right)\right\}\partial^{\xi} \phi - \left(F \theta + \theta F\right)^{\xi\mu}\partial_{\mu}\phi \right] \nonumber\\
&& \qquad \qquad - \frac{e}{2}\epsilon^{\mu \nu}\left[\left\{ 1 + \frac{1}{2} {\rm{Tr}}\left(F \theta\right)\right\}F_{\mu\nu} + \left(F \theta F\right)_{\mu \nu}\right] = 0 \label{eqmS1} \\
&&\partial_{\xi}\left[- \theta^{\xi\alpha}{\mathcal {L}}_{c} + \left\{ 1 + \frac{1}{2} {\rm{Tr}}\left(F \theta\right)\right\}\left(e \epsilon^{\xi\alpha}\phi - F^{\xi\alpha}\right) - \theta^{\alpha\mu}\partial_{\mu}\phi\partial^{\xi}\phi + \theta^{\xi\mu}\partial_{\mu}\phi\partial^{\alpha}\phi \right. \nonumber\\
&&\qquad \left.
- e\phi \left\{\epsilon^{\xi\mu}\left(\theta F\right)^{\alpha}{}_{\mu} - \epsilon^{\alpha\mu}\left(\theta F\right)^{\xi}{}_{\mu} \right\} + \left(F F \theta + \theta F F\right)^{\xi \alpha} + \left(F \theta F\right)^{\xi \alpha}
\right] = 0
\label{eqmS2} 
\end{eqnarray}
We work out the canonical momenta conjugate to $\phi$ and $A_{\alpha}$ respectively as
\begin{eqnarray}
\pi_{\phi} &=& \left\{ 1 + \frac{1}{2} {\rm{Tr}}\left(F \theta\right) - \left(F \theta + \theta F\right)^{00}\right\}\dot\phi - \left(F \theta + \theta F\right)^{0i}\partial_{i}\phi \label{momentaS1}
\end{eqnarray}
\begin{eqnarray}
\pi^{\alpha} &=& - \theta^{0\alpha}{\mathcal {L}}_{c} + \left\{ 1 + \frac{1}{2} {\rm{Tr}}\left(F \theta\right)\right\}\left(e \epsilon^{0\alpha}\phi - F^{0\alpha}\right) - \theta^{\alpha\mu}\partial_{\mu}\phi\partial^{0}\phi \nonumber\\
&& \qquad\qquad + \theta^{0\mu}\partial_{\mu}\phi\partial^{\alpha}\phi - e\phi \left[\epsilon^{0\mu}\left(\theta F\right)^{\alpha}{}_{\mu} - \epsilon^{\alpha\mu}\left(\theta F\right)^{0}{}_{\mu} \right] \nonumber\\
&& \qquad\qquad + \left(F F \theta + \theta F F\right)^{0 \mu} + \left(F \theta F\right)^{0 \mu}
\label{momentaS}
\end{eqnarray}
From (\ref{momentaS1}) and (\ref{momentaS}) we get after a few steps
\begin{eqnarray}
\pi_{\phi} = \dot{\phi} + \theta F_{01}\dot{\phi} \quad; \quad \pi^{0} = 0 \quad; \quad 
\pi^{1} = F_{01} +e\phi + \frac{\theta}{2}\left[{\dot{\phi}}^{2} - \left(\partial_{1}\phi\right)^{2} + 3 F_{01}^{2} \right]
\label{1stmomentaS}
\end{eqnarray}
 The commutative equivalent Hamiltonian follows as
\begin{eqnarray}
\int dx {\cal H}_{CE}  & = &  \int dx \left[{\cal H}_{CS} + \frac{\theta}{2}
\left\{\pi^{1}\left(\phi^{\prime}{}^{2} - \pi_{\phi}{}^{2}\right) + e \phi \left(\pi_{\phi}{}^{2} - \phi^{\prime}{}^{2}\right)\right.\right.\nonumber\\
 &&
\qquad \quad \left. \left.
+ 
e^{3} \phi^{3} - \left(\pi^{1}\right)^{3} + \frac{3}{2} e \phi \pi^{1}\left(\pi^{1} - e\phi\right)\right\}\right]
\end{eqnarray}
where ${\cal H}_{C}$ is given by the Hamiltonian of the ordinary commutative theory
\begin{equation}
{\cal  H}_{C}=  {1\over  2}\left[\pi_\phi^2+ (\pi^{1}){}^{2} + \phi'^2 + e^2\phi^2\right]
+\pi^1A_0'- e\pi^{1}\phi
\label{hcs}
\end{equation}

From  (\ref{momentaS}) we get one primary constraint. Conserving it in time a secondary constraint
emerges. They are given by 
\begin{eqnarray}
\pi^{0}  & \approx &  0, \quad {\rm (Primary)} 
\label{1stc}\\
\partial_{1}\pi^{1}  & \approx &   0, \quad {\rm (Secondary)}
\label{2ndc}
\end{eqnarray}
These constraints (\ref{1stc}, \ref{2ndc}) have vanishing Poission brackets with the Hamiltonian as well as between themselves. No new constraint, therefore, is obtained. It is interesting to note that the constraint structure is identical with the commutative Schwinger model. This structural similarity is remarkable because the gauge field in our commutative equivalent theory is the SW map of a NC gauge field belonging to the Groenewold--Moyal deformed $C^{\star}$ algebra.

 To proceed further we require to eliminate the gauge redundancy in the equations of motion (\ref{eqmS1}, \ref{eqmS2}) by invoking appropriate gauge fixing conditions. The structure of the constraints (\ref{1stc}, \ref{2ndc}) suggests the choice:
\begin{eqnarray}
A_{0} & = & 0 \label{gf1} \\
\partial_{1}A_{1} & = & 0 \label{gf2}
\end{eqnarray}
The constraints (\ref{1stc}, \ref{2ndc}) and (\ref{gf1}, \ref{gf2}) now become second class. We require to substitute all Poisson brackets by the respective Dirac brackets in order to impose the constraints strongly. A straightforward calculation shows that the brackets in the $\left\{\phi, \pi_{\phi}\right\}$ sector does not change, i.e. the corresponding Dirac brackets are the same as their Poisson brackets. The brackets in the other sector change but since these will not be required here, the explicit forms are not given. Once we strongly impose the constraints, we can integrate the relations (\ref{2ndc}) and (\ref{gf2}). Since the fields should vanish at spatial infinity, we get 
\begin{eqnarray}
\pi^{1}& = & 0 \label{gf1i} \\
A_{1} & = & 0 \label{gf2i}
\end{eqnarray}
We thus have the relations $\pi_{0} = \pi^{1} = 0$ and $A_{0} = A_{1} = 0$.
The Hamiltonian density in the reduced phase space then becomes
\begin{eqnarray}
{\cal H}_R = \left[ \frac{1}{2} \left(\pi_\phi^2  + \phi^{\prime}{}^{2} + \phi^{2}\right) + \frac{e\theta}{2}\phi\left(\pi_{\phi}^{2} -  \phi^{\prime 2} + e^{2}\phi^{2}\right)\right]
\label{HR}
\end{eqnarray}
The Reduced Hamiltonian (\ref{HR}) along with the Dirac brackets leads to the following equations of motion.
\begin{eqnarray}
\dot\phi  =  \left(1 + e \theta \phi\right) \pi_{\phi} \quad {\rm and} \quad 
\dot\pi_{\phi}  =  \phi^{\prime \prime} - e^{2}\phi - \frac{e\theta}{2} \left(\pi_{\phi}^{2} + \phi^{\prime}{}^{2} + 2 \phi \phi^{\prime \prime} + 3 e^{2}\phi^{2}\right) 
\label{EQ2}
\end{eqnarray}
The equations in (\ref{EQ2}) after a little algebra reduced to
\begin{equation}
\left( \Box + e^{2} \right)\phi  =  \frac{e\theta }{2} \left(\dot{\phi}^{2} - \phi^{\prime}{}^{2} - 5 e^{2}\phi^{2} \right)
\label{box}
\end{equation}
Equation (\ref{box}) is the relevant equation of motion obtained by removing the gauge arbitrariness of the theory. To zero order in $\theta$ it gives
\begin{equation}
(\Box+e^2)\phi=0
\label{spec}
\end{equation}
The equation (\ref{box}) looks complicated but since our theory is of the order $\theta$, in the right hand side we can substitute $\phi$ to $0$-order where it satisfies (\ref{spec}). The solution to (\ref{spec}) can easily be expanded in terms of the plane wave solutions
\begin{equation}
\phi= \int \frac{d\bar{p}}{\left(2\pi\right)}\frac{1}{\sqrt{2 p_{0}}}\left[a(\bar{p})e^{-ip^{0}x^{0} + i\bar{p}\bar{x}} + a^{\dag}(\bar{p})e^{ip^{0}x^{0}- i\bar{p}\bar{x}} \right] \\
\label{planewave}
\end{equation}
where  $x^{\mu} \equiv \left(x^{0}, \bar{x} \right)$, $p^{\mu} \equiv \left(p^{0}, \bar{p} \right)$ and $p_{0} = \sqrt{\bar{p}^{2} + e^{2}}$. Hence from (\ref{box}) we get 
\begin{eqnarray}
(\Box+e^2)\phi & = & j(x) \label{boxf} \\
{\rm where \qquad}
j(x) & = & \frac{e \theta}{2} \int \frac{d\bar{p} d\bar{q}}{\left(8\pi^2\right)}\frac{e^{i\left(\bar{p}+\bar{q}\right)\bar{x}} }{\sqrt{p^{0}q^{0}}} \left[ \left( - p_{0} q_{0} + \bar{p}\bar{q} - 5 e^{2}\right)\right. 
\nonumber\\ 
&&
 \left. \times\left\{a(\bar{p})a(\bar{q})e^{- i\left(p^{0}+q^{0}\right) x^{0}} \right. \right.
\left.\left.+ 
a^{\dagger}(-\bar{p})a^{\dagger}(-\bar{q})e^{ i\left(p^{0}+q^{0}\right) x^{0}} \right\}\right. \nonumber\\
&&\left.+ 
\left( p_{0} q_{0} - \bar{p}\bar{q} + 5e^{2}\right) \left\{a(\bar{p})a^{\dag}(- \bar{q})e^{- i\left(p^{0} - q^{0}\right) x^{0}}\right. \right. 
\left.\left.+ a^{\dagger}(-\bar{p})a(\bar{q})e^{ i\left(p^{0} - q^{0}\right) x^{0}}\right\}\right] \nonumber \\
\label{j}
\end{eqnarray}
We thus have a bosonic field $\phi$ interacting with a source $j(x)$. It is easy to recognise that the equations (\ref{boxf}) represents the Klien--Gordon (KG) theory with a classical source.

A remarkable observation is inherent in (\ref{boxf}). To understand the proper perspective we have to briefly review the results from the corresponding commutative theory. There we end up with (\ref{spec}) and interpret that the photon has acquired mass and the fermion has disappeared from the physical spectrum \cite{SCH, old1, old2, old3,old4, old5}. This means the fermion is confined. The introduction of noncommutativity changes the scenario in a fundamental way. The gauge boson again acquires mass but this time it is interacting with a background. The origin of this background interaction is the fuzziness of space-time. This is a physical effect carrying NC signature. In this context it may be mentioned that generation of interactions by casting non-interacting theories in NC coordinates have been observed in other contexts also \cite {ref1, ref2, ref3}. 

We observe that the present analysis reveals the presence of a background interaction term which is manifest only in the tiny length scale $\sim \sqrt \theta$. Therefore a natural curiosity arises about the strength of the interaction. In other words, we need at least a order of magnitude estimation of the time-space NC parameter involved in the interaction. Although there are several estimates of the spatial NC parameter \cite{carol, cst, mpr} as well as the momentum space NC parameters \cite{bert0, bert1, RB} in the literature, the same for the time-space NC parameters is lacking. In Appendix {\bf A} we present a quantum mechanical  analysis of the NC gravitational well problem to obtain an upper-bound estimation of the time-space NC parameter.

It is easy to formulate the present theory guided by (\ref{boxf}) as a quantum theory. Note that the NC parameter $\theta$ is a small number and we can treat the interaction term as a perturbation. The resulting $S$-matrix can easily be written down as 
\begin{eqnarray}
S \sim T\left\{\exp\left[ -i \int d^{2}x j(x)\phi(x)\right]\right\}
\label{s}
\end{eqnarray}
Since $j(x)$ is real $S$ is unitary. The theory in the reduced phase space can thus be formulated as a perturbative quantum field theory which is formally similar to the KG theory with a classical source. Consequently, the requirements of unitarity and causality are satisfied. These are good news in view of the presence of time-space noncommutativity and justifies our proposition based on \cite{Daiy} that usual Hamiltonian analysis is applicable in our commutative equivalent model.
\section{A short investigation of the noncommutative correction to General Relativity}
\label{ncgr}
\pagestyle{myheadings}
\markright{A short investigation of the noncommutative correction to General Relativity}
Likewise the other field theories, there have been attempts to fit General Relativity (GR) in an NC framework. Construction of a theory of NC gravity remains a topic of considerable current interest in the literature and various authors have approached the problem from different angles. In \cite{Chamseddine:2000si}, for example, a  deformation of Einstein's gravity was studied using a construction based on gauging the noncommutative SO(4,1) de Sitter group and the SW map \cite{SW} with subsequent contraction to ISO(3,1). Another construction of a noncommutative gravitational theory was proposed in \cite{Aschieri:2005yw}. Very recently noncommutative gravity has been connected with stringy perspective \cite{ag}. In all these works the leading order noncommutative effects appear in the second order in the NC parameter $\theta$.

 Again, the commutative equivalent approach has been used to analyse many gauge theories in the recent past \cite{all, carol, bcsgas, pmas1, pmas2, asarpm1}. Since gravity can be viewed effectively as a gauge theory the commutative equivalent approach seems to be a promising one. Indeed, a minimal theory of NC gravity \cite{cal} has been constructed recently based on this approach where the NC correction appears as a series expansion in the NC parameter. The leading order correction is reported to be linear in $\theta$ in this work. Thus, it seems that the result of \cite{cal} is in contradiction with others existing in the literature \cite{Chamseddine:2000si, Aschieri:2005yw, ag}. In this section we extend the commutative equivalent formulation of \cite{cal} to show that actually there is no such controversy. To begin with, we present a brief review of the results of \cite{cal}. This will be useful as the starting point of our calculation as well as to fix the notations.
\subsection{A review of the minimal formulation of noncommutative general relativity}
\label{ncgr1}
\pagestyle{myheadings}
\markright{A review of the minimal formulation of noncommutative general relativity}
The main problem of implementing GTR on NC platform is that the algebra (\ref{ncgometry}) is not invariant under general coordinate transformation.
However, we can identify a subclass of general coordinate transformations,
\begin{equation}
\hat x^{\mu \prime}=\hat x^{\mu}+\hat \xi^{\mu}(\hat x),
\label{c}
\end{equation}
which are compatible with the algebra given by (\ref{ncgometry}). This imposes a restriction on $\xi^{\mu}$
\begin{eqnarray}
\theta^{\mu \alpha} \hat \partial_{\alpha} \hat  \xi^\nu(\hat x) = \theta^{\nu \beta} \hat \partial_{\beta} \hat \xi^\mu(\hat x).
\end{eqnarray}
and the theory corresponds to the version of General Relativity based on volume-preserving diffeomorphism known as the unimodular theory of gravitation \cite{UNI}. Thus the symmetries of canonical noncommutative space time naturally lead to the noncommutative version of unimodular gravity \cite{cal}. With the symmetries preserved in this manner the extension of GTR to noncommutative perspective is done using the tetrad formalism and invoking the enveloping algebra method \cite{Jurco:2000ja}. The theory is then cast in the commutative equivalent form by the use of appropriate Seiberg--Witten (SW) maps. The final form of NC GR action obtained is
\begin{eqnarray}
S & = & \int  d^4 x \frac{1}{2 \kappa^2} \hat R(\hat x) \label{NCgaction} \\
{\rm with} \qquad \hat R & = & \hat R_{ab}{}^{ab} \label{R}
\end{eqnarray}
where $\hat R$ is the noncommutative version of the Ricci scalar and $\hat {R}_{ab}{}^{cd}$ are the components of the NC Riemann tensor appearing in
\begin{eqnarray}
\hat R_{ab}(\hat x) =\frac{1}{2}\hat R_{ab}^{\ \ cd}(\hat x)
\Sigma_{cd}, \label{Rab}
\end{eqnarray}
The latin indices refer to the vierbein and $\Sigma_{cd}$ are the
generators of the local Lorentz algebra $SO\left(3,1\right)$. $\hat R_{ab}(\hat x)$ can be expanded as \cite{cal}
\begin{eqnarray} 
\hat R_{ab} & = & R_{ab} + R^{(1)}_{ab} + { \cal  O}(\theta^2) \label{rab}\\
{\rm with} \qquad
R^{(1)}_{ab} & = & \frac{1}{2} \theta^{cd} \{R_{ac},R_{bd} \}
-\frac{1}{4} \theta^{cd} \{\omega_c, (\partial_d + D_d) R_{ab} \}
\label{R1ab}
\end{eqnarray}
where $\omega_a$ are given by the spin connection fields antisymmetric in $b$ and $c$
\begin{eqnarray}
 \omega_a(x)  &=&  \frac{1}{2}\omega_a^{\ bc}\Sigma_{bc},
\label{w} \\
 {\rm and \, \, the \, \, covariant \, \, derivative, \qquad}  D_a  &=&   \partial_{a} + \frac{i}{2} \omega_a^{\ bc} \Sigma_{bc}.
\label{covariant1}
\end{eqnarray}
Note that all quantities appearing on the rhs of (\ref{R1ab}) are ordinary commutative functions. Using the above expansion we can write from (\ref{NCgaction})
\begin{eqnarray}
S = \int  d^4 x \frac{1}{2 \kappa^2} \left (R(x)+
R^{(1)}(x)\right)+{\cal O}( \theta^2).
\end{eqnarray}
where $R(x)$ is the usual Ricci scalar and $R^{(1)}(x)$ is its first order correction. In the following section we will explicitly compute this correction term.
\subsection{Explicit computation of the first order correction term}
\label{ncgr2}
\pagestyle{myheadings}
\markright{Explicit computation of the first order correction term}
For the computation of the first order term we need an explicit form of $\Sigma_{cd}$, the generators of the local
Lorentz algebra $SO\left(3,1\right)$. This is given by \cite{weinberg} 
\begin{eqnarray}
\left[\Sigma_{cd}\right]^{a}{}_{b} = \delta^{a}{}_{c} \eta_{db} - \delta^{a}{}_{d} \eta_{cb}, 
\label{sigmat}
\end{eqnarray}
where $\eta_{ab} = \rm{diag}\left(-, +, +, +\right)$. Using these expressions we compute the first order correction to the Ricci tensor (\ref{R1ab}). Since by finding the $ab$-component of $\hat R_{ab}$ we get $\hat R$ on contraction, the corresponding first order correction to the NC action can now be calculated.

We  now proceed to compute the correction term $R^{(1)}\left(x\right)$. First note that $R^{(1)}\left(x\right) = R^{\left(1\right)}_{ab}{}^{ab}$. From (\ref{Rab}) and (\ref{rab}) this is equal to  $\left[R^{\left(1\right)}_{ab}\right]^{ab}$. We thus have to calculate the corresponding matrix element of the rhs of (\ref{R1ab}) and contract. For convenience we write the result as
$
\left[R^{\left(1\right)}_{ab}\right]^{ab}= T_{1} + T_{2},
$
where  $T_{1}$ and $T_{2}$ denote respectively the contributions coming from the first and the second terms of (\ref{R1ab}). After some computation we get
\begin{eqnarray}
T_{1} & = & 2 \theta^{cd} \left[R_{acg}{}^{a}R_{bd}{}^{bg} + R_{ac}{}^{b}{}_{g}R_{bd}{}^{ga}\right]
\label{T1}
\end{eqnarray}
To compute the second term we first compute the part containing the covariant derivative as
\begin{eqnarray}
\left[\left(\partial_{d}+ D_{d}\right)R_{ab}\right]^{e}{}_{f} = 2\partial_{d}R_{ab}{}^{e}{}_{f} + i \omega_{d}{}^{eg}R_{abgf}
\label{covariant}
\end{eqnarray}
Using this expression for the derivative term we compute $T_{2}$ 
\begin{eqnarray}
T_{2} & = & -\theta^{cd} \left[\frac{1}{2}\left(\omega_{c}{}^{aj}\partial_{d}R_{abj}{}^{b} - \omega_{c}{}^{aj}\partial_{d}R_{ba}{}^{b}{}_{j}\right) + \frac{i}{4}\left(\omega_{c}{}^{aj}\omega_{dj}{}^{g}R_{abg}{}^{b} - \omega_{d}{}^{aj}\omega_{c}{}^{bg}R_{abjg}\right) \right] \nonumber\\
\label{T2}
\end{eqnarray}
One can easily see that the first two terms cancel remembering the fact that the indices of the Riemann tensor refer to the tetrad and hence raised or lowered by $\eta_{ab}$. Collecting all the non-vanishing terms from (\ref{T1}, \ref{T2}) we get the correction term $R^{(1)}\left(x\right)$ as
\begin{eqnarray}
R^{(1)}\left(x\right)  &=&  \theta^{cd} \left[ 2\left(R_{acg}{}^{a}R_{bd}{}^{bg} + R_{ac}{}^{b}{}_{g}R_{bd}{}^{ga}\right) - \frac{i}{4}\left(\omega_{c}{}^{aj}\omega_{dj}{}^{g}R_{abg}{}^{b} - \omega_{d}{}^{aj}\omega_{c}{}^{bg}R_{abjg} \right)\right] \nonumber\\
\label{T}
\end{eqnarray}
This concludes our computation.
 Now one can show that all the terms of the above equation (\ref{T}) individually vanishes
 exploiting the antisymmetry of $\theta^{ab}$ and the various symmetry properties of the Riemann tensor and the spin connection fields.

Hence, the apparent contradiction in the results reported in the literature is resolved.
We have shown that this is due to the symmetries of the various factors involved in the correction term. It appears  that in the perturbative framework the order $\theta$ correction must vanish because the zero order theory carries full local Lorentz symmetry.

\section{Conclusion}
\pagestyle{myheadings}
\markright{Conclusion}
In this chapter we have dealt with some noncommutative (NC) gauge field theories. We exploit the gauge-equivalent behaviour of NC gauge symmetry with the ordinary gauge symmetry to analyse these NC models in a commutative space-time. We call this mode of analysis ``the commutative equivalent approach".

In section \ref{chernmax} we constructed an exact commutative equivalent description of NC Chern--Simons and NC Maxwell--Chern--Simons gauge field coupled adjoint matter system. Such theories have exhibited NC solitons which diverge as $\theta \to 0$. Our motivation to pursue such models is to investigate any possible existence or otherwise of NC soliton solutions which possess smooth commutative limits. To this end we computed different types of energy-momentum (EM) tensors for these theories and singled out the gauge invariant ones. These are identified with the physical EM tensors of the theories and the corresponding energy functionals are worked out. In the static limit these energy functionals are found to be positive definite and trivially minimised. Thus the possibility of existence of non-trivial B--P--S solitons with smooth commutative limits are ruled out.

In section \ref{ncsch} we formulated and analysed an NC theory of $\left(1+1\right)$-dimensional bosonised vector Schwinger model. The single space dimension ensured that the time-space noncommutativity was essential. As we have discussed, the presence of time-space noncommutativity may lead to problems such as non-unitarity \cite{gomis, gaume} and non-causal development \cite{sei}. However, since we restricted ourselves only to first-order in the NC parameter, problems regarding the non-unitarity or non-causality could be avoided. The analysis exhibits a massive boson interacting with a small background of the order of the NC parameter. In Appendix {\bf A} we present an analysis of a quantum mechanical model which led to an upper-bound estimation of such time-space NC parameter.

We also present a short calculation in section \ref{ncgr} to show that in a minimal theory of NC general relativity the first order correction to the Ricci scalar \cite{cal} obtained via S--W map actually vanishes. This settles an apparent contradiction of Calmet {\it et.al.}'s result \cite{cal} with others existing in the literature \cite{Chamseddine:2000si, Aschieri:2005yw, ag}.


\clearpage
\label{sec:ncq}
\chapter{Space-time symmetries in a non-relativistic noncommutative model}
\label{ncq}
\pagestyle{myheadings}
\markright{Space-time symmetries in a non-relativistic noncommutative model}
\section{Introduction}
\pagestyle{myheadings}
\markright{Introduction}
So far in this thesis we have either analysed or exploited the internal symmetries in the context of various classical field and string theoretic models. In this chapter, however, we shall investigate the space-time symmetries of a non-relativistic system in a background $U(1)_{\star}$ gauge field on a NC plane where the coordinates follow the algebra (\ref{ncgometry}). Since this is a NC field theory with a background gauge field we can again take a commutative equivalent approach to present the NC model in a commutative space-time. Once the effective commutative field theory is established we can go over to the quantum mechanical domain whenever necessary since first and second quantised formalisms are equivalent as far as non-relativistic models are concerned \cite{Nair}.
For the present we assume time-space noncommutativity is absent, i.e. $(\theta^{0i}=0)$ to avoid any possible nonunitarity or higher order time derivative terms in the action. It can be shown easily that this condition is Galilean invariant. The presence of a constant NC structure spoils Lorentz invariance manifestly. But in case of non-relativistic models this issue is not apparently clear \cite{hms, hms1}. It is therefore interesting to look for any possible violation in Galilean symmetry of our system. We shall therefore study the Galilean symmetry of the commutative equivalent model obtained by first order SW map \cite{SW, bichl, vic}.
\section{The commutative equivalent picture}
\pagestyle{myheadings}
\markright{The commutative equivalent picture}
We consider NC Schr\"{o}dinger field $\hat \psi$ coupled with $U(1)_{\star}$ background gauge field $\hat A_{\mu}(x)$ in the noncommutative plane, the corresponding $U(1)_{\star}$ gauge invariant action is given by 
\begin{eqnarray}
\hat S = \int d^{3}x \, \hat \psi^{\dag}\star(i\hat D_{0} + \frac{1}{2m}\hat D_{i}\star \hat D_{i})\star \hat \psi 
\label{2}
\end{eqnarray}
where the variables $\hat \psi$ is assumed to be Schwartzian \cite{szabo} and compose through the star product defined by equation (\ref{star}). The NC covariant derivative is in the fundamental representation, defined by $\hat D_{\mu}\star = \partial_{\mu} - i\hat A_{\mu}\star$. The equation of motion for the fundamental field $\hat \psi(x)$ is
\begin{eqnarray}
(i \hat D_{0} + \frac{1}{2m} \hat D_{i} \star \hat D_{i}) \star \hat \psi = 0 .
\label{16}
\end{eqnarray}
The $\star$--gauge invariant matter (probability) current density $\hat j_{\mu}$ following from equation (\ref{16}) is 
\begin{eqnarray}
\hat j_{0} = \hat \rho = \hat \psi^{\dag} \star \hat \psi, \qquad 
\hat j_{i} = \frac{1}{2mi} \left[\hat \psi^{\dag} \star \left(\hat D_{i} \star \hat \psi \right) - \left(\hat D_{i} \star \hat \psi \right)^{\dag}\star \hat \psi  \right]
\label{18}
\end{eqnarray}
 which satisfy the usual continuity equation $\partial_{t}\hat j_{0} + \partial_{i}\hat j_{i} = 0$; ($i=1,2$).


\noindent We derive a commutative equivalent description from the action (\ref{2}) by using the first order SW map \cite{SW, bichl, vic} given by the equations (\ref{1stordmp}). Note that at this point we impose the $\theta^{0i} = 0$ condition to do away with any noncommutativity in the time-space sector\footnote{Note that in Appendix {\bf A} we shall deal with a similar situation where NC Schr\"{o}dinger matter coupled to background Newtonian gravitational field will be analysed in a first quantised formalism. There, we shall specifically focus on the effect of time-space noncommutativity on the system and therefore deal with a $(\theta^{0i} \neq 0)$-type of NC geometry.}.
We find a manifest $U(1)$ gauge invariant commutative equivalent action, which when written in a hermitian form reads
\begin{eqnarray}
\hat{S} = \int d^{3}x && \left[\left( 1 - \frac{\theta B}{2}\right)(\frac{i}{2}\psi^{\dag}\stackrel{\leftrightarrow}{D}_{0}{\psi}) - \frac{1}{2m}\left( 1 + \frac{\theta B}{2}\right)\right.
\left.\times(D_{i}{\psi})^{\dag}(D_{i}{\psi}) + \frac{i}{4}\theta^{mj}({\psi^{\dag}}\stackrel{\leftrightarrow}{D_{j}}{\psi})F_{m0}\right.\nonumber \\  
&& \qquad 
+ \left.\frac{1}{8m} \theta^{mj}\left({\psi^{\dag}}\stackrel{\leftrightarrow}{D_{j}}{\psi}\right) \partial_{i} F_{mi} + ...\right]
\label{43A}
\end{eqnarray}
Here the dots indicate missing terms involving $\partial_{\mu}F_{\nu\lambda} $, which are not written down explicitly, as they play no role in the simplectic structure of the theory. Since this action is not in the canonical form, the field $\psi$ in second quantised formalism does not have a canonical structure for the equal time commutation relation between $\psi$ and $\psi^{\dag}$ as
\begin{eqnarray}
\left[\psi(x), \psi^{\dag}(y)\right] = \left(1 + \frac{\theta B}{2}\right) \delta^{2}(x - y)
\label{4300}
\end{eqnarray}
This non-standard form of the commutation relation indicates that $\psi$ cannot represent the basic field variable or the wave function in the corresponding first quantised formalism. We scale $\psi$ as 
\begin{eqnarray}
\psi \mapsto \tilde{\psi} = \sqrt{1 - \frac{\theta B}{2}}\psi
\label{4304}
\end{eqnarray}
so that the commutation relation (\ref{4300}) can be cast as
\begin{eqnarray}
\left[\tilde \psi(x), \tilde \psi^{\dag}(y)\right] =  \delta^{2}(x - y)
\label{4305}
\end{eqnarray}
and $\tilde \psi$ and $\tilde \psi^{\dag}$ can now be interpreted as annihilation and creation operators in second quantised formalism. So it becomes clear that it is $\tilde \psi$, rather than $\psi$, which corresponds to the basic field variable in the action. It is therefore desirable to re-express the action (\ref{43A}) in terms of $\tilde\psi$ and ensure that it is in the standard form in the first pair of terms. Clearly this can be done only for a constant $B$- field. Such a constant magnetic field can only arise from an appropriate background gauge field. We shall therefore consider a constant background for field strength tensor $F_{\mu\nu}$. 

\noindent In this case, the above action (\ref{43A}) can be cast in the form, 
\begin{eqnarray}
\hat{S} =&& \int d^{3}x \left[(\frac{i}{2}\tilde\psi^{\dag}\stackrel{\leftrightarrow}{D}_{0}{\tilde\psi}) - \frac{1}{2\tilde m}(D_{i}{\tilde\psi})^{\dag}(D_{i}{\tilde\psi})\right.
+ \left.\frac{i}{4}\theta^{mj}({\tilde\psi^{\dag}}\stackrel{\leftrightarrow}{D_{j}}{\tilde\psi})F_{m0}\right]
\label{4307}
\end{eqnarray}
where, $\tilde m = (1 - \theta B)m$ and $\tilde \psi$ 
 can now be regarded as rescaled mass and wave function respectively. Note that we did a similar mass and field rescaling in Appendix {\bf A}. We would like to mention that the expression for $\tilde m$ indicates that the external magnetic field $B$ has a critical value  $B_{c}= \frac{1}{\theta}$ as was also observed in \cite{SW}. Incidentally, this relation for $\tilde m$ was also obtained earlier in the literature \cite{horvs}. The equation of motion for the fundamental field $\tilde{\psi}$ from the action (\ref{4307}) is
\begin{eqnarray}
(iD_{0} + \frac{1}{2\tilde{m}} D_{i}D_{i} + \frac{i}{2}\theta^{mj}F_{m0}D_{j})
\tilde{\psi} \equiv K \tilde \psi = 0 
\label{48}
\end{eqnarray}
Now substituting (\ref{1stordmp}) in (\ref{18}), we obtain,
\begin{eqnarray}
\hat j_{0} &=& {\psi}^{\dag}{\psi} + \frac{i}{2}\theta^{mj} \left(D_{m}{\psi}\right)^{\dag}\left(D_{j}{\psi}\right)\nonumber\\
\hat j_{i} &=& \frac{1}{2\tilde{m}i}\left[\left\{{\psi}^{\dag}\left(D_{i}{\psi}\right) - {\mathrm{c.c}}
\right\}\right.
\left.+\frac{i}{2}\theta^{mj}\left\{\left(D_{m}\tilde{\psi}\right)^{\dag}\left(D_{i}D_{j}\tilde{\psi}\right) + {\mathrm{c.c}}\right\} \right]
\label{50}
\end{eqnarray}
where c.c denotes the complex conjugate part. Note that $ \hat j_{0}$ does not have the standard form due to the presence of the $\theta$-dependent term. However, when rewritten in terms of rescaled wave function $\tilde \psi$ (\ref{4304}), $\hat j_{0}$ can be brought to an almost canonical form, up to a $\left( 1 - \frac{\theta B}{2}\right)$ factor (assuming to be positive), by dropping a total divergence term
\begin{eqnarray}
\int d^{2}x \hat j_{0} = \left(1 - \frac{\theta B}{2} \right)\int d^{2}x \psi^{\dag}\psi = \int d^{2}x \tilde \psi^{\dag}\tilde \psi
\label{51}
\end{eqnarray}
Since the above expression corresponds to the total charge of a single particle, it can be set to unity ($\int d^2 x \tilde \psi^{\dag}\tilde \psi =1$). With this normalisation condition, it now becomes clear that $\tilde{\psi}^{\dag}\tilde{\psi}$ has to be identified as the probability density which is manifestly positive definite at all points. It immediately follows that the spatial components of $\hat j_{\mu}$, i.e $\hat j_{i}$ {\it must} correspond to the spatial component of the probability current, as $\hat j_{\mu}$ satisfies the continuity equation $\partial_{\mu}\hat j_{\mu}=0$.
\section{The Galilean generators and their algebra}
\pagestyle{myheadings}
\markright{The Galilean generators and their algebra}
In this section we shall construct all the Galilean symmetry generators for the model defined by the action (\ref{4307}).
 The canonically conjugate momenta corresponding to $\tilde{\psi}$ and $\tilde \psi^{\dag}$ are $\Pi_{\tilde{\psi}}  = \frac{i}{2} \tilde \psi^{\dag}$ and
$\Pi_{\tilde{\psi}^{\dag}}  = -\frac{i}{2} \tilde \psi$.
The  Hamiltonian computed by a Legendre transform reads, 
\begin{eqnarray}
H = \int d^{2}x \left[\frac{1}{2\tilde{m}}(D_{i}\tilde{\psi})^{\dag}(D_{i}\tilde{\psi}) - \frac{i}{4}\theta^{mj}(\tilde{\psi^{\dag}}\stackrel{\leftrightarrow}{D_{j}}\tilde{\psi})F_{m0}\right.
-\left. A_{0}(\tilde{\psi^{\dag}}\tilde{\psi})\right]
\label{G3}
\end{eqnarray}
 It is clear that the system contains second-class constraints which can be strongly implemented by Dirac scheme \cite{dir} to obtain the following bracket $\left\{\tilde \psi(x), \tilde \psi^{\dag}(y)\right\} = -i \delta^{2}(x - y)$
which in turn can be elevated to obtain the quantum commutator (\ref{4305}).
Now it can be easily checked using (\ref{4305}), the above Hamiltonian (\ref{G3}) generates appropriate time translation
 $\dot {\tilde\psi}(x) = \left\{\tilde\psi(x), H\right\}$.

We can now easily construct the generator of spatial translation and $SO(2)$ rotation by using Noether's theorem and the above mentioned constraints to get, 
\begin{eqnarray}
P_{i} =\int d^{2}x\frac{ i}{2} \tilde{\psi}^{\dag}(x)\stackrel{\leftrightarrow}{\partial_{i}}\tilde{\psi}(x), \qquad 
J = \frac{i}{2}\int d^{2}x \epsilon_{ij}x_{i} \tilde{\psi}^{\dag}(x)\stackrel{\leftrightarrow}{\partial_{j}}\tilde{\psi}(x)
\label{G5}
\end{eqnarray}
which generates appropriate translation and rotation:
\begin{equation}
\left\{\tilde\psi(x), P_{i}\right\}= \partial_{i} {\tilde\psi}(x); 
\left\{\tilde\psi(x),J\right\}=\epsilon_{ij} x_{i}\partial_{j} {\tilde\psi}(x)
\label{G7}
\end{equation}
Note that $J$ consists of only the orbital part of the angular momentum as in our simplistic treatment we have ignored the spin degree of freedom for the field $\tilde\psi$, so that it transforms as an $SO(2)$ scalar. Using the Dirac bracket between $\tilde\psi$ and $\tilde\psi^{\dag}$, one can verify the following algebra:
\begin{eqnarray}
\left\{P_{i},P_{j}\right\} = \left\{P_{i},H\right\} = \left\{J,H\right\} =0 ; \left\{P_{k},J\right\} = \epsilon_{kl}P_{l}
\label{G81}
\end{eqnarray} 
This shows that $P_{k}$ and $J$ form a closed $E(2)$ (Euclidian) algebra. 
Now coming to the boost, we shall try to analyse the system from first principle and shall check the covariance of (\ref{48}) under Galileo boost. For this, we essentially follow \cite{bcasm}. To that end, consider an infinitesimal Galileo boost along the $X$-direction, $t \mapsto t^{\prime} = t,\quad x^{1}{\mapsto} x^{1}{}^{\prime} = x^{1} - vt,\quad x^{2} \mapsto x^{2}{}^{\prime} = x^{2}$,
with an infinitesimal velocity parameter ``$v$''.
The canonical basis corresponding to unprimed and primed frames are thus given as $(\partial / \partial t, \partial / \partial x^{i} )$ and $(\partial / \partial t^{\prime}, \partial / \partial x^{i}{}^{\prime}) $, respectively. They are related as 
\begin{eqnarray}
\frac{\partial }{\partial t^{\prime}} = \frac{\partial }{\partial t} + v \frac{\partial }{\partial x^{1}},\qquad \frac{\partial }{\partial x^{i}{}^{\prime}} = \frac{\partial }{\partial x^{i}}
\label{G10}
\end{eqnarray}
Now, note that in the first quantised version $\tilde \psi$ is going to represent probability amplitude and  $\tilde\psi^{\dag}\tilde\psi$ represents the probability density. Hence in order that $\tilde\psi^{\dag}\tilde\psi$ remains invariant under Galileo boost ($\tilde\psi^{\prime \dag}(x^{\prime},t^{\prime})\tilde\psi^{\prime}(x^{\prime},t^{\prime}) =\tilde \psi^{\dag}(x,t)\tilde\psi(x,t)$), we expect $\tilde \psi$ to change at most by a phase factor. This motivates us to make the following ansatz :
\begin{eqnarray}
\tilde\psi\left(x,t\right) \mapsto \tilde\psi^{\prime}\left(x^{\prime},t^{\prime}\right) = e^{i v \eta(x,t)}\tilde \psi \left(x,t\right) \simeq \left(1 + i v \eta \left(x,t\right) \right)\tilde\psi \left(x,t\right)
\label{G11}
\end{eqnarray}
for the transformation of the field $\tilde \psi$ under infinitesimal Galileo boost ($v<< 1$). Further the gauge field $A_{\mu}(x)$ should transform like the basis $\frac{\partial }{\partial x^{\mu}}$ (\ref{G10}).
 This is because $A_{\mu}(x)$'s can be regarded as the components of the one-form $A(x) = A_{\mu}(x)dx^{\mu}$.
 It thus follows that
\begin{eqnarray}
A_{0}(x)\mapsto  A_{0}{}^{\prime}(x^{\prime}) =A_{0}(x)+ v A_{1}(x), \qquad 
A_{i}(x)\mapsto  A_{i}{}^{\prime}(x^{\prime}) = A_{i}(x)
\label{G12}
\end{eqnarray}
under Galileo boost.
 Now demanding that
 the equation of motion (\ref{48}) remains covariant implies that the following pair of equations $K\tilde\psi = 0 \quad; K^{'}\tilde\psi^{'} = 0 $
must hold in unprimed and primed frames respectively. Now making use of (\ref{G10},\ref{G11}) in the above equations and then using (\ref{G12}), we get the following condition involving $\eta$ :
\begin{eqnarray}
D_{1}\tilde \psi + i \partial_{0}\eta \tilde \psi = \left[-\frac{1}{\tilde m} \partial_{j} \eta - \frac{\theta}{2} \epsilon^{ij} F_{i1}\right] D_{j} \tilde \psi 
+ \left[- \frac{1}{2\tilde m} \nabla^{2}\eta - \frac{\theta}{2} \epsilon^{ij} F_{i0}\partial_{j} \eta\right]\tilde \psi
\label{G15}
\end{eqnarray}
Since we have considered the boost along the $x$-axis the variable $\eta$ occurring in the phase factor in (\ref{G11}) will not have any $x^{2}$ dependence ($\partial_{2}\eta = 0$). Also since we have taken the background electric field $F_{i0}=E_{i}$ to be constant, we have to consider here two independent possibilities : $\vec{E}$ along the direction of the boost and $\vec{E}$ perpendicular to the direction of the boost. Let us consider the former possibility first. Clearly in this case the term $\epsilon^{ij} E_{i}\partial_{j} \eta $ in the right hand side of (\ref{G15}) vanishes and the above equation becomes 
\begin{eqnarray}
D_{1}\tilde \psi + i \left(\partial_{0}\eta\right)\tilde \psi = \left[-\frac{1}{\tilde m} \partial_{1} \eta - \frac{\theta B}{2} \right] D_{1} \tilde \psi 
- \frac{1}{2\tilde m}\left(\partial_{1}^{2}\eta\right) \tilde \psi
\label{G16}
\end{eqnarray}
Equating the coefficients of $D_{1}\tilde \psi$ and $\psi$ from both sides we get the following conditions on $\eta$.
\begin{eqnarray}
\left[\frac{1}{\tilde m} \partial_{1} \eta + \frac{\theta B}{2} \right] = -1;\quad i\partial_{0}\eta = - \frac{1}{2\tilde m}\partial_{1}^{2}\eta
\label{G17}
\end{eqnarray}
It is now quite trivial to obtain the following time-independent ($\partial_{0}\eta = 0$) real solution for $\eta$ :
\begin{eqnarray}
\eta = - \tilde m \left(1 + \frac{\theta B}{2}\right) x^{1}
\label{G19}
\end{eqnarray}
This shows that boost in the direction of the electric field is a symmetry for the system. This is, however, not true when electric field is perpendicular to the direction of the boost. This can be easily seen by re-running the above analysis for this case, when one gets 
\begin{eqnarray}
\left[\frac{1}{\tilde m} \partial_{1} \eta + \frac{\theta B}{2} \right] = -1; i\partial_{0}\eta = - \frac{1}{2\tilde m}\partial_{1}^{2}\eta + \frac{\theta E}{2}\partial_{1}\eta
\label{G20}
\end{eqnarray} 
 Clearly this pair does not admit any real solution. In fact, the solution can just be read off as 
\begin{eqnarray}
\eta = - \tilde m \left(1 + \frac{\theta B}{2}\right) x^{1} + \frac{i}{2} \theta E \tilde m t
\label{G21}
\end{eqnarray}
This complex solution of $\eta$ implies the wave function (\ref{G11}) does not preserve its norm under this boost transformation as this transformation is no longer unitary. This demonstrates that the boost in the perpendicular direction of the applied electric field is not a symmetry of the system. Clearly this is a noncommutative effect as it involves the NC parameter $\theta$. This violation of boost symmetry rules out the possibility of ordinary Galilean symmetry, although, exotic Galilean symmetry obtained by \cite{hms, hms1} may not be impossible.
\section{Conclusion}
\pagestyle{myheadings}
\markright{Conclusion}
In this chapter we analysed the space-time symmetries of non-relativistic matter in presence of a background gauge field in NC space-time. We have obtained an effective $U(1)$ gauge invariant Schr\"{o}dinger action for the NC model by applying our commutative equivalent approach. A physically irrelevant wave-function and mass rescaling was employed to cast the effective theory as a standard Schr\"{o}dinger theory. The effect of non-commutativity on the mass parameter appears naturally in our analysis. Interestingly, we observed that the external magnetic field has to be static and uniform in order to get a canonical form of Schr\"{o}dinger equation upto $\theta$-corrected terms, so that a natural probabilistic interpretation could emerge. We investigated the Galilean symmetry of the model where the translation and the rotation generators are seen to form a closed Euclidean sub algebra of Galilean algebra. However, the boost is not found to be a symmetry of the system, even though the condition $\theta^{0i} = 0$ that we assumed throughout the calculation is Galilean invariant.
\clearpage
\label{sec:conclusion}
\chapter{Concluding remarks}
\label{conclusion}
\pagestyle{myheadings}
\markright{Concluding remarks}
The present thesis deals with different aspects of symmetries in connection with various field and string theoretic models. Mainly, the manifestations and impacts of gauge symmetries have been discussed. In Chapters \ref{string} and \ref{gravity} the focus was on different generally covariant models like string, membrane, $p-$brane and metric gravity. Here the main concern was to explore the gauge symmetries of the models including their interconnection with the diffeomorphism (dif.) invariances. In Chapter \ref{nc} the emphasis was on the impact of gauge symmetries on the field theories defined over noncommutative (NC) space-time. There is a correspondence between the NC gauge symmetry and ordinary gauge symmetry manifest in the existence of maps between the NC gauge fields and gauge parameter with ordinary gauge fields and gauge parameter \cite{SW}. Using the Seiberg--Witten (SW) maps we can construct a equivalent commutative theory describing the original NC gauge theory. Using this commutative equivalent approach we have analysed several NC gauge field theories of current interest.

Though gauge symmetry constitutes the focal theme of this thesis we have not, altogether, neglected space-time symmetries. In the penultimate chapter we have discussed Galilean symmetry in a planer NC model subject to background interaction.

The following is a brief summary of the methodology and the results of the works presented in the thesis.

In chapter \ref{string} we have applied a specific Hamiltonian algorithm \cite{brr} to work out the gauge generators of different stringy models in terms of the independent gauge parameters. These independent parameters are connected with the diffeomorphism (diff.) parameters through certain maps. These maps are used in turns to identify the gauge variation of different fields of the models with their corresponding diff. variations, thus establishing the underlying unity of reparametrisation and gauge invariance \cite{bms1, bms2}. The first class constraints of the N-G theory which enforces the gauge transformations of the model are used to develop a novel Lagrangean which serves as a pathway from Nambu-Goto (N--G) to Polyakov type actions. We, therefore, have named it the interpolating Lagrangean and have shown that this interpolating Lagrangean provides a suitable platform to analyse the internal symmetries of the theory. The process of generating the Polyakov metric from interpolating Lagrangean automatically decomposes the metric components in analogy with the Arnowitt-Deser-Misner (ADM) formulation of metric gravity. We have also analysed the second order metric gravity theory employing the same Hamiltonian algorithm \cite{brr} in chapter \ref{gravity}. A purely Dirac approach is employed to establish the interconnection of gauge symmetry and reparametrisation invariance in this connection. Although this is a known result this exercise serves to further demonstrate the generality of the Hamiltonian algorithm that we have applied earlier in the stringy context. The same can also be applied to other gravity theories and for that matter any theory with first-class constraints only.

Chapter \ref{nc} dealt with various NC gauge theories where the gauge-equivalent behavior of NC gauge symmetry with the ordinary gauge symmetry is exploited to work out the commutative equivalent versions of the original NC gauge field theories. While working in a commutative equivalent approach one mostly works in a perturbative framework where the star product is expanded and Seiberg--Witten (SW) maps are employed to certain order in the NC parameter. The resulting theories are therefore unsuitable to investigate non-perturbative effects {\it viz.} the existence of solitonic solutions. We have applied a closed form SW map to NC Chern--Simons (C--S) and Maxwell-coupled adjoint scalar field theories to explore such a possibility. These models exhibited NC solitons when treated in a operator approach but they were singular in the commutative limit $\theta \to 0$. On the other hand a smooth commutative limit is inherent in SW maps, a fact which properly justifies our approach to the problem from a commutative equivalent platform. We found that no B--P--S solitons are possible in our framework, however non-B--P--S solitons can not be ruled out. A possible extension in this direction may be to formulate the closed form SW maps for fundamental scalar fields and investigate the NC solitons generated by complex scalar field coupled to NC gauge fields.

We also employed the commutative equivalent approach in a perturbative framework to introduce and analyse a NC extension of the bosonised vector Schwinger model in $(1+1)-$dimensional NC space-time. The choice of the model essentially involved time-space noncommutativity and we presented argument supporting the fact that an order by order perturbative approach in such cases can avoid non-unitarity and non-causality. The analysis of the model exhibited a massive gauge boson interacting with a classical background proportional to the time-space NC parameter.

We ended Chapter \ref{nc} with a brief discussion of our work on a commutative equivalent version of NC general relativity (GR). Our work involves an extension of the work of Calmet {\it et.al.} on the commutative equivalent formulation of NC GR. We settled an apparent contradiction of the commutative equivalent theory with other NC theories of GR. The essential result of our computation is the fact that canonical NC corrections in GR starts from second order in the NC parameter. Subsequently this feature has been shown to follow from a more general NC structure of space-time \cite{liegrav}. This is a very relevant result in connection with the recent trend of investigating NC effects in the field of cosmology {\it viz.} inflation \cite{inf}, CMB \cite{cmbr} and Black Hole physics \cite{chichian, pmas4, bhp}.

Finally in Chapter \ref{ncq} we have analysed the space-time symmetries of a nonrelativistic NC model. This type of analysis is particularly relevant in the sense that although a canonical NC structure manifestly violates Lorentz symmetry it may not do so in case of Galilean symmetry. We considered a NC Schr\"{o}dinger field coupled to background gauge field in a commutative equivalent approach and constructed the Galilean generators of the model. Explicit computation revealed a violation of Galilean symmetry in the boost sector. Non-relativistic field theories are very important, mainly for their applications in the condensed matter physics. Our method of analysis in Chapter \ref{ncq} can be extended to such models in general.

The NC effects following from our various analysis depends on the size of the NC parameter. Particularly, a background interaction developed in our study of $\left(1+1\right)$-dimensional bosonised Schwinger model turned out to be proportional to the time-space NC parameter. Though there are several estimates of the spatial NC parameters in the literature \cite{cst, mpr, bert0, RB} similar estimates for the time-space NC parameter is lacking. In Appendix {\bf A} we present our work on finding an order of magnitude estimate of this time-space NC parameter.

\clearpage
\addcontentsline{toc}{chapter}{Appendix}
\appendix
\chapter{Time-space noncommutative parameter: An order of magnitude estimation}
\label{appendix}
\label{nctp}
\pagestyle{myheadings}
\markright{Appendix A. Time-space noncommutative parameter: An order of magnitude estimation}
Apart from studying the formal aspects of the NC geometry certain possible phenomenological consequences have also been investigated in the literature \cite{jabbari1, jabbari2, jabbari3, cs, rs1, rs2, rs3, rs4, rs5, rs6, rs7, rs8, rs9, rs10}. A part of this endeavor is spent in finding the order of the NC parameters appearing in the spatial sector of the NC geometry \cite{carol, cst, mpr, stern2} and in exploring its connection with observations\footnote{Note that the main result of \cite{cst} is off by a factor of 1000 and it has been pointed out in \cite{stern2}.}. A particular piece of the scenario is the quantum well problem which has emerged in recent GRANIT experiments by Nesvizhevsky {\it et al.} \cite{nes1, nes2, nes3}  who detected the quantum states of the neutrons trapped in earth's gravitational field. Their results have been used by Bertolami {\it et al.} \cite{bert0, bert1} and Banerjee {\it et al.} \cite{RB} to set an upper bound on the momentum space NC parameters. These works have been done on the level of quantum mechanics (QM) where noncommutativity is introduced among the phase space variables. Naturally noncommutativity in the time-space sector have been left out in this picture since in QM as such space and time could not be treated on an equal footing. We, on the other hand, would like to get an estimate particularly of the time-space NC parameter.

To introduce time-space noncommutativity in this quantum well scenario a second quantized theory is required. We propose to discuss the NC quantum well problem reducing it from a NC Schr\"{o}dinger field theory. This is a reasonable starting point since single particle quantum mechanics can be viewed as the one-particle sector of quantum field theory in the very weakly coupled limit where the field equations are essentially obeyed by the Schr\"{o}dinger wave function \cite{Nair, ban1, bcsgas}. This allows us to examine the effect of the whole sector of space time noncommutativity in an effective noncommutative quantum mechanical (NCQM) theory. We shall then study the effect of time-space NC (if any) on the energy spectrum of a cold neutron trapped in a gravitational quantum well by restricting ourselves to first order perturbative treatment\footnote{Note that we do not consider momentum space NC effects as have been done by \cite{bert0, bert1, RB}.}. Our motivation is to study the effect of noncommutativity on the level of quantum mechanics when time-space noncommutativity is accounted for. We briefly review the ordinary quantum well problem, its solutions and the experimental results \cite{nes1, nes2} to fix the notations and make the platform to discuss the NC extension.

\section{The Gravitational quantum well scenario}
\label{nctp1}
\pagestyle{myheadings}
\markright{Appendix A. The Gravitational quantum well scenario}
The gravitational well problem describes the quantum states of a particle with mass $\tilde{m}$ trapped in a gravitational potential well. The system's wave function can be separated into two parts, corresponding to each of the coordinates $x$ and $y$. Since the particle is free to move in $y$-direction its energy spectrum is continuous along $y$ and the corresponding wave function can be written as collection of plane waves 
\begin{eqnarray} \tilde{\psi}(y)=\int_{-\infty}^{+\infty}g(k)e^{i k y}dk~,
\label{Airy_5}
\end{eqnarray}
where the function $g(k)$ determines the group's shape in phase space. The analytical solution of the Schr\"{o}dinger equation in $x$ direction i.e. the eigenvalue equation, $H_{0}\tilde{\psi}_n=E_n\tilde{\psi}_n$, are well known \cite{Landau}. The eigenfunctions corresponding to $x$ can be expressed in terms of the Airy function $\phi(z)$,
\begin{eqnarray}
\psi_n(x)=A_n\phi(z)~,
\label{Airy_1}
\end{eqnarray}
with eigenvalues determined by the roots of the Airy function, $\alpha_n$, with $n=1,2\ldots$,
\begin{eqnarray}E_n=-\bigg({\tilde{m}g^2\hbar^2\over2}\bigg)^{1/3}\alpha_n~.
\label{Airy_2}
\end{eqnarray}
The dimensionless variable $z$ is related to the height $x$ by means of the following linear relation:
\begin{eqnarray}z=\bigg({2m^2g\over\hbar^2}\bigg)^{1/3}\bigg(x - {E_n\over\tilde{m}g}\bigg)~.
\label{Airy_3}
\end{eqnarray}
The normalization factor for the $n$-th eigenstate is given by:
\begin{eqnarray} A_n=\Bigg[\bigg({\hbar^2\over 2m^2g}\bigg)^{1\over3}\int_{\alpha_n}^{+\infty}dz\phi^2(z)\Bigg]^{-{1\over2}}~.
\label{Airy_4}
\end{eqnarray}
The wave function for a particle with energy $E_{n}$ oscillates below the classically allowed hight $x_{n} = \frac{E_{n}}{\tilde{m}g}$ and above $x_{n}$ it decays exponentially. This was realized experimentally by Nesvizhevsky {\it{et al}} \cite{nes1, nes2, nes3} where they observed the lowest quantum state of neutrons in the earth's gravitational field. The idea of the experiment was to let cold neutrons flow with a certain horizontal velocity ($6.5\:\mathrm{ms^{-1}}$) through a horizontal slit formed between a mirror below and an absorber above. The number of transmitted neutrons as a function of absorber hight is recorded and the classical dependence is observed to change into a stepwise quantum-mechanical dependence at a small absorber hight.
Their results and a comparison with the theoretical values are given below.
The experimentally found value of the classical height for the first quantum state is
\begin{eqnarray} 
x_1^{exp}&=&12.2\pm1.8(\mathrm{syst.})\pm0.7(\mathrm{stat.})\ (\mathrm{\mu m})
\label{Nesvizhevsky_1}
\end{eqnarray}
The corresponding theoretical value from Eq. (\ref{Airy_2}) and (\ref{Airy_3}) (for $\alpha_1=-2.338$) gives
\begin{eqnarray} 
x_1=13.7\:\mathrm{\mu m}
\label{Nesvizhevsky_1a}
\end{eqnarray}
This value is contained in the error bars and allow for maximum absolute shift of the first energy level with respect to the predicted values:
\begin{eqnarray} 
\Delta E_1^{exp}&=&6.55\times10^{-32}\ \mathrm{J}=0.41\ \mathrm{peV}
\label{Nesvizhesky_2}
\end{eqnarray}
The values of the constants taken in this calculations are as follows:
\begin{eqnarray} 
\hbar = 10.59\times 10^{-35}~\mathrm{Js}, \quad g = 9.81~ \mathrm{ms^{-2}} \quad
\tilde{m} = 167.32 \times 10^{-29}~\mathrm{Kg}.
\label{constants}
\end{eqnarray}

\section{The noncommutative extension : A field theoretic approach}
\label{nctp2}
\pagestyle{myheadings}
\markright{Appendix A. The noncommutative extension : A field theoretic approach}
Let us consider a NC field theory of a nonrelativistic system with a constant background gravitational interaction. Our starting point is to write down the action in the deformed phase space by replacing the ordinary product by the star product.The action for the system reads
\begin{eqnarray} 
\hat S = \int dx \hspace{0.5mm} dy\hspace{0.5mm} dt \hspace{1.0mm} \hat \psi^{\dag}\star \left[i \hbar \partial_{0} + \frac{{\hbar}^{2}}{2m} \partial_{i}\partial_{i} - m g\hat{x} \right] \star \hat \psi
\label{NCaction} 
\end{eqnarray}
Equation (\ref{NCaction}) describes a system in a vertical $x-y$ ($i = 1, 2$) plane where the external gravitational field is taken parallel to the $x$-direction. Under $\star $ composition the Moyal bracket between the coordinates is
\begin{eqnarray} 
\left[\hat x^{\mu},\hat x^{\nu}\right]_{\star} = i\Theta^{\mu\nu} = \left(\begin{array}{ccc}
0 & -\eta & -\eta^{\prime} \\
\eta & 0 & \theta\\
\eta^{\prime}  & -\theta & 0\\
\end{array}\right)
\label{NCpara}
\end{eqnarray}
where $\mu, \nu$ take the values $0, 1, 2$. Spatial noncommutativity is denoted by $\Theta^{12} = \theta$ and noncommutativity among time and the two spatial directions are denoted by the parameters $\Theta^{10} = \eta$ and $\Theta^{20} = \eta^{\prime}$.
Expanding the $\star$-product to first order in the NC parameters (\ref{NCpara}) we get
\begin{eqnarray} 
\hat S = \int  dx \hspace{0.5mm} dy\hspace{0.5mm} dt \hspace{1.0mm} \psi^{\dag}\hspace{-4.0mm}  && \left[i \hbar \left( 1  - \frac{1}{2\hbar} m g \eta \right) \partial_{t}  +  \frac{{\hbar}^{2}}{2m} \partial_{i}{}^{2} - m g x  - \frac{i}{2} m g \theta \partial_{y} \right] \psi
\label{Caction} 
\end{eqnarray} 
where everything is in terms of commutative variables and NC effect is manifest by the presence of $\theta$ and $\eta$ terms. Clearly the standard form of the kinetic term of Schr\"{o}dinger action is deformed due to time-space noncommutativity. We rescale the mass and the field variable by
\begin{eqnarray}
\psi \mapsto \tilde{\psi} = \sqrt{\left(1 - \frac{\eta}{2 \hbar} m g\right)} \psi, \qquad
\tilde m = \left(1 - \frac{\eta}{2 \hbar} m g \right) m.
\label{scal1}
\end{eqnarray}
which gives conventionally normalized kinetic term. Such physically irrelevant rescalings of fields have been done earlier \cite{bcsgas, carol}. We can interpret $\tilde {m}$ as the observable mass. A similar charge rescaling of NC origin in context of NC QED was shown in \cite{carol}. It becomes clear that it is $\tilde \psi$, rather than $ \psi$, which corresponds to the  basic field variable in the action (\ref{Caction}) once we re-express it in terms of $\tilde\psi$ and see that it is in the standard form in the first pair of terms.
\begin{eqnarray} 
\hat S = \int  dx \hspace{0.5mm} dy\hspace{0.5mm} dt \hspace{1.0mm} \tilde{\psi}^{\dag}\hspace{-4.0mm}  && \left[i \hbar \partial_{t}  +  \frac{{\hbar}^{2}}{2\tilde{m}} \partial_{i}{}^{2} - \tilde {m}\left( 1  +  \frac{\tilde{m} g \eta}{\hbar} \right)  g x   - \frac{i}{2} \tilde{m} g \theta \partial_{y} \right] \tilde{\psi}
\label{Caction1} 
\end{eqnarray} 
 The last term in (\ref{Caction1}) can be absorbed in the $\partial_{y}{}^{2}$ by rewriting
\begin{eqnarray}
\partial_{y} = \left(\partial_{y} -  \frac{i \theta}{2 \hbar^{2}} \tilde {m}{}^{2} g \right) .
\label{scal3}
\end{eqnarray}
and the final effective NC Schr\"{o}dinger action reads
\begin{eqnarray} 
\hat S = \int  dx \hspace{0.5mm} dy\hspace{0.5mm} dt \hspace{1.0mm} \tilde{\psi}^{\dag}\hspace{-4.0mm}  && \left[i \hbar \partial_{t}  +  \frac{{\hbar}^{2}}{2\tilde{m}} \left(\partial_{x}{}^{2} + \partial_{y}{}^{2}\right)- \tilde {m}g x  -  \eta \left(\frac{\tilde{m}{}^{2} g^{2}}{\hbar}\right)x \right] \tilde{\psi}
\label{Caction2} 
\end{eqnarray} 
The Lagrange equation of motion for the fundamental field $\tilde {\psi}(x)$ is
\begin{eqnarray} 
\left[i \hbar \partial_{t}  +  \frac{{\hbar}^{2}}{2\tilde{m}} \left(\partial_{x}{}^{2} + \partial_{y}{}^{2}\right) - \tilde {m}g x  -  \eta \left(\frac{\tilde{m}{}^{2} g^{2}}{\hbar}\right) x \right] \tilde{\psi} = 0
\label{eqm} 
\end{eqnarray} 
Note that owing to the field and mass redefinition (\ref{scal1}) everything but the last term in (\ref{eqm}) takes the form of standard Schr\"{o}dinger field equation.

\section{The first quantised picture and the noncommutative energy spectrum}
\label{nctp3}
\pagestyle{myheadings}
\markright{Appendix A. The first quantised picture and the noncommutative energy spectrum}
Dealing with the second quantized formalism where $\tilde{\psi}$ was the basic field variable of the theory we found out that the only nontrivial change in the Schr\"{o}dinger equation shows up only in the direction of the external gravitational field ${\bf{g}} = -\rm{ g {\bf {e_{x}}}}$. This result is in conformity with \cite{bert0, bert1, RB} where along with spatial noncommutativity, momentum space noncommutativity has been included as well and it was shown that it is the latter that shows up in first order computations.
However their treatment essentially left a gap in the analysis which we fill in here. Since first and second quantized formalisms are equivalent as far as Galilean systems are concerned, we carry out a equivalent NC quantum mechanical analysis in the first quantized formalism.

The Schr\"{o}dinger field equation (\ref{eqm}) will now be treated as the quantum mechanical equation of motion and the field variable $\tilde{\psi}$ will be interpreted as wave function. This is a quick and simple but standard procedure to reduce the field theoretic setup to one-particle quantum mechanics as has been illustrated in \cite{Nair} for a general external potential. 

 We begin by checking that $\tilde {\psi}$ does have an interpretation of probability amplitude and satisfies the continuity equation 
\begin{eqnarray}
\partial_{0} j_{0} + \partial_{i} j_{i} = 0 ; \qquad (i=1,2) 
\label{continuity}
\end{eqnarray}
with the usual expressions of probability density $j_{0} $ and probability current $j_{i}$ in terms of $\tilde {\psi}$. From equation (\ref{eqm}) we easily read off the Hamiltonian as
\begin{eqnarray} 
H = H_{0} + H_{1} = \frac{1}{2\tilde{m}} \left(p_{x}{}^{2} + p_{y}{}^{2}\right) + \tilde {m}g x  +  \eta \frac{\tilde{m}{}^{2} g^{2}}{\hbar} x 
\label{H} 
\end{eqnarray} 
Note that the the NC effect in the ordinary part $H_{0}$ is hidden in the mass and field redefinition (\ref{scal1}). Such rescalings are only viable in a region of space time where variation of the external field is negligible. Since the results derived here are to be compared with the outcome of a laboratory-based experiment we can safely assume a constant external gravitational field throughout.

Before proceeding with the Hamiltonian (\ref{H}) note that even if the variables are commuting it is not obvious that the usual Hamiltonian procedure could produce dynamics with respect to noncommuting time. In section \ref{ncsch2} we have discussed this issue and presented favourable arguments 
which has been shown to lead to reasonable outcome there \cite{asarpm1}. Following the same argument we propose to assume the applicability of the usual Hamiltonian dynamics for the present model and carry out our analysis to first order in $\eta$ . Since we expect the time-space NC parameter is rather small at the quantum mechanical level, the last term in equation-(\ref{H}) represents a perturbation $H_{1}$ in the usual gravitational quantum well scenario described by $H_{0}$.
\section{Upper-bound estimation of the time-space noncommutative parameter}
\label{nctp4}
\pagestyle{myheadings}
\markright{Appendix A. Upper-bound estimation of the time-space noncommutative parameter}
The perturbative potential is given by
$H_{1} = \eta\left(\frac{{\tilde{m}}^{2}{g}^{2}}{\hbar}  \right)x$.
Since it is a direct manifestation of time-space noncommutativity it enables us to work out a upper bound for the corresponding NC parameter. Following the prescription of \cite{bert0} we can demand that the correction due to (\ref{H}) in the energy spectrum should be smaller or equal to the maximum energy shift allowed by the experiment \cite{nes1}. We work out the theoretical value of the energy shift in three independent ways.

\noindent {\bf {Numerical Method:}} \quad 
First we take the numerical approach and calculate the leading order energy shift of the first quantum state. It is just the expectation value of the perturbation potential, given by
\begin{eqnarray}
\Delta E_{1}&=&\eta \frac{\tilde{m}^{2}g^{2}}{\hbar}
\int_0^{+\infty} dx~\tilde{\psi}_{1}^{*}(x)~x~\tilde{\psi}_{1}(x)
= \eta \frac{\tilde{m}^{2} g^{2}}{\hbar} \Bigg[\bigg({2\tilde{m}{}^2g\over\hbar^2}\bigg)^{-\frac{2}{3}}A_{1}^2I_{1}+ {E_{1}\over \tilde{m}g}\Bigg]~
\label{energy}
\end{eqnarray}
where $I_{1}$ is defined as
$I_{1}\equiv\int_{\alpha_{1}}^{+\infty}dz\phi(z)z\phi(z)~$.
The values of the first unperturbed energy level is determined from (\ref{Airy_2}) with $\alpha_{1} = -2.338$
\begin{eqnarray} 
E_{1}  =  2.259 \times 10^{-31} \ \mathrm{(J)}  = 1.407 \ \mathrm{(peV)}
\label{energy_1}
\end{eqnarray}
The normalization factor $A_{1}$ is calculated from (\ref{Airy_4}). The integrals in (\ref{Airy_4}) and $I_{1}$ were numerically determined for the first energy level, which give
$A_{1}  =  588.109\ ,\ I_{1}  =  -0.383213. $
Using these values the first order correction in the energy level $\Delta E_{1}$ is given by
\begin{eqnarray} 
\Delta E_{1} = 2.316\times10^{-23} \eta\quad\mathrm{(J)~,}
\label{energy_corrections}
\end{eqnarray}
Comparing with the experimentally determined value of the energy level from (\ref{Nesvizhesky_2}) we found the bound on the time--space NC parameter is
\begin{eqnarray} 
|\eta| & \lesssim & 2.83\times 10^{-9}\ \mathrm{m^{2}}
\label{eta_bounds_1}
\end{eqnarray}
\noindent {\bf {WKB Method:}} \quad
Alternatively, we analyze the energy spectrum using a quasiclassical approximation. The potential term in the unperturbed Hamiltonian $H_{0}$ in (\ref{H}) is linear, hence a simple WKB method suffices. The first energy level is given by the Bohr--Sommerfeld formula
\begin{eqnarray} 
E_{1} & = & \left(\frac{9m}{8}[\pi\hbar g(1-\frac{1}{4})]^2\right)^{\frac{1}{3}}
\label{WKB_energy_1} = \alpha_{1} g^{\frac{2}{3}} \ ; \ n=1, \ 2, \ 3...
\label{WKB_energy_2}
\end{eqnarray}
with $\alpha_{1}=\left(\frac{9m}{8}[\pi\hbar(1-\frac{1}{4})]^2\right)^{\frac{1}{3}}$. This approximation gives nearly exact value for the first energy level,
\begin{eqnarray}
E_{1}  = 2.23\times 10^{-31} \ \mathrm{(J)}  = 1.392 \ \mathrm{(peV)}
\label{WKB_energy_3}
\end{eqnarray}
as compared to equation (\ref{energy_1}). Since the perturbation term $H_{1}$ in (\ref{H}) is also linear in $x$ we can combine it with the potential term and rewrite the potential term as
\begin{eqnarray}
V(x) = \tilde{m} g^{\prime} x = \tilde{m} g \left(1 - \frac{\eta \tilde{m}}{\hbar} \right) x 
\label{WKB_energy_4}
\end{eqnarray}
Now using the modified acceleration $g^{\prime}$ from (\ref{WKB_energy_4}) in (\ref{WKB_energy_2}) the approximate shift in the energy value is obtained by first order expansion in $\eta$ as
\begin{eqnarray}
E_{1} + \Delta E_{1} 
=  \alpha_{1} g^{\prime}{}^{\frac{2}{3}}
= \alpha_{1} g^{\frac{2}{3}} \left(1 - \frac{\eta \tilde{m} g}{\hbar}\right)^{\frac{2}{3}}  =
\alpha_{1} g^{\frac{2}{3}} \left(1 - \frac{2 \eta \tilde{m} g}{3\hbar}\right) =
E_{1} - \eta \left(\frac{2 E_{1} \tilde{m} g}{3\hbar}\right)
\label{WKB_energy_5}
\end{eqnarray}
Using the values of $\tilde{m}, g, \hbar\  \mathrm{and}\ E_{1}$ from (\ref{constants}) and (\ref{WKB_energy_3}) we calculate the energy shift $\Delta E_{1}$
\begin{eqnarray}
\Delta E_{1}  = 2.304\times 10^{-23}\eta \ \mathrm{(J)}
\label{WKB_energy_6}
\end{eqnarray}
Again this is comparable with (\ref{energy_corrections}). So we get nearly the same upper bound on the time-space NC parameter as in (\ref{eta_bounds_1}) by comparison with the experimental value (\ref{Nesvizhesky_2})
\begin{eqnarray} 
|\eta| & \lesssim & 2.843\times 10^{-9}\ \mathrm{m^{2}}
\label{eta_bounds_2}
\end{eqnarray}
\noindent {\bf {Virial Theorem Method:}} \quad
Yet another simple analytical approach to calculate the energy shift $\Delta E_{1}$ is to use the virial theorem \cite{brau} which implies $\langle T \rangle =\frac{1}{2}\langle V \rangle$ where $T$ and $V$ are kinetic and potential energies, respectively. Hence total energy is given by $ E = \frac{3}{2}\langle V \rangle$. The gravitational potential is $V = \tilde{m} g x$, which gives
\begin{eqnarray}
\langle x\rangle=\frac{2E}{3\tilde{m}g}
\label{virial_1}
\end{eqnarray}
Now the perturbation term is
$H_{1} = \eta\left(\frac{{\tilde{m}}^{2}{g}^{2}}{\hbar}  \right)\langle x \rangle$.

\noindent Here using (\ref{virial_1}) we find the energy shift in the first energy level as
\begin{eqnarray}
\Delta E_{1} = - \eta \left(\frac{2 E_{1} \tilde{m} g}{3\hbar}\right)
\label{virial_3}
\end{eqnarray}
which reproduces the same expression for $\Delta E_{1}$ as derived in (\ref{WKB_energy_5}). Hence the upper bound on $\eta$ using the virial theoram method is exactly same as in (\ref{eta_bounds_2}).
\section{Comparison with the existing results}
\label{nctp5}
\pagestyle{myheadings}
\markright{Appendix A. Comparison with the existing results}
It will be instructive to enquire whether our order of magnitude estimation for the time space NC parameter is in conformity with the estimates of other NC parameters reported earlier \cite{bert0, bert1, RB}. In \cite{bert0} the upper bound on the fundamental momentum scale was calculated to be 
\begin{eqnarray}
\Delta p \lesssim  4.82 \times 10^{-31}\ \mathrm{kg\ m\ s^{-1}} \qquad \qquad \qquad \qquad \qquad
\label{bert_scale1}\\
{\rm Since} \quad E \approx \frac{p_{y}{}^{2}}{2\tilde{m}}, \quad {\rm so} \quad \Delta E \approx \frac{p_{y}}{\tilde{m}}\ \Delta p_{y} = v_{y} \Delta p_{y} \lesssim 31.33 \times 10^{-31} {\mathrm kg\ m^{2}\ s^{-2}}
\label{bert_scale2}
\end{eqnarray}
Here we have used the value of $v_{y} = 6.5\ \mathrm{m\ s^{-1}}$ used by the GRANIT experiment group. Using this value of $\Delta E$ in the time energy uncertainty relation $\Delta E \Delta t \geq \hbar$, we find 
\begin{eqnarray}
\Delta t \geq \frac{\hbar }{\Delta E} = 3.38 \times 10^{-4}\ s
\label{bert_scale3}
\end{eqnarray}
Hence uncertainty in time-space sector can be calculated using the results of \cite{bert0} as
\begin{eqnarray}
\Delta x \ \Delta t \sim  3.38 \times 10^{-18}\ m\ s
\label{bert_scale4}
\end{eqnarray}
where following \cite{bert0} we have taken $\Delta x \simeq 10^{-15} \ \mathrm{m}$.
On the other hand in the present paper we have derived the upper bound on the parameter $\eta$ as 
\begin{eqnarray}
\eta = - i \ \left[x^{1}, x^{0}\right] & \lesssim & 2.843\times 10^{-9}\ \mathrm{m^{2}}
\label{bert_scale5}
\end{eqnarray}
Restoring the $c$-factor in (\ref{bert_scale5}) we write the commutator in terms of $x$ and $t$ variables
\begin{eqnarray}
 - i \ \left[x , t \right] = \frac{\eta}{c} = & \lesssim & 9.51 
 \times 10^{-18} \mathrm{m \ s}
\label{bert_scale6}
\end{eqnarray}
Using the generalised uncertainty theorem \cite{jjs} for the commutation relation in (\ref{bert_scale6})
we can write 
\begin{eqnarray}
 \Delta x \ \Delta t \geq \frac{1}{2}\frac{\eta}{c} \sim 4.75 \times 10^{-18} \mathrm{m \ s}
\label{bert_scale7}
\end{eqnarray}
Interestingly the value of the upper bound on the time-space NC parameter as derived here turned out to be consistent with the results of \cite{bert0, bert1, RB}. However, one should keep in mind that this value is only in the sense of an upper bound and not the value of the parameter itself.
\clearpage


\pagestyle{myheadings}
\markright{Bibliography}

\end{document}